\documentclass[11pt,a4paper,twoside]{report}

\usepackage{graphicx}
\usepackage{verbatim}
\usepackage{enumerate}
\usepackage[width=.8\textwidth,font=small,format=plain,labelfont=bf,up]{caption}
\usepackage{multirow}
\usepackage{dsfont}
\usepackage{amssymb}
\usepackage{slashed}
\usepackage{mathtools}
\usepackage{nicefrac}
\usepackage{axodraw}
\usepackage{parskip, setspace, fullpage}
\usepackage{appendix}
\usepackage[a4paper,left=1in,right=1in,top=1in,bindingoffset=0.7in,bottom=1.2in]{geometry}

\raggedright
\newcommand{\ie}{{i.e.}}

\newcommand{\rr}[1]{\mathrm{#1}}
\newcommand{\cc}[1]{\mathcal{#1}}
\newcommand{\bb}[1]{\mathbf{#1}}

\newcommand{\dd}[1]{\mathds{#1}}
\newcommand{\kk}[1]{\mathfrak{#1}}

\newcommand{\f}[2]{\dfrac{#1}{#2}}
\newcommand{\nf}[2]{\nicefrac{#1}{#2}}

\newcommand{\uz}[1]{^{#1}}
\newcommand{\dz}[1]{_{#1}}
\newcommand{\ud}[2]{^{#1}_{#2}}
\newcommand{\dvp}[2]{^{\vphantom{#1}}_{#2}}
\newcommand{\uvp}[2]{^{#1}_{\vphantom{#2}}}

\newcommand{\uu}[1]{\underline{\smash[b]{#1}}}

\newcommand{\be}{\begin{eqnarray}}
\newcommand{\nn}{\nonumber}
\newcommand{\ee}{\end{eqnarray}}
\newcommand{\ba}{\begin{array}}
\newcommand{\ea}{\end{array}}
\newcommand{\bea}{\begin{equation}\begin{array}}
\newcommand{\eea}{\end{array}\end{equation}}

\newcommand{\E}{E\dz{6}}
\newcommand{\EZ}{E\dz{6}\dd{Z}\ud{S}{2}}
\newcommand{\eff}{{\rr{eff}}}
\newcommand{\eq}{{\rr{eq}}}
\newcommand{\GSM}{G\dz{\rr{SM}}}

\newcommand{\NeffZ}{N\ud{\rr{LEP}}{\eff}}
\newcommand{\NeffZexp}{N\ud{\rr{LEP}\,\smash{\rr{exp}}}{\eff}}
\newcommand{\Neff}{N\dz{\eff}}
\newcommand{\rhoCrit}{\rho\dz{\rr{c}}}

\newcommand{\ct}[1]{\rr{c}\dz{#1}}
\newcommand{\st}[1]{\rr{s}\dz{#1}}
\newcommand{\mass}[1]{m_{#1}}
\newcommand{\masss}[2]{m_{#2}^{#1}}
\newcommand{\ZZ}[1]{\dd{Z}\ud{#1}{2}}

\newcommand{\DM}{\rr{DM}}

\newcommand{\Yukawa}{\rr{Yukawa}}

\newcommand{\tr}{\rr{tr}}

\begin{document}

\pagestyle{empty}
\begin{center}
{\Large\bf UNIVERSITY OF SOUTHAMPTON}\\[3mm]
{\Large FACULTY OF PHYSICAL AND APPLIED SCIENCES\\[3mm]
School of Physics and Astronomy\\[5mm]
\textit{Southampton High Energy Physics Group}}\\[30mm]
{\Huge\bf Phenomenological Aspects\\[5mm]
of the $\E$SSM}\\[8mm]
{\huge\it by}\\[8mm]
{\huge\bf Jonathan Peter Hall}\\[30mm]
{\Large\textit{Presented for the degree of}\\[3mm]
\textbf{Doctor of Philosophy}}\\[20mm]
\vfill
November 2011
\end{center}
\cleardoublepage

\setcounter{page}{1}
\pagenumbering{roman}
\pagestyle{plain}
\onehalfspacing
\setlength{\parindent}{5mm}
\begin{center}
{\Large UNIVERSITY OF SOUTHAMPTON}\\
\vspace{\baselineskip}
\underline{\Large\bf ABSTRACT}\\
FACULTY OF PHYSICAL AND APPLIED SCIENCES\\
SCHOOL OF PHYSICS AND ASTRONOMY\\
\vspace{\baselineskip}
\underline{Doctor of Philosophy}\\
\texttt{PHENOMENOLOGICAL ASPECTS OF THE E(6)S.S.M.}\\
by Jonathan Peter Hall
\end{center}
\vspace{\baselineskip}

The work in this thesis explores various phenomenological aspects of the
$\E$SSM with a particular focus on the inert neutralino sector of the model
and on the dark matter implications. The $\E$SSM is a string theory inspired supersymmetric
extension to the Standard Model with an $\E$ grand unification group. The model provides a solution to
the hierarchy problem of the Standard Model, provides an explanation for neutrino mass,
and has automatic gauge anomaly cancellation.

The inert neutralino sector of the $\E$SSM and the dark matter that naturally arises from
this sector is studied for the first time. Limits on the parameter space from experimental and cosmological observations relating to
the inert neutralino dark matter are explored and the consequences for Higgs boson phenomenology are investigated.
In plausible scenarios it is found that the couplings of the lightest
inert neutralinos to the SM-like Higgs boson are always rather large. This has major implications for
Higgs boson collider phenomenology and leads to large spin-independent LSP-nucleon cross-sections.
Because of the latter, scenarios in which $\E$SSM inert neutralinos account for all of the observed dark matter are
now severely challenged by recent dark matter direct detection experiment analyses.
In plausible scenarios consistent with observations from both cosmology and LEP
the lightest inert neutralino is required to have a mass around half of the $Z$ boson mass if it contributes to
cold dark matter and this means that $\tan(\beta)$ cannot be too large.

A new variant of the $\E$SSM called the $\EZ$SSM is also presented in which the dark matter scenario
is very different to the inert neutralino cold dark matter scenario and in which the presence of
supersymmetric massless states in the early universe modifies the expansion rate of the universe prior to Big Bang Nucleosynthesis.
The new dark matter scenario is consistent with current observations and the modified expansion rate provides a better explanation
of various data than the SM prediction. The prospects for a warm dark matter scenario
in the $\E$SSM are also briefly discussed.

\cleardoublepage

\doublespacing
\setlength{\parindent}{0mm}
\phantom{~}
\vfill
\begin{center}
\textit{Dedicated to my friends and family\\
and to those whom we have lost.}
\end{center}
\vfill
\vfill
\cleardoublepage

\tableofcontents
\cleardoublepage
\addcontentsline{toc}{chapter}{\numberline{}List of Figures}
\listoffigures
\cleardoublepage
\addcontentsline{toc}{chapter}{\numberline{}List of Tables}
\listoftables
\cleardoublepage
\chapter*{Author's Declaration}
\addcontentsline{toc}{chapter}{\numberline{}Author's Declaration}

I declare that this thesis
has been composed by myself and constitutes work completed by myself
wholly while I was in candidature for a research degree at the University of Southampton.
Where the published work of others has been consulted or quoted from this is always clearly attributed
and all main sources of help have been acknowledged.\\
\vspace{.5\baselineskip}
I make no claim of originality for the work in chapters~\ref{chap:sm}--\ref{ref:dm} and appendices~\ref{ap:spinors} and~\ref{ap:pseudoreality}.
These chapters present background information compiled from a variety of sources that have been referenced in the text.
Chapter~\ref{chap:ndmin} contains work that was previously published in \textbf{paper~I}.
This work was carried by myself out under the supervision of my Ph.D. supervisor Steve King.
Chapter~\ref{chap:nhde} contains work that was previously published in \textbf{paper~II}.
This work was a collaborative effort between the authors. I was directly responsible for the calculation
of the dark matter relic density using \texttt{micrOMEGAs} and for the generation of
the benchmark points in tables~\ref{tab:tab1} and~\ref{tab:tab2}
and of the plots in figure~\ref{fig:fig2} and directly worked on
the RG and direct detection analyses.
Chapter~\ref{chap:ezssm} contains work that was previously published in \textbf{paper~III}
with the exception of section~\ref{ref:wdm} which contains work that is original to this thesis.
This work was carried out by myself under the supervision of Steve King.

\vspace{2\baselineskip}

\textbf{Signed:}

\vspace{\baselineskip}

\textbf{Date:}

\clearpage
\begin{tabular}{p{0.15\linewidth}p{0.75\linewidth}}
\textbf{Paper~I}   & J. P. Hall and S. F. King,
                     \textit{Neutralino dark matter with inert}\\&\textit{higgsinos and singlinos},
                     \textit{Journal of High Energy Physics} \textbf{2009}\\&(Aug., 2009) 088–088 [arXiv/0905.2696].\hfill\cite{Hall2009}\\
\textbf{Paper~II}  & J. P. Hall, S. F. King, R. Nevzorov, S. Pakvasa and M. Sher,\\&
                     \textit{Novel Higgs decays and dark matter in the exceptional}\\&\textit{supersymmetric standard model},
                     \textit{Physical Review D} \textbf{83} (Apr.,\\&2011) 39 [\texttt{arXiv/1012.5114}].\hfill\cite{Hall2010}\\
\textbf{Paper~III} & J. P. Hall and S. F. King,
                     \textit{Bino dark matter and big bang}\\&\textit{nucleosynthesis in the constrained E6SSM with massless inert}\\&\textit{singlinos},
                     \textit{Journal of High Energy Physics} \textbf{2011} (June, 2011)\\&24 [\texttt{arXiv/1104.2259}].\hfill\cite{Hall2011}
\end{tabular}

\cleardoublepage

\chapter*{Acknowledgements}
\addcontentsline{toc}{chapter}{\numberline{}Acknowledgements}

First of all I would like to thank my supervisor Steve King for his knowledge, time, encouragement, and motivation.\\
\vspace{.5\baselineskip}
I am grateful to Jonathan Roberts for his help with
the writing of {\tt LanHEP} code in 2008.
The {\tt LanHEP} codes used for the work detailed in this thesis are extended from his
USSM {\tt LanHEP} code which was used for the study in ref.~\cite{Kalinowski2009}
and I would like to thank Jonathan Roberts and Jan Kalinowski for donating this code and
for critically reading a draft of \textbf{paper~I}.
I would like to thank Peter Athron for donating his c$\E$SSM RG code which was
used for the study in ref.~\cite{Athron2009}.\\
\vspace{.5\baselineskip}
I would also like to thank Alexander Belyaev for valuable discussions.\\
\vspace{.5\baselineskip}
I am thankful to the STFC for providing studentship funding.\\
\vspace{.5\baselineskip}
I am also thankful to David Miller and Stefano Moretti for suggesting corrections to this thesis.\\
\clearpage
I would like to thank everyone in the group for making my time at Southampton such an enjoyable part of my life.
I am glad to have known you all.
I would particularly like to thank James French and Thomas Rae for all of the support and comradeship
over the years; Colin Whaymand, Matthew Brown, and James Lyon
for being such great housemates; George Weatherill for amusing me greatly;
and Iain Cooper, Jason Hammett, Shane Drury, and Thomas Rae again for
the all of the many enjoyable lunches and games of Hobo Blackjack.
I would also like to thank Jad Marouche and Iain Cooper again for the fun times in
California and Germany respectively.\\
\vspace{.5\baselineskip}
I would like to thank my dad and brother for their constant support
and my grandparents for giving me a quiet place to live while writing up much of this thesis.\\
\vspace{.5\baselineskip}
Finally, I would like to thank my secondary school maths teacher Trevor Phillips for
teaching me in his own time and for the inspiration and encouragement.

\cleardoublepage

\chapter*{Abbreviations and Conventions}
\addcontentsline{toc}{chapter}{\numberline{}Abbreviations and Conventions}
\begin{center}
\begin{tabular}{p{0.15\linewidth}p{0.75\linewidth}}
ADM & Asymmetric Dark Matter\\
BBN & Big Bang Nucleosynthesis\\
CDM & Cold Dark Matter\\
CKM & Cabibbo-Kobayashi-Maskawa\\
CMB & Cosmic Microwave Background\\
DMP & Dark Matter Particle\\
EM & ElectroMagnetism \\
EWSB & ElectroWeak Symmetry Breaking\\
FCCCs & Flavour Changing Charged Currents\\
FCNCs & Flavour Changing Neutral Currents\\
GUT & Grand Unified Theory\\
GWS & Glashow-Weinberg-Salam\\
LEP & Large Electron-Positron (collider)\\
LH & Left-Handed\\
LHC & Large Hadron Collider\\
LSP & Lightest Supersymmetric Particle\\
NLSP & Next-to-Lightest Supersymmetric Particle\\
PQ & Peccei-Quinn\\
PMNS & Pontecorvo-Maki-Nakagawa-Sakata\\
QCD & Quantum ChromoDynamics\\
QED & Quantum ElectroDynamics\\
QFT & Quantum Field Theory\\
RG & Renormalisation Group\\
\end{tabular}
\end{center}
\vfill
\pagebreak
\begin{center}
\begin{tabular}{p{0.15\linewidth}p{0.75\linewidth}}
RGEs & Renormalisation Group Equations\\
RH & Right-Handed\\
SSB & Soft Supersymmetry Breaking\\
VEV & Vacuum Expectation Value\\
WDM & Warm Dark Matter\\
WMAP & Wilkinson Microwave Anisotropy Probe\\
\vphantom{A}\\
\hline
\vphantom{A}\\
SM & Standard Model (of particle physics)\\
SSM & Supersymmetric (extension to the) SM\\
MSSM & Minimal SSM\\
NMSSM & Next-to-Minimal (SM-singlet extended) SSM\\
USSM & $U(1)'$ extended SSM (NMSSM with a gauged PQ symmetry)\\
$\E$SSM & Exceptional SSM ($\E$ grand unified SSM)\\
$\EZ$SSM & $\E$SSM with massless inert singlinos\\
\vphantom{A}\\
\hline
\vphantom{A}\\
cMSSM & GUT scale constrained MSSM\\
c$\E$SSM & GUT scale constrained $\E$SSM\\
c$\EZ$SSM & GUT scale constrained $\EZ$SSM\\
\vphantom{A}\\
\hline
\end{tabular}
\end{center}
\raggedright
\vspace{\baselineskip}
We work in the natural system of units throughout where what is written is what is meant multiplied
by factors of $c$ and $\bar{\vphantom{a}\smash{h}}$ until it has the dimensions displayed.
We consistently refer to the Lagrangian density as the Lagrangian.
The conventions for spinor notation are found in appendix~\ref{ap:spinors}.

\cleardoublepage

\setcounter{page}{0}
\clearpage
\pagenumbering{arabic}
\setcounter{page}{1}
\setlength{\parindent}{5mm}
\newpage
\chapter*{Overview}
\addcontentsline{toc}{chapter}{\numberline{}Overview}

In chapter~\ref{chap:sm} we present an introduction to the SM
with a focus on EWSB and other aspects relevant for subsequent chapters such as
gauge anomaly cancellation and the invisible decay width of the $Z$ boson.
Motivations for extensions of
the SM such as the hierarchy problem and neutrino mass are explored and various notation is fixed.

In chapter~\ref{chap:super} we motivate TeV scale softly broken supersymmetry as a possible extension to
the SM. A summary of supersymmetric Lagrangians is presented and
various notation is fixed. Further concepts such as grand unification and universality of
soft mass parameters are introduced.

In chapter~\ref{chap:essm} the $\E$SSM is motivated and introduced. This chapter contains
previous work that has been carried out on the subject of the $\E$SSM and provides background information relevant for
chapters~\ref{chap:ndmin}, \ref{chap:nhde}, and~\ref{chap:ezssm}.

In chapter~\ref{ref:dm} the subject of dark matter is introduced. Information about the thermal dark matter
relic density calculation relevant for chapters~\ref{chap:ndmin}, \ref{chap:nhde}, and~\ref{chap:ezssm}
is provided. We also provide an introduction to thermal relic dark matter in supersymmetric models.

Chapter~\ref{chap:ndmin} contains work that was first published in \textbf{paper I}. This work represents
a first study of the inert neutralino sector of the $\E$SSM and the dark matter that naturally arises from
this sector.

Chapter~\ref{chap:nhde} contains work that was first published in \textbf{paper II}. This work represents
a more in-depth study of the inert neutralino and chargino sectors with a particular focus on physics
relating to the Higgs boson. In plausible scenarios it is found that the couplings of the lightest
inert neutralinos to the SM-like Higgs boson are always rather large. This has major implications for
Higgs boson collider phenomenology and leads to large spin-independent LSP-nucleon cross-sections.
Because of the latter, scenarios in which $\E$SSM inert neutralinos account for all of the observed dark matter are
now severely challenged by recent dark matter direct detection experiment analyses.
In plausible scenarios consistent with observations from both cosmology and LEP
the lightest inert neutralino is required to have a mass around half of the $Z$ boson mass if it contributes to
cold dark matter and this means that $\tan(\beta)$ cannot be too large.

Chapter~\ref{chap:ezssm} contains work that was first published in \textbf{paper III}
with the exception of section~\ref{ref:wdm} which contains work that is original to this thesis.
In this chapter a new variant of the $\E$SSM called the $\EZ$SSM is presented in which the dark matter scenario
is very different to the inert neutralino CDM scenario and in which the presence of
supersymmetric massless states in the early universe modifies the expansion rate of the universe prior to BBN.
The dark matter scenario is consistent with current observations and the modified expansion rate provides a better explanation
of various data than the SM prediction. In section~\ref{ref:wdm} the prospects for a warm dark matter scenario
in the $\E$SSM are briefly discussed.

Summary and conclusions are found in sections~\ref{ndmin:conclusions}, \ref{novel:conclusions}, and~\ref{ezssm:conclusions}
and in chapter~\ref{chap:conclusions}.

Notation relating to Weyl, Majorana, and Dirac Spinors and to the doublet representation of $SU(2)$ is fixed in
appendices~\ref{ap:spinors} and~\ref{ap:pseudoreality}.

\cleardoublepage

\newpage
\chapter{The Standard Model}
\label{chap:sm}

The SM is an effective QFT
describing the known particles and their interactions with the known forces of
Nature excluding gravity. It is not currently known how to construct a
consistent theory that unifies quantum field theory with our current best
understanding of gravity which is the classical theory of general relativity. One
candidate for the fully quantum description of gravity describing Nature is
string theory, but whatever the correct description
corrections due to the effects of quantum gravity are not expected to become
relevant unless the energies involved in a process approach the Planck scale
$M\dvp{1}{\rr{P}}\sim 10\uz{18}$~GeV, or alternatively unless one wishes to consider length
or time intervals as small as $M\ud{-1}{\rr{P}}$. We therefore expect to be able to
use QFT, neglecting the effects of quantum gravity, at energies
far below the Planck scale. Whilst the general
framework of QFT is not expected to be valid above the Planck scale, the SM
is itself only an effective QFT and is expected only to be valid below roughly the TeV
scale~--- the energy scale currently being probed at the LHC. The reasons for this
are outlined in section~\ref{ref:hierarchy}.

The SM contains our current best understanding of the observed particles and
forces excluding gravity. The
observed mesons and baryons that we observe are bound states of SM quarks,
which are charged under the strong nuclear force described by QCD and there are
also SM leptons which are free fundamental particles
such as the electron. In terms of forces, the SM comprises the strong force of
QCD as well the GWS theory of EWSB
describing both QED and the weak force
responsible for nuclear decay. The description of the SM given in this chapter is
largely based on the one given in ref.~\cite{Peskin1995}.

\section{Gauge Symmetry and Matter Content}
\label{ref:ghajsdg}

The SM is a Yang-Mills QFT with a gauge symmetry
group
\be
\GSM &=& SU(3)\dz{c}\otimes SU(2)\dz{L}\otimes U(1)\dz{Y}.
\ee
Is it is a direct product of the $SU(3)$ gauge
symmetry describing QCD and the $SU(2)\times U(1)$ gauge symmetry of
the electroweak theory~--- the unified theory describing both electromagnetism (QED)
and the weak nuclear force. The observed fermionic matter of the SM can be
described by LH Weyl spinors in $3 + 1$ dimensions forming chiral
representations of $\GSM$ as shown in table~\ref{tab:sMCharges}. A RH
Weyl spinor may be expressed as a LH one using the $CP$ conjugation (charge conjugation and parity)
operation. The notation for spinors used throughout is explained in
appendix~\ref{ap:spinors}.

The SM also includes a fundamental complex scalar doublet field known as the
Higgs doublet whose VEV is responsible for the spontaneous breaking of
$SU(2)\dz{L}\times U(1)\dz{Y}$ down to the $U(1)\dz{\rr{EM}}$ of QED, as per
the GWS theory of EWSB~\cite{Glashow1961,Weinberg1967,Glashow1970},
and for the generation of fermion masses. Although the
evidence for ESWB is overwhelming (for a review see ref.~\cite{ParticleDataGroupCollaboration2010}),
the mechanism for this
symmetry breaking is currently unknown, although it must have the 
correct custodial symmetry leading to the observed mass relation between the
heavy electroweak $W$ and $Z$ gauge bosons. The SM assumes the GWS theory in
which electroweak symmetry is spontaneously broken by the VEV of a
fundamental complex scalar field~--- the Higgs doublet
$H$~\cite{Englert1964,Higgs1964,Guralnik1964,Higgs1966}.
The VEV of this field is also able
to generate masses for all of the SM fermions.

\begin{table}
\begin{center}
\begin{tabular}{cc|ccc|}
&& $SU(3)\dz{c}$ & $SU(2)\dz{L}$ & $U(1)\dz{Y}$ \\\hline
LH quark doublet & $Q\dvp{c}{L}$ & $3$ & $2$ & $+\nf{1}{6}$\\
RH down-type quark & $d\ud{c}{R}$ & $\bar{3}$ & $1$ & $+\nf{1}{3}$\\
RH up-type quark & $u\ud{c}{R}$ & $\bar{3}$ & $1$ & $-\nf{2}{3}$\\
LH lepton doublet & $L\dvp{c}{L}$ & $1$ & $2$ & $-\nf{1}{2}$\\
RH charged lepton & $e\ud{c}{R}$ & $1$ & $1$ & $+1$\\\hline
Higgs doublet & $H$ & $1$ & $2$ & $+\nf{1}{2}$\\\hline
\end{tabular}
\caption{The $SU(3)\dz{c}$ and $SU(2)\dz{L}$ representations
and the $U(1)\dz{Y}$ charges
(hypercharges) of the SM matter fields, as LH Weyl spinors, and of the SM Higgs
doublet $H$.\label{tab:sMCharges}}
\end{center}
\end{table}

The three components of the fundamental ($3$) and antifundamental ($\overline{3}$)
antitriplets of $SU(3)\dz{c}$ are known as colours (red, green, and blue) and
anticolours (antired, antigreen, and antiblue) respectively.
Since the EWSB vacuum respects
$SU(3)\dz{c}$, redefinitions of the three colours and three anticolours by
$SU(3)\dz{c}$ transformations does not change the description of the
physics. The $SU(2)\dz{L}$ gauge symmetry, however, is
spontaneously broken by the vacuum and it makes sense to label the
components separately. We define the third direction of weak isospin $T\uz{3}$
such that the Higgs VEV is an eigenstate of $\tau\uz{3}$ with eigenvalue
$-\nf{1}{2}$. Here we use $T\uz{a}$ for generators of a
general $SU(2)_L$ representation and $\tau\uz{a}$ for the generators of the 2
representation specifically.
Since the direction in $SU(2)\dz{L}$ space of the Higgs VEV
defines which direction will be uncharged under the unbroken $U(1)\dz{\rr{EM}}$,
electric charge will then commute with
$T\uz{3}$. This choice of direction for the Higgs VEV is arbitrary and has no
effect on the physics, since any other equivalent choice would be related by a
$SU(2)\dz{L}$ gauge transformation that leaves the Lagrangian invariant.
Choosing the Higgs VEV to be an eigenstate of $\tau\uz{3}$, the upper and
lower components of the doublet in the standard basis
$\tau\uz{a} = \sigma\uz{a}/2$ are then eigenstates of electric charge.
We write the quark doublet $Q\dz{L}$ and lepton doublet $L\dz{L}$ as
\be
Q\dz{L} = \left(\ba{c}u\dz{L}\\d\dz{L}\ea\right) & \mbox{\quad and\quad} &
L\dz{L} = \left(\ba{c}\nu\\e\dz{L}\ea\right).
\ee
The upper component is an eigenstate of $\tau\uz{3}$ with eigenvalue
$+\nf{1}{2}$ and the lower component an eigenstate of $\tau\uz{3}$ with
eigenvalue $-\nf{1}{2}$. The charge under the unbroken $U(1)\dz{\rr{EM}}$ of a
field can be written
\be
Q &=& T\uz{3} + Y,
\ee
where $T\uz{3}$ is understood to stand for the relevant eigenvalue. This is
because it is the gauge transformation with this combination of generators
$H \rightarrow (1+i\rr{d}\alpha(T\uz{3} + Y))H$ that leaves the Higgs VEV
$\langle H\rangle$ invariant, \ie~$\langle H\rangle$ is uncharged under this
combination of generators which must then represent the unbroken $U(1)$.

With one copy of each of the fields listed in table~\ref{tab:sMCharges}
we can describe what is known as the first generation of SM matter.
This comprises the strongly interacting up and down quarks~--- two
colour triplet Dirac fermions that are formed from the four Weyl spinor colour
triplets of one copy of $Q\dz{L}$, $u\dz{R}$, and $d\dz{R}$~--- and also the
electron~--- a Dirac fermion formed from $e\dz{L}$ and $e\dz{R}$~--- and a LH
neutrino $\nu$, or equivalently the RH antineutrino $\nu\uz{c}$.

The Lagrangian of the SM for this first generation, including all possible
renormalisable, gauge invariant, and Lorentz invariant 
terms\footnote{We do not address the strong $CP$ problem and will
consistently neglect terms of the form
$\tilde{\cc{A}}\uz{\mu\nu}\cc{A}\dz{\mu\nu}$. The contribution
to this term from electroweak gauge bosons is always a total derivative
and has no effect on the observable physics. However, the contribution to this
term from QCD eventually, after chiral matter phase rotations removing 5 of the
6 arbitrary complex phases of CKM matrix, has an independent and arbitrary
coefficient.
This coefficient should be very close to zero in order for the theory to be
consistent with the non-observation of $CP$-violating effects from the QCD
sector, such as an electric dipole moment for the neutron, but theoretically
the smallness of this coefficient is not explained in the SM in a natural
way. This is known as the strong $CP$ problem~--- an unsolved problem in particle
physics.} is
\be
\cc{L} &=& -\f{1}{4}\cc{A}\uz{a\mu\nu}\cc{A}\ud{a}{\phantom{a}\mu\nu}
+ \psi\ud{\dagger}{i}i\bar{\sigma}\uz{\mu}\cc{D}\dz{\mu}\psi\dvp{\dagger}{i}
+ \cc{L}\dz{\rr{Yukawa}} + \cc{L}\dz{\rr{Higgs}},
\label{eq:sMLagrangian}\ee
where
\be
\cc{L}\dz{\rr{Yukawa}} &=& -h\uz{D}d\ud{\dagger}{R}H\uz{\dagger}Q\dvp{\dagger}{L}
-h\uz{U}u\ud{\dagger}{R}H.Q\dvp{\dagger}{L}
-h\uz{L}e\ud{\dagger}{R}H\uz{\dagger}L\dvp{\dagger}{L} + \mbox{c.c.}
\label{eq:LYukawa}
\ee
and $\cc{L}\dz{\rr{Higgs}}$ contains the gauge invariant kinetic term and
scalar potential of the Higgs scalar field shown in section~\ref{ref:Higgs}.
$\psi\ud{\dagger}{i}i\bar{\sigma}\uz{\mu}\cc{D}\dz{\mu}\psi\dvp{\dagger}{i}$
is the gauge invariant kinetic term
for all LH Weyl spinors $\psi\dz{i}$, with $\cc{D}\dz{\mu}$ the relevant gauge
covariant derivative for each field $\psi\dz{i}$. For the gauge kinetic
term
\be
\cc{A}\ud{a}{\mu\nu} &=& \partial\dvp{a}{\mu}A\ud{a}{\nu} - \partial\dvp{a}{\nu}A\ud{a}{\mu} + g\uz{(a)}f\uz{abc}A\ud{b}{\mu}A\ud{c}{\nu}
\ee
and the adjoint index $a$ runs over all of the generators of $\GSM$. The gauge coupling
$g\uz{(a)}$ can have a different value for each of the three simple subgroups of
$\GSM$ and the gauge group structure function $f\uz{abc}$ vanishes when $a$, $b$, and $c$ do
not belong to the same simple subgroup.
The dot stands for the $SU(2)$ invariant contraction of two $SU(2)$ doublets
given in (\ref{eq:pseudoreality})
\be
\left(\ba{c}\uparrow\dz{1} \\ \downarrow\dz{1}\ea\right).
\left(\ba{c}\uparrow\dz{2} \\ \downarrow\dz{2}\ea\right)
&=& \downarrow\dz{1}\uparrow\dz{2} - \uparrow\dz{1}\downarrow\dz{2}.
\ee

The vacuum state of the Higgs potential is supposed to spontaneously break
$SU(2)\dz{L}\times U(1)\dz{Y}$ so that classically expanding around the
true electroweak vacuum, rather than $H=0$, we can write
\be
H &=& \langle H\rangle + \phi,
\label{eq:ESBWVExpansion}\ee
where the Higgs VEV
\be
\langle H\rangle &=& \f{1}{\sqrt{2}}\left(\ba{c}0 \\ v\ea\right)
\label{eq:higgsVEV}\ee
and is an eigenstate of
$\tau\uz{3}$ with eigenvalue $-\nf{1}{2}$ ($Q=0$).

It is important to note that the unbroken gauge symmetry of the SM does not
allow for any fermion mass terms~--- either Dirac or Majorana. However, in the
EWSB vacuum the VEV of the Higgs doublet
will generate Dirac fermion mass terms proportional to the Yukawa
couplings $h$ in (\ref{eq:LYukawa}) for all fermions other than the
LH neutrino
\be
\cc{L}\dz{\Yukawa} &=& \left(-\f{h\uz{D}v}{\sqrt{2}}d\ud{\dagger}{R}d\dvp{\dagger}{L}
-\f{h\uz{U}v}{\sqrt{2}}u\ud{\dagger}{R}u\dvp{\dagger}{L}
-\f{h\uz{L}v}{\sqrt{2}}e\ud{\dagger}{R}e\dvp{\dagger}{L} + \mbox{c.c.}\right) + \cdots.
\label{eq:1stGenerationMasses}\ee
In the SM the Higgs VEV generates Dirac masses for all of the observed Dirac
fermions, but does not induce any neutrino masses.

Below the EWSB scale, one can integrate out the $W\uz{\pm}$ and $Z$ electroweak gauge bosons
that acquire masses from EWSB and write an effective theory with the gauge symmetry
$SU(3)\dz{c}\times U(1)\dz{\rr{EM}}$.
The Lagrangian for one generation contains the EWSB-induced mass terms of
(\ref{eq:1stGenerationMasses}).
Each mass term couples a LH and RH spinor together into a massive Dirac state.
With both the LH and RH component of each Dirac spinor taken together,
each Dirac spinor forms a real representation of
$SU(3)\dz{c}\times U(1)\dz{\rr{EM}}$. The LH neutrino forms a real
representation on its own since it is a singlet~--- completely uncharged under the
effective gauge group. Since this effective theory is non-chiral,
containing pairs of LH and RH spinors that are equally charged under the
gauge group, it is invariant under parity $P$ and charge
conjugation $C$ separately. At low energy left- and right-handedness are
only distinguished fundamentally in weak nuclear decay processes
which violate separately both $C$ and $P$ maximally since the massive
$W\uz{\pm}$ bosons only couple to LH states.

\section{Gauge Anomaly Cancellation, Generations, and the Invisible $Z$ Boson
Decay Width as Measured at LEP}

In Nature we observe that there are in fact at least three complete copies,
known as generations, of the Weyl fields listed in table~\ref{tab:sMCharges}.
The particle
content of three complete generations has now been directly observed and there
exists evidence, outlined in this section, that indicates that there are no more
generations of any of the SM representations beyond these three.
We shall label these three generations with the Roman indices $i$, $j$, etc., with
the notation for the particles in these generations as complied in
table~\ref{tab:sMGenerations}.

\begin{table}
\begin{center}
\begin{tabular}{c|c|c|}
\hline
Basis & Flavour & Mass\\
\hline
LH down quark        & $d\ud{\prime}{L1}=d\ud{\prime}{L}$ & $d\dvp{\prime}{L1}=d\dvp{\prime}{L}$ \\
LH strange quark     & $d\ud{\prime}{L2}=s\ud{\prime}{L}$ & $d\dvp{\prime}{L2}=s\dvp{\prime}{L}$ \\
LH bottom quark      & $d\ud{\prime}{L3}=b\ud{\prime}{L}$ & $d\dvp{\prime}{L3}=b\dvp{\prime}{L}$ \\\hline
RH down quark        & \multicolumn{2}{|c|}{$d\dvp{\prime}{R1}=d\dvp{\prime}{R}$}  \\
RH strange quark     & \multicolumn{2}{|c|}{$d\dvp{\prime}{R2}=s\dvp{\prime}{R}$}  \\
RH bottom quark      & \multicolumn{2}{|c|}{$d\dvp{\prime}{R3}=b\dvp{\prime}{R}$}  \\\hline
LH up quark          & \multicolumn{2}{|c|}{$u\dvp{\prime}{L1}=u\dvp{\prime}{L}$}  \\
LH charm quark       & \multicolumn{2}{|c|}{$u\dvp{\prime}{L2}=c\dvp{\prime}{L}$}  \\
LH top quark         & \multicolumn{2}{|c|}{$u\dvp{\prime}{L3}=t\dvp{\prime}{L}$}  \\\hline
RH up quark          & \multicolumn{2}{|c|}{$u\dvp{\prime}{R1}=u\dvp{\prime}{R}$}  \\
RH charm quark       & \multicolumn{2}{|c|}{$u\dvp{\prime}{R2}=c\dvp{\prime}{R}$}  \\
RH top quark         & \multicolumn{2}{|c|}{$u\dvp{\prime}{R3}=t\dvp{\prime}{R}$}  \\\hline
LH electron          & \multicolumn{2}{|c|}{$e\dvp{\prime}{L1}=e\dvp{\prime}{L}$}  \\
LH muon              & \multicolumn{2}{|c|}{$e\dvp{\prime}{L2}=\mu\dvp{\prime}{L}$}\\
LH tau lepton        & \multicolumn{2}{|c|}{$e\dvp{\prime}{L3}=\tau\dvp{\prime}{L}$}\\\hline
RH electron          & \multicolumn{2}{|c|}{$e\dvp{\prime}{R1}=e\dvp{\prime}{R}$}  \\
RH muon              & \multicolumn{2}{|c|}{$e\dvp{\prime}{R2}=\mu\dvp{\prime}{R}$}\\
RH tau lepton        & \multicolumn{2}{|c|}{$e\dvp{\prime}{R3}=\tau\dvp{\prime}{R}$}\\\hline
LH electron neutrino & $\nu\ud{\prime}{1}=\nu\dvp{\prime}{e}$ & \\
LH muon neutrino & $\nu\ud{\prime}{2}=\nu\dvp{\prime}{\mu}$ & \\
LH tau neutrino & $\nu\ud{\prime}{3}=\nu\dvp{\prime}{\tau}$ & \\
Light neutrino 1 & & $\nu\dvp{\prime}{1}$ \\
Light neutrino 2 & & $\nu\dvp{\prime}{2}$ \\
Light neutrino 3 & & $\nu\dvp{\prime}{3}$ \\\hline
\end{tabular}
\caption{The notation for the three generations of fermionic matter of the SM.
Where the flavour and mass eigenstate columns are combined the flavour and mass
eigenstates are equal by definition.
In the down quark sector the mass eigenstates are then rotated
with respect to the flavour eigenstates by the CKM matrix.
In the neutrino sector the mass eigenstates are rotated with respect to the
flavour eigenstates by the PMNS matrix, which is analogous to the CKM matrix
of the quark sector. The CKM matrix is relatively close to the identity, whereas
the PMNS is close to tribimaximal form, meaning that the
mass and flavour eigenstate bases are very different from each other.\label{tab:sMGenerations}}
\end{center}
\end{table}

\subsection{Gauge anomalies}

Anomalies are quantum mechanical effects that violate one or more
symmetries of the classical Lagrangian. A gauge anomaly is a quantum mechanical
effect that violates some gauge symmetry. QFTs with gauge anomalies are
inconsistent since gauge symmetry is required to cancel the unphysical degrees
of freedom of the massless gauge bosons~--- the longitudinal space-like and
time-like polarisations. In 4-dimensional QFTs such as the SM, gauge anomalies
arise at one-loop level via triangle diagrams of the form
\be
\left[
\begin{picture}(140,70)(-10,0)
\SetWidth{1}
\Photon(0,0)(40,0){2}{4}
\Line(-3,3)(3,-3)
\Line(-3,-3)(3,3)
\Line(40,0)(80,30)
\Line(40,0)(80,-30)
\Line(80,30)(80,-30)
\Photon(80,30)(120,30){2}{4}
\Line(117,33)(123,27)
\Line(117,27)(123,33)
\Photon(80,-30)(120,-30){-2}{4}
\Line(117,-33)(123,-27)
\Line(117,-27)(123,-33)
\Text(20,5)[b]{$a$}
\Text(100,35)[b]{$b$}
\Text(100,-25)[b]{$c$}
\end{picture}\right]
& \propto & \cc{A}\uz{abc} = \tr[T\uz{a}
\{T\uz{b},T\uz{c}\}],
\label{eq:anomalies}\ee
where $T\uz{a}$ is the group generator corresponding to the adjoint index $a$ of
the gauge boson labelled $a$. The trace is a sum over all LH Weyl spinors
running around the loop, each in some representation, and also a trace over the
generator indices of the relevant representation.
The anticommutator comes from considering each Weyl
fermion running around the loop in both directions. Such diagrams must sum to zero
for all combinations of different gauge boson external legs
in order for the QFT to be consistent.

Instead of any given RH spinor in the representation $r$, one may consider the
$CP$ conjugate state which is a LH spinor in the representation $\bar{r}$
(see appendix~\ref{ap:spinors}).
This gives a contribution to $\cc{A}\uz{abc}$ equal to $\tr[T\ud{a}{\bar{r}}
\{T\ud{b}{\bar{r}},T\ud{c}{\bar{r}}\}]$. Since
$T\ud{a}{r} = -T\ud{a\rr{T}}{\bar{r}}$ this contribution is in fact equal to
$-\tr[T\ud{a}{r}\{T\ud{b}{r},T\ud{c}{r}\}]$, which is minus the contribution
from a LH spinor in the representation $r$. So, if one has an equal number of
LH spinors in each of the representations $r$ and $\bar{r}$ of the entire
gauge group, or equivalently
an equal number of LH and RH spinors in the representation $r$, then these
states, which taken together from the real representation $r \oplus \bar{r}$,
do not contribute to the gauge anomaly. If the representation $r$ is explicitly
real then this condition is automatically satisfied. This is the case in the
$SU(3)\dz{c}\times U(1)\dz{\rr{EM}}$ effective theory.

However, since the SM is a chiral theory, meaning that the fermionic matter
cannot be written as above in terms of their representations under $\GSM$,
it takes some more work to compute
the gauge anomalies in (\ref{eq:anomalies}) associated with the various
combinations of SM gauge bosons. Eventually one computes that if one has a
complete generation of the chiral fermions listed in table~\ref{tab:sMCharges}
then all of the gauge anomalies do indeed cancel,
but this is not the case for just the leptons or for just the
quarks separately~\cite{Peskin1995}. We therefore conclude that the SM is gauge-anomaly-free
as long as it contains only complete generations of matter, \ie~it
contains the same number of generations of quarks and leptons.

\subsection{The effective number of neutrinos contributing to the invisible $Z$
boson decay width}

The best evidence for the number of generations comes from the number of
neutrinos as inferred from the invisible decay width of the $Z$ boson measured at LEP,
\ie~the partial decay width of the $Z$ boson into particles that do
not show up in the detector.
The effective number of neutrinos at LEP $\NeffZ$ is defined by
\be
\Gamma(Z\rightarrow\mbox{invisible}) &=&
\NeffZ\Gamma(Z\rightarrow\bar{\nu}\nu)
\label{eq:NeffZ}\ee
where the decay width on the left is measured and the decay width on the right
is calculated assuming that $\nu$ is a massless LH neutrino. $\bar{\nu}$ is the
corresponding RH antineutrino $\nu\uz{c}$. The result from LEP~\cite{TheALEPHCollaboration2006} is
\be
\NeffZ &=& 2.984 \pm 0.008\mbox{\quad(1-sigma)},\label{eq:NeffZ}
\ee
leading to the conclusion that the number of neutrinos is 3, this being the
closest integer to the central measured value and
2-sigma away. This means that there is no
fourth generation neutrino with a mass lower than about half of the $Z$ boson
mass. This is taken as evidence for there being only three generations of
leptons and, since the numbers of generations of quarks and leptons should be
equal in order to have gauge anomaly cancellation, only three generations of quarks
also.

\section{The Higgs Potential and GWS EWSB}
\label{ref:Higgs}

In the SM EWSB is caused by the non-zero VEV of a single Higgs scalar
doublet $H$. The most general gauge invariant and renormalisable form of
$\cc{L}\dz{\rr{Higgs}}$ appearing in (\ref{eq:sMLagrangian}) is
\be
\cc{L}\dz{\rr{Higgs}} &=& (\cc{D}\uz{\mu}H)\uz{\dagger}(\cc{D}\dz{\mu}H) - V(H),
\ee
with
\be
V(H) &=& m\uz{2}H\uz{\dagger}H + \lambda(H\uz{\dagger}H)\uz{2}.\label{eq:HiggsPot}
\ee
The parameter $\lambda$ must be positive in order for the Higgs potential to
be bounded from below.
If the mass-squared parameter $m\uz{2}$ is also positive, or zero, then
classically
$V(H)$ has a minimum at $H=0$. In this case $H=0$ is the true vacuum and
$\GSM$ remains unbroken. If, however, $m\uz{2}$ is negative, then the degenerate
minima of $V(H)$ occur on the surface given by
\be
H\uz{\dagger}H &=& \f{-m\uz{2}}{2\lambda},
\ee
points on which are related by arbitrary $SU(2)\dz{L}$ gauge transformations.
Using our earlier definition of the Higgs VEV (\ref{eq:higgsVEV}) we can then
identify
\be
v &=& \sqrt{\f{-m\uz{2}}{\lambda}}.
\ee
In this case in the EWSB vacuum $\GSM$ is spontaneously broken to
$SU(3)\dz{c}\times U(1)\dz{\rr{EM}}$. $SU(3)\dz{c}$ remains unbroken
since $H$ is a singlet under this group. Since it is a scalar particle,
it is also a singlet under the Lorentz group and its VEV therefore does not
spontaneously break Lorentz symmetry.

The covariant derivative acting on $H$ is given by
\be
\cc{D}\dz{\mu}H &=& (\partial\dz{\mu} - ig\dz{2}W\ud{a}{\mu}\tau\uvp{a}{\mu}
- \nf{1}{2}g\uz{\prime}B\dz{\mu})H
\ee
where $W\uz{a}$ are the three $SU(2)\dz{L}$ gauge bosons, $\tau\uz{a}$ are
the generators of $SU(2)\dz{L}$ ($T\uz{a}$) for the fundamental doublet
representation ($\tau\uz{a} = \sigma\uz{a}/2$), and $B\dz{\mu}$
is the gauge boson for $U(1)\dz{Y}$ under which $H$ has charge $+\nf{1}{2}$.
$g\dz{2}$ and $g\uz{\prime}$ are the $SU(2)\dz{L}$ and $U(1)\dz{Y}$ gauge
coupling constants respectively.
Expanding the kinetic term around the Higgs VEV as in (\ref{eq:ESBWVExpansion})
we find that
in the EWSB breaking vacuum we generate the gauge boson mass terms
\be
\cc{L}\dz{\rr{Higgs}} &=& \f{1}{2}\f{v\uz{2}}{4}\left[
g\ud{2}{2}(W\ud{1}{\mu})\uz{2} + g\ud{2}{2}(W\ud{1}{\mu})\uz{2}
+ \left(g\dz{2}W\ud{3}{\mu} - g\uz{\prime}B\dvp{3}{\mu}\right)\uz{2}\right]
+ \cdots.
\ee
It is useful to define the positively and negatively charged $SU(2)\dz{L}$
gauge bosons
\be
W\ud{\pm}{\mu} &=& \f{1}{\sqrt{2}}(W\ud{1}{\mu} \mp iW\ud{2}{\mu})
\ee
as well as the usual $T\uz{\pm} = T\uz{1} \pm iT\uz{2}$. In the EWSB vacuum
we read off that these particles have a mass
\be
m\dz{W} &=& g\dz{2}\f{v}{2}.
\ee
The (correctly normalised) mass eigenstate 
\be
Z\dz{\mu} &=& \f{1}{\bar{g}}
(g\dz{2}W\ud{3}{\mu} - g\uz{\prime}B\dvp{3}{\mu}),
\ee
where
\be
\bar{g} &=& \sqrt{g\ud{2}{2} + g\uvp{\prime 2}{2}},
\ee
has a mass
\be
m\dz{Z} &=& \bar{g}\f{v}{2},
\ee
leaving the orthogonal combination
\be
A\dz{\mu} &=& \f{1}{\bar{g}}
(g\uz{\prime}W\ud{3}{\mu} + g\dz{2}B\dvp{3}{\mu})
\ee
massless. This is the photon~--- the gauge boson of the unbroken
$U(1)\dz{\rr{EM}}$ which corresponds to the the unbroken combination of generators
$Q = T\uz{3} + Y$.

The general covariant derivative, neglecting the $SU(3)\dz{c}$ gluon terms,
\be
\cc{D}\dz{\mu} &=& \partial\dz{\mu} - ig\dz{2}W\ud{a}{\mu}T\uz{a}
- Yg\uz{\prime}B\dz{\mu}
\ee
can then be written in terms of the gauge boson mass eigenstates as
\be
\cc{D}\dz{\mu} &=& \partial\dz{\mu} - i\f{g\dz{2}}{\sqrt{2}}W\ud{+}{\mu}T\uz{+}
- i\f{g\dz{2}}{\sqrt{2}}W\ud{-}{\mu}T\uz{-}\nn\\
&& \vphantom{A} - i\f{g\ud{2}{2}T\uz{3} - g\uz{\prime 2}Y}
{\bar{g}}Z\dz{\mu}
- i\f{g\dvp{2}{2}g\uz{\prime}}{\bar{g}}
\Big(T\uz{3} + Y\Big)B\dz{\mu}\nn\\\nn\\
&=& \partial\dz{\mu} - i\f{g\dz{2}}{\sqrt{2}}W\ud{+}{\mu}T\uz{+}
- i\f{g\dz{2}}{\sqrt{2}}W\ud{-}{\mu}T\uz{-}\nn\\
&& \vphantom{A} - i\f{g\dz{2}}{\ct{W}}\Big(T\uz{3} - \rr{s}\ud{2}{W}Q\Big)Z\dz{\mu}
- ieQB\dz{\mu},\label{eq:covdermas}
\ee
where
\be
e &=& \f{g\dz{2}g\uz{\prime}}{\bar{g}},\nn\\\nn\\
\ct{W} \equiv \cos(\vartheta\dz{W}) &=&
\f{g\dz{2}}{\bar{g}}\mbox{,\quad and}\\\nn\\
\st{W} \equiv \sin(\vartheta\dz{W}) &=&
\f{g\uz{\prime}}{\bar{g}},\nn
\ee
implying that $m\dz{Z} = m\dz{W}/\ct{W}$.

In the unbroken theory the Higgs complex scalar doublet has four real degrees of
freedom. After EWSB the three massive gauge bosons each acquire one extra
degree of freedom from the complex scalar doublet, corresponding to the
longitudinal polarisation that exists for a massive vector boson, but not for a
massless one. The remaining one degree of freedom belongs to a real scalar,
known as the SM Higgs boson.

We can work in the unitarity gauge in which the three Goldstone
modes of the Higgs doublet are set to zero and we expand around the EWSB
vacuum
\be
H &=& \f{1}{\sqrt{2}}\left(\ba{c}0 \\ v + h\ea\right),
\ee
where $h$ is the (canonically normalised) real scalar known as the Higgs boson.
Since in the basis that we have chosen $v$ appears in the real part of the 
lower component of $H$, this is also the direction corresponding to the
massive boson state $h$. The other directions are flat and correspond to the
massless Goldstone modes whose degrees of freedom contribute to those of the
massive gauge bosons.

We can expand $V(H)$ in the unitarity gauge in order to find the mass of the
Higgs boson $h$. We find
\be
V(H) &=& \f{m\uz{2}}{2}h\uz{2} + \f{3\lambda v\uz{2}}{2}h\uz{2}
+ \cdots\nn\\\nn\\
&=& \f{1}{2}(-2m\uz{2})h\uz{2} + \cdots,
\ee
from which we read off a mass-squared for the real scalar $h$
\be
\mass{h}\uz{2} &=& -2m\uz{2} = 2\lambda v\uz{2}.\label{eq:higgsMassLambda}
\ee
Like the induced fermion masses, the mass of the Higgs boson itself is
proportional to the Higgs VEV $v$, but also to an unknown coupling constant
$\lambda$. The value of $v = 246$~GeV is determined from the masses of the
$W\uz{\pm}$ and $Z$ bosons, but although in the GWS theory this combination
of $m\uz{2} < 0$ and $\lambda > 0$ are determined, the individual values of
these parameters are not determined unless the Higgs boson mass is known. At the
time of writing the
Higgs boson is currently the only particle of the SM yet to be discovered.
By looking for the process $e^+e^- \rightarrow Zh$ at LEP, a lower limit
on the SM Higgs boson mass of 114.4~GeV is obtained~\cite{ParticleDataGroupCollaboration2010}.
Recent LHC analyses from CMS~\cite{cmsLp11Higgs} and ATLAS~\cite{atlasLp11Higgs1,atlasLp11Higgs2,atlasLp11Higgs3,atlasLp11Higgs4,atlasLp11Higgs5}
between them exclude the existence of a SM Higgs boson with a mass between 145 and 288~GeV or between 296 and 466~GeV at a 95\% confidence level.

\section{Induced Dirac Fermion Masses, the CKM Matrix, and Neutrino Mass}

Including all three generations of SM matter, (\ref{eq:1stGenerationMasses})
becomes
\be
\cc{L}\dz{\Yukawa} &=& -\f{v}{\sqrt{2}}\left(h\ud{D}{ij}d\ud{\dagger}{Ri}d\dvp{\dagger}{Lj}
+ h\ud{U}{ij}u\ud{\dagger}{Ri}u\dvp{\dagger}{Lj}
+ h\ud{E}{ij}e\ud{\dagger}{Ri}e\dvp{\dagger}{Lj} + \mbox{c.c.}\right) + \cdots.
\label{eq:3GenerationMasses}
\ee
The Yukawa coupling matrices $h\uz{D}$, $h\uz{U}$, and $h\uz{E}$ may be made diagonal
if one performs unitary transformations on the fermion fields in flavour space, \ie
\be
\psi\dz{i} &\rightarrow& U\dz{ij}\psi\dz{j}\label{eq:gentrans}
\ee
for each of the fields $d\dz{Ri}$, $d\dz{Li}$, $u\dz{Ri}$, $u\dz{Li}$, $e\dz{Ri}$, and $e\dz{Li}$.
Specifically for the LH quarks we write
\be
d\dvp{D}{Li} \rightarrow U\ud{D}{ij}d\dvp{D}{Lj} &\mbox{\quad and\quad}&
u\dvp{U}{Li} \rightarrow U\ud{U}{ij}u\dvp{U}{Lj}.\label{eq:gentransleft}
\ee
This is biunitary diagonalisation of each of the three Yukawa coupling matrices
and the basis where these matrices are diagonal is the mass eigenstate basis.
In the gauge invariant fermion kinetic term in the Lagrangian these transformations
leave everything invariant apart from the couplings of the fermions to the
heavy $W\uz{\pm}$ bosons coming from the covariant derivative (\ref{eq:covdermas}).
If one begins with non-diagonal Yukawa coupling matrices and then transforms to
the mass eigenstate basis that diagonalises them, the Lagrangian term coupling
quarks to $W\uz{\pm}$ bosons transforms
\be
u\ud{\dagger}{Li}i\bar{\sigma}\uz{\mu}\f{g\dz{2}}{\sqrt{2}}W\uz{+}d\dvp{\dagger}{Li} + \mbox{c.c.}
&\rightarrow&
u\ud{\dagger}{Li}i\bar{\sigma}\uz{\mu}\f{g\dz{2}}{\sqrt{2}}W\uz{+}V\dvp{\dagger}{ij}d\dvp{\dagger}{Lj} + \mbox{c.c.},\label{eq:RHS}
\ee
producing a non-diagonal, unitary flavour mixing matrix
\be
V &=& U\uz{U\dagger}U\uz{D\vphantom{\dagger}}
\ee
known as the CKM~\cite{Cabibbo1963,Kobayashi1973} matrix.

As well as the mass eigenstate basis we also define a
flavour eigenstate basis in which the couplings to the heavy $W\uz{\pm}$ are diagonal. By
convention we choose the up-quark flavour basis to be equal to the mass eigenstate
basis. The down-type quark flavour basis is then
\be
d\ud{\prime}{Li} &=& V\dvp{\prime}{ij}d\dvp{\prime}{j}.
\ee
This convention is summarised in table~\ref{tab:sMGenerations}. The most general
form of $V$ can contain three angles and six complex phases. However, complex phases
in (\ref{eq:gentrans}) cancel out of (\ref{eq:3GenerationMasses}) and therefore
complex phases in (\ref{eq:gentransleft}) can be defined to remove five of these
six phases. (One of the six phases in (\ref{eq:gentransleft}) can be parametrised
as an overall phase for all six transformations which cancels out of the right hand side
of (\ref{eq:RHS}).) The SM CKM matrix can therefore be parametrised by three angles
and one complex phase. This complex phase, which is responsible for $CP$-violating effects,
is quite small. Although the CKM matrix is close to being the identity,
the flavour eigenstates of the quark sector are not quite equal to the
mass eigenstates. This means that there is a non-zero probability amplitude for
a $W\uz{\pm}$ boson to couple together quark mass eigenstates of different generations.
The $W\uz{\pm}$ bosons therefore contribute to FCCCs. The $Z$ boson and the photon (and also the gluons)
do not contribute to FCNCs since the transformations (\ref{eq:gentrans}) leave terms
coupling neutral bosons to fermions, from (\ref{eq:covdermas}), invariant.

\subsection{Neutrino mass and the type-I see-saw mechanism}

In the SM, which does not include RH neutrinos, LH neutrinos are exactly
massless since both explicit mass terms and renormalisable terms coupling them
to the Higgs VEV are forbidden by the gauge symmetry. However, in Nature
we now know that neutrinos oscillate~\cite{ParticleDataGroupCollaboration2010,King2007b}~---
a mechanism that requires them to have
different masses, with the mass eigenstates being rotated with respect to the
flavour eigenstates. There should be a mixing matrix for the lepton
sector, analogous to the CKM matrix of the quark sector, known as the
PMNS~\cite{Pontecorvo1957,Pontecorvo1968,Maki1962} matrix.

If we define the charged lepton mass eigenstates to be the eigenstates of
flavour then the neutrino mass eigenstates will be rotated with respect to the
flavour eigenstates by the PMNS matrix. Neutrinos that are produced in some
flavour eigenstate will then be in a superposition of mass eigenstates. If the
mass differences involved are small enough, as they must be since the masses
themselves are small, then the neutrino will propagate coherently as this
superposition, but with each mass eigenstate component evolving at a different
rate, causing interference.
Therefore a neutrino that is produced as one flavour and propagates for some
distance may, when it eventually participates
in another charged weak current interaction, be measured as a different flavour
with some probability.

Whereas the CKM matrix is relatively close to the identity, neutrino oscillation
data indicates that the PMNS matrix is close to tribimaximal form~\cite{Harrison2002a,King2007b},
meaning that the mass and flavour eigenstate bases are very different from each other.

One may in principle add to the SM model matter content some number of RH neutrinos $\kk{N}$ that couple
to the Higgs field and lepton doublet, inducing Dirac mass terms after
EWSB. Such RH neutrinos would have to be uncharged under $\GSM$ in order
for the term $\kk{N}\uz{\dagger}H.L\dz{L}$ to be gauge invariant. This in
turn means that Majorana mass terms for the RH neutrino could also be added to
the SM Lagrangian. These RH neutrino masses would be unrelated to EWSB~---
their scale associated with some new physics. Let us assume that these Majorana masses
are much larger than the Dirac neutrino masses induced by EWSB.
For one generation we may write a neutrino mass term
\be
\cc{L}\dz{\kk{N}\nu} &=&
\left(\ba{cc}\kk{N}\uz{\dagger} & \nu\uz{c\dagger}\ea\right)
\left(\ba{cc}\kk{M} & \kk{m} \\ \kk{m} & 0\ea\right)
\left(\ba{c}\kk{N}\uz{c} \\ \nu\ea\right),
\ee
where $\kk{M}$ is the
RH neutrino Majorana mass, and $\kk{m}$ is the Dirac mass equal to
some Yukawa coupling times $v$. For $\kk{M} \gg \kk{m}$ there is one eigenvalue
approximately equal to $\kk{M}$ and another approximately equal to
$\kk{m}\uz{2}/\kk{M}$. In this case there is then a light mass eigenstate that is
almost, but not quite, $\nu$ and that has a Majorana mass that is suppressed
relative to the EWSB scale. This is the type-I see-saw
mechanism~\cite{Minkowski1977,Mohapatra1980}. The principle
still holds for three generations of RH and LH neutrinos, with three
of the states arising from the $6\times 6$ mass matrix
having non-zero, but suppressed, masses.

\section{Baryon and Lepton Number Conservation}

The renormalisable SM Lagrangian, without RH neutrinos, is invariant under
two extra $U(1)$ global symmetries known as $U(1)\dz{B}$ and $U(1)\dz{L}$,
corresponding to baryon and lepton number conservation respectively.
Quark fields $Q\dz{L}$, $d\dz{R}$, and $u\dz{R}$ have baryon number $B=+\nf{1}{3}$ and
lepton number $L=0$, with the $CP$ conjugate antiquark fields having $B=-\nf{1}{3}$
and $L=0$. The lepton fields $L\dz{L}$ and $e\dz{R}$ have $B=0$ and $L=+1$, with
antileptons having $L=-1$. Both $B$ and $L$ are conserved by the classical
renormalisable Lagrangian, but are anomalous if gauged and are violated
non-perturbatively.
However, the global symmetry $U(1)\dz{B-L}$, corresponding to the conservation
of the combination $B-L$, happens to be anomaly free in the SM if gauged
and globally is conserved even non-perturbatively.

\begin{figure}\begin{center}
\begin{picture}(140,140)(-20,-70)
\SetWidth{1}
\Line(0,60)(40,40)
\Line(0,-60)(40,-40)
\Photon(40,40)(60,20){-2}{4}
\Photon(40,-40)(60,-20){2}{4}
\Line(40,40)(80,60)
\Line(40,-40)(80,-60)
\Line(60,20)(60,-20)
\Line(57,3)(63,-3)
\Line(57,-3)(63,3)
\Line(60,20)(100,20)
\Line(60,-20)(100,-20)
\Text(-5,60)[r]{$d\dz{L}$}
\Text(-5,-60)[r]{$d\dz{L}$}
\Text(68,0)[l]{Majorana neutrino mass}
\Text(85,60)[l]{$u\dz{L}$}
\Text(85,-60)[l]{$u\dz{L}$}
\Text(105,20)[l]{$e\dz{L}$}
\Text(49,29)[rt]{$W\uz{-}$}
\Text(49,-29)[rb]{$W\uz{-}$}
\Text(105,-20)[l]{$e\dz{L}$}
\end{picture}
\caption{A diagram for a neutrinoless double beta decay process induced by the existence
of a Majorana neutrino mass.\label{fig:0nbb}}
\end{center}\end{figure}
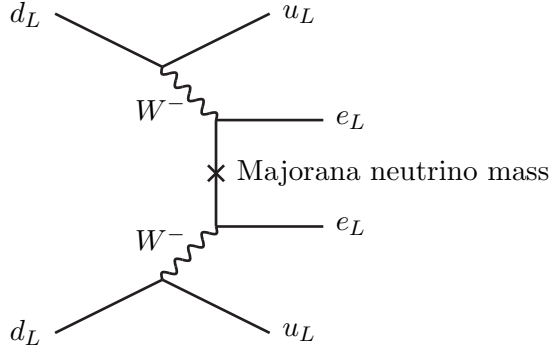

$U(1)\dz{B-L}$ is, however, broken explicitly by Majorana neutrino mass terms.
Majorana mass terms for the light neutrino mass eigenstates imply (and are
implied by) the existence of
neutrinoless double beta decay~\cite{Avignone2008,Gomez-Cadenas2011} (see figure~\ref{fig:0nbb})~---
a process in which baryon number remains
unchanged but lepton number is changed by 2.
Experiments searching for neutrinoless double beta decay (see for example
CUORE~\cite{Ardito2005}, EXO~\cite{EXOCollaboration2010}, GERDA~\cite{Abt2004},
MAJORANA~\cite{Aalseth2011}, NEXT~\cite{NEXTCollaboration2011}, and SNO+~\cite{Kraus2010})
will eventually determine the nature of neutrino mass~---
Dirac or Majorana. Because of the smallness of physical neutrinos masses,
this $U(1)\dz{B-L}$-breaking effect if the neutrino is Majorana in nature
would be corresponding rather small, with the
physical Majorana mass appearing in the matrix element of any such process.

To date no processes violating
either baryon or lepton number have ever been directly observed. In addition
to neutrinoless double beta decay, another example
of such a process would be proton decay~--- the decay of a proton with $B=+1$ into
a final state with $B=0$. This is a process which, unlike neutrinoless double beta
decay, needs not necessarily violate $U(1)\dz{B-L}$.

\section{The Hierarchy Problem of the SM}
\label{ref:hierarchy}

As previously stated, the Higgs boson mass in the SM is not determined by the
other known parameters of the model. There do, however, exist various theoretical
bounds~\cite{Hambye1997}. From (\ref{eq:higgsMassLambda}) we see that the Higgs boson mass-squared
is proportional to the self-coupling $\lambda$. The running coupling $\lambda$
in the loop-corrected potential is required to remain positive in order for
the EWSB vacuum to be stable. For low values of the running coupling $\lambda$
the coupling decreases with increasing energy scale. Depending on the cut-off energy scale $\Lambda$ that
one requires the model to be valid up to the vacuum stability requirement
puts a lower bound on the SM Higgs boson mass~--- a bound that increases with $\Lambda$~\cite{Sher1989}.
At the same time, for larger values of $\lambda$ the coupling increases with energy.
Too large values of $\lambda$ below some cut-off energy scale $\Lambda$ therefore
render the perturbation theory invalid. Here there exist uncertainties associated
with the using of perturbation theory to try to assess where perturbation remains
valid, but nonetheless the requirement that there is no Landau pole in $\lambda$ below $\Lambda$
puts an upper bound on the Higgs mass~--- a bound that decreases with $\Lambda$.
This large Higgs mass effect can also be seen non-perturbatively in Lattice calculations~\cite{Luscher1988,Luscher1989}.
Furthermore, too large Higgs boson masses lead to a non-unitarity of the S-matrix
for certain processes
where unitarity is preserved via cancellations between divergent diagrams involving
virtual Higgs bosons and divergent diagrams involving virtual longitudinal polarisations
of massive weak gauge bosons such as $WW$ scattering.
Unitarity bounds should not be violated in renormalisable theories~\cite{Cornwall1974}.

Because of these various upper bounds the SM Higgs boson should have a mass
below around the TeV scale
in order for the theory to be valid. This however introduces a naturalness problem
into the theory~--- the unnatural hierarchy between the EWSB scale and the
Planck scale which is about 16 orders of magnitude greater. The reason why this is
considered unnatural is because in the SM the Higgs doublet is a doublet of
fundamental complex scalars. (They are fundamental scalars as opposed to
composite scalars which would be expected
only to appear in some effective theory of the constituent particles that one
would only expect to be valid up to some energy scale associated with the
confinement.) Fundamental scalars are a problem in non-supersymmetric theories,
because their masses receive radiative corrections proportional to the masses
of any particles that they couple to~\cite{Susskind1979,Martin1997}.
Their self-energy Feynman diagrams
are quadratically sensitive to the highest mass scales in the theory.

For example, let us consider the one-loop contribution to the self-energy diagram of
a fundamental scalar that couples to a fermion of mass $m\dz{\rr{F}}$ with a Yukawa coupling
$\lambda\dz{\rr{F}}$ using dimensional regularisation with a mass scale parameter $\mu$
\be
-iA\ud{2}{\rr{F}} &=&
\left[
\begin{picture}(140,60)(-10,0)
\SetWidth{1}
\DashLine(0,0)(40,0){4}
\CArc(60,0)(20,0,360)
\DashLine(80,0)(120,0){4}
\Line(-3,3)(3,-3)
\Line(-3,-3)(3,3)
\Line(117,3)(123,-3)
\Line(117,-3)(123,3)
\Text(39,2)[rb]{$-i\lambda\dz{\rr{F}}$}
\Text(20,-5)[t]{$q$}
\end{picture}
\right]
\label{eq:scalarSelfEnergy}\\
\nn\\
&=&
-i\lambda\ud{2}{\rr{F}}\int\!\rr{d}\alpha\f{1}{(4\pi)\uz{2}}\Delta\ud{2}{\rr{F}}
\left[
-\f{2}{\varepsilon} - \gamma + 1
-\ln\!\left(\f{4\pi\mu\uz{2}}{\Delta\ud{2}{\rr{F}}}\right) + \cc{O}\varepsilon
\right],
\ee
where $\Delta\ud{2}{\rr{F}} = m\ud{2}{\rr{F}} - \alpha(1-\alpha)q\uz{2}$ and
the number of dimensions $d = 4 - \varepsilon$.
This contribution contains a part
that has a pole at the 4-dimensional limit $\varepsilon \rightarrow 0$ and
additional finite parts including a part proportional to $m\ud{2}{\rr{F}}$.
The coefficient of $\Delta\ud{2}{\rr{F}}$ in $A\ud{2}{\rr{F}}$ is proportional
to the logarithm which is order one.

If we define the renormalised
scalar propagator to have a pole where the energy equals the renormalised
mass $q\uz{2} = M\ud{2}{\rr{R}}$ then we have
\be
0 = \Sigma\uz{2}\bigg|_{q\uz{2}=M\ud{2}{\rr{R}}} &=&
\bigg[A\uz{2} + (Z\dz{M}-1) + (Z-1)q\uz{2}\bigg]_{q\uz{2}=M\ud{2}{\rr{R}}},
\ee
where $\Sigma^2$ is the total correction to the scalar mass-squared in the propagator from loop corrections to
and counter term insertions in the scalar propagator at one-loop order.
$Z\dz{M}$ and $Z$ are the scalar mass term and wavefunction
renormalisations respectively and $-iA\uz{2}$ is the total one-loop correction
to the scalar self-energy diagram.

If the SM were valid up to arbitrarily large energy scales, with no new physics
existing at higher energy scales, then there would be no problem. The renormalisation constants
may be defined to cancel the poles in
$\varepsilon$ of the constant and $q\uz{2}$ coefficients and, since the theory is
renormalisable, such poles would then be cancelled by counter terms at all orders
in perturbation theory.
The additional finite corrections would be at most of order the top quark mass.

However, this is not the case. Even if no physics comes in earlier the SM
cannot be valid above the Planck scale where contributions from quantum gravity should
become important. This being the case, it is not clear that using dimensional regularisation
and integrating momenta up to infinity is a valid prescription, but if one
alternatively uses a cut-off regulator, with
a momentum cut-off at some energy scale where the theory ceases to be valid, then
one still obtains the generic result
\be
-iA\ud{2}{F} &=& -i\lambda\ud{2}{F}\left[\mbox{pole} + C\dvp{2}{F}m\ud{2}{F} + \cdots\right],
\ee
where $C\dz{F}$ is some order one coefficient and the `pole' is now of order
the cut-off scale squared. If the scalar also couples
to another scalar of mass $m\dz{\rr{S}}$ with a coupling constant
$\lambda\dz{\rr{S}}$ then this also gives a contribution of the generic form
\be
-iA\ud{2}{S} &=& i\lambda\dvp{2}{S}\left[\mbox{pole} + C\dvp{2}{S}m\ud{2}{S} + \cdots\right].
\ee
Regardless of the physical interpretation, it is still the case that if these pole parts are
cancelled by the renomalisation constants at this level then, since the theory
is renormalisable, the poles will still be cancelled by counter terms at all orders.
This, however, cannot be said of any large finite corrections proportional to boson or
fermion masses associated with some new high energy physics~\cite{Martin1997}.

If one requires the renormalised mass to be much smaller than such large additional
finite contributions to $A\uz{2}$ then one one may also define the
renomalisation constants to almost completely cancel these additional finite
terms, leaving the small desired mass, at fixed order.
The unnaturalness arises when one then goes to higher order in perturbation theory.
While the new poles that arise at
this order will be exactly cancelled if one also includes all diagrams
containing
counter terms up the relevant order, new large finite
corrections will also arise that will not in general be cancelled by the finite parts of the
counter terms. The renormalised mass is therefore expected to be of order
the largest mass scale in the theory unless the finite parts of the
counter terms are carefully retuned at every order in perturbation theory.

The SM Higgs mass is thus sensitive to any new physics that might exist
at or below the Planck scale. Since we know that QFT itself is not expected
valid at the Planck scale, it is unreasonable to assume that there
is not some new physics at some scale far higher than the EWSB scale.
If the Higgs boson couples to this new physics at all then the Higgs boson mass
should be at least of order this scale unless one is willing to accept large tunings at
every order in perturbation theory to make it such that the large
contributions cancel, leaving a Higgs mass of order the EWSB scale. This is the
hierarchy problem of the SM.

If, for
example, we include radiative contributions from RH neutrinos with masses of
order $10\uz{14}$~GeV, then the additional finite corrections to the
Higgs boson self-energy of order
$10\uz{28}$~GeV$\uz{2}$ should be tuned to almost cancel leaving a physical
Higgs boson mass-squared 24 orders of magnitude smaller.

This hierarchy problem leads us to conclude that some new physics must exist
at or around the TeV scale to stabilise the Higgs mass. The most common
theories motivated to solve the hierarchy problem have solved it by assuming
that the Higgs boson is
a composite, as in the case of technicolour theories; by assuming that the
Planck scale is in fact around the TeV scale, as in the case of large extra
dimensions; or by assuming that the Higgs mass is stabilised by supersymmetry,
as in the theories that we shall introduce in chapters~\ref{chap:super} and~\ref{chap:essm}. All of these
theories involve the existence of new physics at the TeV scale.

\section{Unsolved Problems in Particle Physics}

Although the main motivation for new physics, particularly at the TeV scale
currently being probed by the LHC, is the hierarchy problem, there are many
other questions left unanswered by the SM. We shall briefly
mention some of them in this section. Neutrino mass has already been
discussed, but many other questions about fundamental fermion mass also remain
unanswered. Although the induced fermion masses are allowed in the SM, the
Yukawa couplings are measured and not predicted. Theories that attempt
to explain the sizes and values
of these Yukawa couplings as well as the striking difference between the CKM
and PMNS matrices are known as theories of flavour.
These typically invoke some new symmetry
known as flavour symmetry with the spontaneous breaking of flavour symmetry
producing the observed patterns of Yukawa couplings (see for example
refs.~\cite{Chen2000,Altarelli2000,Chkareuli2002}).

In the SM there are also three independent and unexplained gauge couplings.
Grand unification (introduced in section~\ref{sec:grand}) proposes that
the SM gauge group is in fact the remnant of some larger spontaneously broken gauge group with a
single gauge coupling. In such a scenario the SM gauge couplings, running up in energy, should unify
to the same value at some energy scale associated the breaking of the larger GUT group.
In the SM the couplings do not in fact unify, but they do in the supersymmetric models
introduced in the next chapter.

The baryon asymmetry of the universe is another problem. It is not known why the
universe appears to be made almost entirely of matter and not antimatter.
Although the SM technically satisfies the Sakharov conditions~\cite{Sakharov1991}~--- conditions
required for the existence of baryogenesis processes that could have created
this asymmetry~--- of baryon number non-conservation (non-perturbatively) and $CP$
violation (via the CKM matrix), the small $CP$-violating phase of the CKM matrix
is not thought to be large enough to have been the origin of the observed baryon
asymmetry.

Although most of the baryonic mass in the universe is to some extent
understood (being mostly due to QCD colour confinement rather than the Higgs
mechanism), dark matter, discussed in chapter~\ref{ref:dm}, and dark energy are
completely unaccounted for in the SM.

The strong $CP$ problem (see footnote~1 in section~\ref{ref:ghajsdg})
is another unsolved problem.

The origin of the gauge group, matter representation, and number of
space-time dimensions is
also not understood, although the theory must be consistent with respect to
gauge anomaly cancellation and the existence of stable atoms
and stable gravitational orbits is obviously necessary for our existence.

\cleardoublepage

\newpage
\chapter{Supersymmetry and Grand Unification}
\label{chap:super}

The description of supersymmetry given in this chapter is
largely based on the descriptions in refs.~\cite{Bailin1994,Martin1997}.
The notation for fermion spinors used is given in appendix~\ref{ap:spinors}.

Supersymmetry is a symmetry relating particles of different spin. In a theory
with some amount of supersymmetry each particle, possessing a given spin
and other internal quantum numbers, necessarily comes as part of what is known
as a supermultiplet~--- an association of particles that have different spins but
all other quantum numbers the same. The particles in these supermultiplets are
then transformed into each other by supersymmetry transformations that leave the
supersymmetric Lagrangian invariant. The size of the supermultiplets describing
the theory depends on the number of conserved supercharges $\cc{N}$
of the supersymmetry algebra. If the supersymmetry is preserved by the vacuum
then the particles of different
spin that make up a supermultiplet are all degenerate in mass as well as having
all other quantum numbers the same.

Plausible low energy models for physics beyond the SM can be constructed using
$\cc{N} = 1$ supersymmetry. This theory contains the following types of
supermultiplet: A chiral supermultiplet containing a complex scalar and a
LH (for a left chiral supermuliplet, RH for a right chiral supermultiplet)
Weyl spinor; a vector supermultiplet containing a spin-1 real vector and a
Weyl spinor; and a graviton supermultiplet containing the spin-2 graviton
and a spin-$\nf{3}{2}$ gravitino. Each of these supermultiplets separately
contains the same number of physical bosonic and fermionic degrees of freedom.
The (as we will see is necessary, multiple) Higgs scalars whose VEVs are
responsible for EWSB must then be part of chiral supermultiplets containing the
same number of fermionic degrees of freedom in the form of spin-$\nf{1}{2}$
fields. These fermions are known as Higgsinos. Each Weyl fermion matter field of
the SM must be part of either a
chiral or vector supermultiplet and in plausible models they are all
contained in chiral supermultiplets. The scalar superpartners of the quarks
and leptons are known as squarks and sleptons respectively. In supersymmetric
gauge theories the
massless gauge bosons form vector supermultiplets along with Weyl fermions known
as gauginos.

Although not the original motivation for supersymmetry itself~\cite{Wess1974},
the main motivation for what is known as TeV scale softly broken supersymmetry
(see for example ref.~\cite{Chung2005})
is that it provides a solution to the hierarchy problem of the SM.
In supersymmetric theories
scalar self-energies do not have the quadratic sensitivity to high energy scales
that are the origin of the SM hierarchy problem. The quadratic terms
due to fermions in loops such as (\ref{eq:scalarSelfEnergy}) are cancelled
by quadratic terms due to the bosons from the same supermultiplet.
These boson terms have the opposite sign since they do not have the extra minus sign
associated with a fermion loop. In another sense, the non-existence of this
quadratic sensitivity comes about because the Higgs scalar itself is part of a
supermultiplet and must remain degenerate in mass with its non-scalar
superpartners. For $\cc{N} = 1$ chiral supermultiplets the complex scalar must
remain degenerate with the Weyl fermion. Since Weyl fermions do not have
the quadratic sensitivity to high energy scales, the quadratic contributions
to the scalar self-energy must necessarily cancel.

Clearly the scenario described contradicts observation if supersymmetry is
preserved by the vacuum since it invokes the existence of many new unobserved
particles that are degenerate in mass with observed particles and have
similar interactions.
In realistic models supersymmetry, and this mass relation, must be broken. We
will see that the TeV scale soft breaking scenario provides a
solution to the hierarchy problem,
but predicts that there should be observable superpartners with masses not too
far above the TeV scale.

\section{Superpotentials}

The renormalisable Lagrangian of an $\cc{N} = 1$ supersymmetric gauge theory is
specified by specifying the gauge group, the gauge group representations
of the chiral supermultiplets, and what is known as the superpotential $\cc{W}$.
The superpotential is a chiral object, being a dimension-3 holomorphic function
of complex scalars
from either purely left or purely right chiral supermultiplets. Here we will
work purely with left chiral supermultiplets as is canonical.

Let a supersymmetric gauge theory contain left chiral supermultiplets, labelled
with $i$,
each containing a complex scalar $\phi\dz{i}$ and a LH Weyl spinor $\psi\dz{i}$.
Furthermore, let the superpotential
\be
\cc{W} &=& \f{1}{2}m\dz{ij}\phi\dz{i}\phi\dz{j}
+ \f{1}{6}\lambda\dz{ijk}\phi\dz{i}\phi\dz{j}\phi\dz{k},
\ee
with $m\dz{ij} = m\dz{ji}$ and $\lambda\dz{ijk}$ similarly symmetric in all of its
indices. Terms in the superpotential with mass dimension greater than 3 are
non-renormalisable.
The renormalisable, supersymmetric, gauge invariant Lagrangian\footnote{We do
not address non-renormalisable operators in supersymmetric theories. Although
the effects of dimension-5 operators are interesting and potentially important,
they have not been systematically studied in the $\E$SSM.} is then
\be
\cc{L} &=& -\f{1}{4}\cc{A}\uz{a\mu\nu}\cc{A}\ud{a}{\phantom{a}\mu\nu}
+ \tilde{A}\uz{ac\dagger}i\sigma\uz{\mu}\cc{D}\dz{\mu}\tilde{A}\uz{ac}
+ \psi\ud{\dagger}{i}i\bar{\sigma}\uz{\mu}\cc{D}\dz{\mu}\psi\dvp{\dagger}{i}
+ (\cc{D}\uz{\mu}\phi\dz{i})\uz{\dagger}(\cc{D}\dz{\mu}\phi\dz{i})\nn\\
&& \vphantom{A} + i\sqrt{2}g\uz{(a)}\left[
\phi\ud{\dagger}{i}T\uz{a}\tilde{A}\uz{ac\dagger}\psi\dvp{\dagger}{i}
- \psi\ud{\dagger}{i}\tilde{A}\uz{ac}T\uz{a}\phi\dvp{\dagger}{i}
\right]
-\f{1}{2}D\uz{a}D\uz{a}\nn\\
&& \vphantom{A} - \f{1}{2}\left[
m\dvp{\dagger}{ij}\psi\ud{c\dagger}{i}\psi\dvp{\dagger}{j}
+ \lambda\dvp{\dagger}{ijk}\psi\ud{c\dagger}{i}\psi\dvp{\dagger}{j}\phi\dvp{\dagger}{k} + \mbox{c.c.}
\right]
- F\ud{\dagger}{i}F\dvp{\dagger}{i},\label{eq:superLagrangian}
\ee
where
\be
F\dz{i} &=& \f{\partial\cc{W}}{\partial\phi\dz{i}} = -m\dz{ij}\phi\dz{j}
- \f{1}{2}\lambda\dz{ijk}\phi\dz{j}\phi\dz{k}
\ee
and
\be
D\uz{a} &=& -g\uz{(a)}\phi\ud{\dagger}{i}T\uz{a}\phi\dvp{\dagger}{i}.
\ee
The gaugino $\tilde{A}\uz{a}$ with adjoint index $a$ is a LH Weyl spinor,
so $\tilde{A}\uz{ac}$ is a RH Weyl spinor. If the gauge group is a direct
product of simple subgroups then the gauge coupling constant $g\uz{(a)}$ can
have a different value for each of these subgroups.

\section{The Matter Content of the MSSM}

The MSSM is minimal in the sense that it introduces as few new particles as
possible to the particles of the SM. To this end one begins by simply assigning all of the
fields in table~\ref{tab:sMCharges} to left chiral supermultiplets. This
immediately creates a number of problems and in the MSSM these are solved
in a way that introduces as few new fields as possible.

Firstly there are two problems
related to the assigning of the Higgs doublet $H$ to a chiral supermultiplet, but
both have the same solution. The first of these problems is that the Weyl fermion
superpartner of the Higgs scalar doublet contributes to the gauge anomaly (\ref{eq:anomalies}). The
inclusion of this field gives extra non-zero contributions to gauge
anomalies and therefore makes the gauge theory anomalous. The second of these two
problems is that, since the superpotential must be a holomorphic function
of complex scalars from purely left (by convention, alternatively right) chiral
supermultiplets, superpotential terms coupling $H$ to down-like squarks and
charged sleptons are forbidden by the $U(1)\dz{Y}$ gauge symmetry. This means that
the mass inducing couplings to down-like quarks and charged leptons that appear
in (\ref{eq:LYukawa}) cannot be present in the supersymmetric Lagrangian. The minimal
solution to both of these problems is the same and it is to have two Higgs scalar
doublets as in table~\ref{tab:mSSMCharges}.

\begin{table}
\begin{center}
\begin{tabular}{ccc|ccc|}
Supermultiplet & Boson & Fermion & $SU(3)\dz{c}$ & $SU(2)\dz{L}$ & $U(1)\dz{Y}$ \\\hline
LH quark doublet chiral
& $\tilde{Q}\dvp{c}{L}$ & $Q\dvp{c}{L}$ & $3$ & $2$ & $+\nf{1}{6}$\\
LH down-type antiquark chiral
& $\tilde{d}\ud{c}{R}$ & $d\ud{c}{R}$ & $\overline{3}$ & $1$ & $+\nf{1}{3}$\\
LH up-type antiquark chiral
& $\tilde{u}\ud{c}{R}$ & $u\ud{c}{R}$ & $\overline{3}$ & $1$ & $-\nf{2}{3}$\\
LH lepton doublet chiral
& $\tilde{L}\dvp{c}{L}$ & $L\dvp{c}{L}$ & $1$ & $2$ & $-\nf{1}{2}$\\
LH charged antilepton chiral
& $\tilde{e}\ud{c}{R}$ & $e\ud{c}{R}$ & $1$ & $1$ & $+1$\\\hline
Down-type Higgs doublet chiral
& $H\dz{d}$ & $\tilde{H}\dz{d}$ & $1$ & $2$ & $-\nf{1}{2}$\\
Up-type Higgs doublet chiral
& $H\dz{u}$ & $\tilde{H}\dz{u}$ & $1$ & $2$ & $+\nf{1}{2}$\\\hline
Gluon vector & $G\uz{\mu}$ & Gluino $\tilde{G}$ & $8$ & $1$ & $0$\\
$SU(2)\dz{L}$ gauge vector & $W\uz{\mu}$ & Wino $\tilde{W}$ & $1$ & $3$ & $0$\\
$U(1)\dz{Y}$ gauge vector & $B\uz{\mu}$ & Bino $\tilde{B}$ & $1$ & $1$ & $0$\\\hline
\end{tabular}
\caption{The $SU(3)\dz{c}$ and $SU(2)\dz{L}$ representations
and the $U(1)\dz{Y}$ charges of the supermultiplets of the
MSSM.\label{tab:mSSMCharges}}
\end{center}
\end{table}

The extra contributions to gauge anomalies from the Higgsinos then cancel since
their charges are opposite and together they from a real representation of
$\GSM$. (We choose to write all $SU(2)$ antidoublets as doublets since they
are equivalent, as shown in appendix~\ref{ap:pseudoreality}.)
The contribution to the gauge anomaly due to gauginos is automatically zero
since gauginos are necessarily in the adjoint representation which is real.
Since $L\dz{L}$ and
$H\dz{d}$ have the same quantum numbers one might think that a more minimal
solution would be to declare these fields to be part of the same supermultiplet,
but in practice such models prove unrealistic.

The most general renormalisable and gauge invariant
superpotential containing the fields in table~\ref{tab:mSSMCharges} is
\be
\cc{W} &=& \mu H\dvp{c}{d}.H\dvp{c}{u}
    + h\ud{U}{ij}\tilde{u}\ud{c}{Ri}H\dvp{c}{u}.\tilde{Q}\dvp{c}{Lj}\nn\\
&& \vphantom{A} - h\ud{D}{ij}\tilde{d}\ud{c}{Ri}H\dvp{c}{d}.\tilde{Q}\dvp{c}{Lj}
    - h\ud{E}{ij}\tilde{e}\ud{c}{Ri}H\dvp{c}{d}.\tilde{L}\dvp{c}{Lk}
+ \Delta\cc{W},\label{eq:mssmW}
\ee
where
\be
\Delta\cc{W} &=&
  \f{1}{2}\xi\ud{LLe}{ijk}\tilde{L}\dvp{c}{Li}.\tilde{L}\dvp{c}{Lj}\tilde{e}\ud{c}{Rk}
        + \xi\ud{LQd}{ijk}\tilde{L}\dvp{c}{Li}.\tilde{Q}\dvp{c}{Lj}\tilde{d}\ud{c}{Rk}\nn\\
&& \vphantom{A} + \zeta\ud{LH}{i}\tilde{L}\dvp{c}{Li}.\tilde{H}\dvp{c}{u}
+ \f{1}{2}\xi\ud{udd}{ijk}\tilde{u}\ud{c}{Ri}\tilde{d}\ud{c}{Rj}\tilde{d}\ud{c}{Rk}.\label{eq:deltaW}
\ee

If both of the Higgs scalars acquire VEVs
\be
\langle H\dz{d}\rangle = \f{1}{\sqrt{2}}\left(\ba{c}v\dz{d} \\ 0\ea\right)
&\mbox{\quad and\quad}&
\langle H\dz{u}\rangle = \f{1}{\sqrt{2}}\left(\ba{c}0 \\ v\dz{u}\ea\right)
\ee
then this superpotential yields Dirac mass terms equivalent to those in (\ref{eq:3GenerationMasses})
\be
\cc{L}\dz{\Yukawa} &=& -\f{1}{\sqrt{2}}\left(h\ud{D}{ij}v\dz{d}d\ud{\dagger}{Ri}d\dvp{\dagger}{Lj}
+ h\ud{U}{ij}v\dz{u}u\ud{\dagger}{Ri}u\dvp{\dagger}{Lj}\right.\nn\\
&&\left.\qquad\qquad\qquad\qquad\quad\vphantom{A}
+ h\ud{L}{ij}v\dz{d}e\ud{\dagger}{Ri}e\dvp{\dagger}{Lj} + \mbox{c.c.}\right) + \cdots.
\ee
In order for the $W\uz\pm$ and $Z$ boson masses to be the same as their SM values we require
\be
v\uvp{2}{d} &=& v\ud{2}{d} + v\ud{2}{u}.
\ee
We therefore define an angle $\beta$ such that
\be
\tan(\beta) &=& \f{v\dz{u}}{v\dz{d}}\\\nn\\
\Rightarrow\quad v\dz{d} &=& v\cos(\beta)\quad\mbox{and}\nn\\
                 v\dz{u} &=& v\sin(\beta).\nn
\ee
The Yukawa coupling matrices $h\uz{D}$, $h\uz{U}$, and $h\uz{L}$ must then be multiplied with respect to those
of the SM by factors of $1/\cos(\beta)$, $1/\sin(\beta)$, and $1/\cos(\beta)$ respectively.
With increasing $\tan(\beta)$ the hierarchy between the top and bottom Yukawa couplings
is lessened, but all of these Yukawa couplings are greater than their SM values for all
angles $\beta$.

\subsection{$R$-parity}

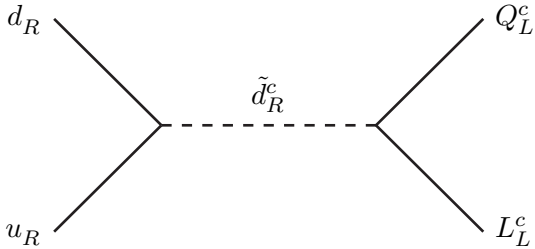
\begin{figure}\begin{center}
\begin{picture}(200,100)(-20,-50)
\SetWidth{1}
\Line(0,40)(40,0)
\Line(0,-40)(40,0)
\DashLine(40,0)(120,0){4}
\Line(120,0)(160,40)
\Line(120,0)(160,-40)
\Text(-5,40)[r]{$d\dvp{c}{R}$}
\Text(-5,-40)[r]{$u\dvp{c}{R}$}
\Text(80,5)[b]{$\tilde{d}\ud{c}{R}$}
\Text(165,40)[l]{$Q\ud{c}{L}$}
\Text(165,-40)[l]{$L\ud{c}{L}$}
\end{picture}
\caption{A proton decay diagram using the couplings $\xi\uz{LQd}$ and
$\xi\uz{udd}$ in (\ref{eq:deltaW}).\label{fig:protonDecay}}
\end{center}\end{figure}

The terms in $\Delta\cc{W}$, however, are dangerous since they all violate
either lepton or baryon number conservation. Most importantly they lead to
Lagrangian
terms that allow protons to decay into final states with zero baryon number. These
decays are mediated by squarks and require both baryon and lepton violating
terms from $\Delta\cc{W}$ (see figure~\ref{fig:protonDecay}).
The most common solution in the MSSM is to impose an
additional discrete $\dd{Z}\dz{2}$ symmetry on the fields in the superpotential.
This is known as $R$-parity, which we denote $\dd{Z}\ud{M}{2}$, and is defined
such that the Higgs scalars are even and all of the matter scalars (squarks and sleptons) are
odd. When interpreted as a symmetry of the Lagrangian, angular momentum
conservation implies that the fermionic superpartners
have opposite $R$-parity. The gauge bosons must be
$\dd{Z}\ud{M}{2}$-even
and the gauginos are then $\dd{Z}\ud{M}{2}$-odd. Oddness under $\dd{Z}\ud{M}{2}$
is the meaning of the
tilde over the squark, slepton, Higgsino, and gaugino fields.

The imposition of $\dd{Z}\ud{M}{2}$ forbids all of the terms in $\Delta\cc{W}$
and makes the renormalisable MSSM Lagrangian invariant under global $U(1)\dz{B}$
and $U(1)\dz{L}$. In the SM only gauge invariance is required in order for these
to be global symmetries since the squarks the sleptons do not exist.
The $\dd{Z}\ud{M}{2}$-odd particles, denoted with a tilde thoughout, are known as the
supersymmetric particles and the imposition of $\dd{Z}\ud{M}{2}$ means that the
lightest supersymmetric particle is absolutely stable. It is therefore
the case that in the MSSM a discrete symmetry imposed in order to prevent
rapid proton decay also leads to the existence of a new stable particle that may
be a plausible candidate for dark matter.

\subsection{The $\mu$ problem of the MSSM}

The other problem with $\cc{W}$ in (\ref{eq:mssmW}) is that it contains the bilinear mass term
$\mu$. This is a supersymmetry respecting parameter that a priori has
no relation to either the EWSB or supersymmetry breaking scales. The problem is
that in order to achieve EWSB, with $v$ of the correct magnitude, the parameter
$\mu$ should be of order the EWSB scale. In the MSSM as written, however, it is
not clear why it should not be either of order the Planck scale or zero. The $\mu$ problem
refers to a fine-tuning that has to be imposed on the $\mu$ parameter once.
Supersymmetry does at least mean that the parameter is stable at the EWSB scale under
radiative corrections, even as it is not explained.

\section{Soft Supersymmetry Breaking}

Although unbroken supersymmetry is easily ruled out,
even in the case of spontaneously broken
supersymmetry the relationship
\be
\tr[\cc{M}\ud{2}{\phi}] &=& 2\tr[\cc{M}\ud{2}{\psi}],
\ee
where $\cc{M}\ud{2}{\phi}$ is the mass-squared matrix for all
real scalars and $\cc{M}\ud{2}{\psi}$ is the mass-squared matrix for all Weyl
spinors in chiral supermultiplets, still holds at tree level in the absence of
gauge anomalies~\cite{Martin1997}.
It is trivially satisfied in the case of unbroken supersymmetry
since the two real scalars and Weyl spinor from each chiral supermultiplet are
degenerate. Because of this relation it has not been possible to create a
realistic model for supersymmetry breaking in the MSSM without introducing extra
physics.

In order to create realistic models one usually invokes
the existence of some other sector, known as the hidden sector, in which
supersymmetry is spontaneously broken. In this hidden sector scenario the
visible sector (containing SM matter, here the MSSM sector) does not itself
cause spontaneous supersymmetry breaking, but
supersymmetry breaking effects are communicated to it somehow from the hidden
sector.

In this hidden sector scenario it is useful to parametrise the kinds of
modifications to the visible sector Lagrangian that spontaneous supersymmetry
breaking in the hidden sector can cause in ignorance of the exact mechanism
of supersymmetry breaking. It is therefore useful to list the gauge invariant
mass terms that may be induced in the visible sector. Firstly one can have SSB
masses for all scalars $\phi$ of the form
\be
-\phi\uz{\dagger}m\uz{2}\phi.
\ee
If there is more than one copy of a scalar with the same quantum numbers,
\ie~there is more than one generation, then there can be more structure to
the mass matrix. For example, for the LH slepton\footnote{Since sleptons are
scalars there is no concept of slepton handedness. The handedness refers
to the handedness of the fermionic superpartner.}
doublets one can have
\be
-\tilde{L}\ud{\dagger}{Li}m\ud{\vphantom{\dagger}2}{Lij}\tilde{L}\dvp{\dagger}{Lj}
\label{LSoftMass}\ee
and for the RH charged sleptons one can have
\be
-\tilde{e}\ud{\dagger}{Ri}m\ud{\vphantom{\dagger}2}{eij}\tilde{e}\dvp{\dagger}{Rj},
\label{eSoftMass}\ee
whereas the only possible soft scalar mass-squared term involving the down-type
Higgs doublet is
\be
-H\ud{\dagger}{d}m\ud{\vphantom{\dagger}2}{H\dz{d}}H\dvp{\dagger}{d}.
\ee
Note that $\tilde{L}\dz{L}$ and $H\dz{d}$ do not have the same quantum numbers
since $\tilde{L}\dz{L}$ is $\dd{Z}\ud{M}{2}$-odd.
Secondly one can have SSB masses for gauginos
\be
-\f{1}{2}\left[M\uz{(a)}\tilde{A}\uz{ac\dagger}\tilde{A}\uz{a} + \mbox{c.c.}\right].
\ee
If the gauge group is a direct product of subgroups, then in general the
gauginos associated with each subgroup can have a different gaugino mass.
For example, for the
SM gauge group one can have soft gaugino mass terms
\be
-\f{1}{2}\left[M\dz{3}\tilde{G}\uz{ac\dagger}\tilde{G}\uz{a}
             + M\dz{2}\tilde{W}\uz{ac\dagger}\tilde{W}\uz{a}
             + M\dz{1}\tilde{B}\uz{ac\dagger}\tilde{B}\uz{a}
             + \mbox{c.c.}\right].
\ee
These mass terms always exist for gauginos since they are necessarily in real
representations of the gauge group. They do not exist for chiral
fermions. Thirdly one can have SSB trilinear terms. For each trilinear term
that is allowed to appear in the superpotential
\be
\lambda\dz{ijk}\phi\dz{i}\phi\dz{j}\phi\dz{k}\mbox{\quad(no sum on $i,j,k$)}
\ee
one can have the soft supersymmetry breaking Lagrangian term
\be
\lambda\dz{ijk}A\dz{\lambda\dz{ijk}}
\phi\dz{i}\phi\dz{j}\phi\dz{k}\mbox{\quad(no sum on $i,j,k$)},
\ee
where $A$ has mass dimension 1. In fact it is the case that for any
term that can appear in the superpotential one can have
a corresponding SSB Lagrangian term that is equal
to the superpotential term multiplied by some new supersymmetry breaking
parameter with mass dimension 1. In the MSSM there then exists the SSB
breaking term corresponding to the $\mu$ term
\be
\mu BH\dz{d}.H\dz{u}.
\ee

There terms are known as soft since they only involve new parameters that have
positive-definite mass dimension. The supersymmetric relationships between
the dimensionless couplings involving bosons and fermions that lead to the
cancellation of the quadratic sensitivity of fundamental scalars to arbitrarily
high scales are preserved. If these new parameters with the dimensions of mass
are roughly of order some scale associated with SSB
then the consequences are the following: Firstly, while the
observed
quarks and leptons only acquire masses proportional to the EWSB scale, the
unobserved gauginos, squarks, and sleptons acquire masses proportional to the
SSB scale, allowing their current non-observation to be naturally explained.
Secondly,
contributions to the finite radiative corrections to the Higgs boson self-energy will be
at most of order the SSB scale since this is the scale of differences between
bosonic and fermionic masses within supermultiplets. Therefore if the SSB
scale is not too far
above the EWSB scale then the hierarchy problem is still solved. It is therefore
believed that if supersymmetry is the solution to the hierarchy problem then the squarks,
sleptons and gauginos, while currently unobserved, should not have masses too
far above the TeV scale and should therefore be discovered at the LHC. This is
what is known as TeV scale softly broken supersymmetry.

If the squarks and sleptons are present at or not too far above the TeV scale then
these soft supersymmetry breaking terms can in general lead to FCNCs in
contradiction with observation. This problem can be avoided if each of the
$3 \times 3$ soft scalar mass-squared matrices, such as those in
(\ref{LSoftMass}) and (\ref{eSoftMass})~--- namely $m\ud{2}{Qij}$, $m\ud{2}{dij}$,
$m\ud{2}{uij}$, $m\ud{2}{Lij}$, and $m\ud{2}{eij}$~---
are proportional to the identity and if for each of the $3 \times 3$
Yukawa matrices~--- $h^U_{ij}$, $h^D_{ij}$, and $h^E_{ij}$~---
all nine associated soft trilinear couplings are equal. Explicitly this means
\be
m\ud{2}{Fij} &=& m\ud{2}{F}\delta\dvp{2}{ij} \quad\forall\; F\in\{Q,d,u,L,e\}\quad\mbox{and}\label{eq:m0}\\
A\dz{h\ud{G}{ij}} &=& A\dz{G} \quad\forall\; i,j \quad\forall\; G\in\{U,D,E\}.\label{eq:A0}
\ee
The SSB gaugino masses and trilinear couplings may also in general
have large phases that lead to large $CP$-violating effects, again in
contradiction with observation. The actual restrictions on and relationships
between these soft mass parameters will depend on the nature of the
SSB mechanism.

\section{Grand Unification}
\label{sec:grand}

Grand unification is the idea that just like $U(1)\dz{\rr{EM}}$ is a remnant of
the spontaneously broken electroweak gauge symmetry group, so the SM gauge
symmetry group $\GSM$ is a remnant of some still larger group that is
spontaneously broken by some mechanism at some GUT scale.
The further assumption is
that this GUT group should not be a direct product of simple groups, as $\GSM$
is, but should itself be simple, with a single gauge coupling. Grand unified
theories can offer an explanation for the charges of the observed SM particles
and for the observed values of the three gauge couplings at low energies.
The three gauge couplings would
be equal to some single GUT group gauge coupling at the GUT scale and below the
GUT scale, where the GUT symmetry is
spontaneously broken, the three couplings would then run, with different beta
functions, to the low energy values that we observe. Grand unification therefore
makes the prediction that, running to higher energy,
the three gauge couplings should unify at some scale.
Since we know the values of the couplings at low energy, if we know the beta
functions in some theory then this prediction can be tested~--- the beta functions
depending on the whatever new physics exists between the EWSB scale and the
GUT scale.

A $U(1)$ gauge theory is of course invariant under a rescaling of the gauge
coupling as long as the charges are also appropriately rescaled, but if the
$U(1)$ is a remnant from some spontaneously broken larger group then the
$U(1)$ charges of the particles, forming some representation under the larger
group, will then be fixed.

In the SM, with no new physics coming in above the EWSB scale, the couplings do
not unify. They do, however, unify if one assumes supersymmetry spontaneously
broken at the TeV scale. In the MSSM the scale of this unification is
around $10\uz{16}$~GeV (see for example ref.~\cite{Martin1997}).

The smallest possible GUT group is $SU(5)$~\cite{Dimopoulos1981a}.
SM matter can arise from $10$ and
$\overline{5}$ representations which decompose under
\be
SU(5) &\rightarrow& SU(3)\dz{c}\otimes SU(2)\dz{L}\otimes U(1)\dz{Y}
\ee
as
\be
10      &\rightarrow& \left(3,2,+\sqrt{\f{3}{5}}\f{1}{6}\right)
               \oplus \left(\overline{3},1,-\sqrt{\f{3}{5}}\f{2}{3}\right)
               \oplus \left(1,1,+\sqrt{\f{3}{5}}\right)\\\nn\\
\overline{5} &\rightarrow& \left(\overline{3},1,+\sqrt{\f{3}{5}}\f{1}{3}\right)
               \oplus \left(1,2,-\sqrt{\f{3}{5}}\f{1}{2}\right).
\ee
This is one generation of SM quarks and leptons as long as one uses the
correctly GUT normalised $U(1)\dz{Y}$ gauge coupling, which we can read off as
\be
g\dz{1} &=& \sqrt{\f{5}{3}}g\uz{\prime}.
\ee
The two Higgs doublets of the MSSM, however, do not form a complete
representation of $SU(5)$. If they are from $\overline{5}$ and $5$ representations
then one must explain why the colour triplets from these representations are not
present at low energy while the Higgs doublets are. This is known as the
doublet-triplet splitting problem.

\subsection{Unification of SSB masses}

At the GUT scale in order for the soft gaugino mass terms to be gauge invariant
under the GUT group all soft gaugino masses, like all gauge couplings,
must be equal. This GUT scale soft gaugino mass is known as $M\dz{1/2}$.
Grand unified gauge symmetry implies that these soft gaugino masses
should be unified at the GUT scale, but the further assumption is sometimes made
that not only should the unifications (\ref{eq:m0}) and (\ref{eq:A0}) be imposed
at the GUT scale for phenomenological reasons, but that all of the soft scalar masses in (\ref{eq:m0})
should be equal to a unified soft scalar mass $m\dz{0}$ and that all of the soft
trilinear couplings in (\ref{eq:A0}) should be equal to a unified trilinear coupling $A\dz{0}$.
This this known as the constrained scenario or sometimes gravity mediated supersymmetry
breaking. In the MSSM this constrained scenario is known as the cMSSM.

Gravity mediated supersymmetry breaking assumes that supersymmetry breaking
is communicated to the visible sector only by non-renormalisable operators that are suppressed
by the Planck mass. If the supersymmetry breaking scale
in the hidden sector is $M\dvp{2}{\rr{S}}$ then the visible sector SSB masses
will be of order the gravitino mass
$m\dvp{2}{3/2} \sim M\ud{2}{\rr{S}}/M\dvp{2}{\rr{P}}$~\cite{Bailin1994}.
A further assumption is that the non-renormaliasable operators should be completely flavour blind
and that the SSB parameters should be unified at the Planck scale.
In the constrained scenario, however, the unification relations are all applied
at the nearby GUT scale.

One success of this scenario is that if the soft Higgs masses start off equal
to $m\dz{0}$ at the high scale then they will typically be driven negative by
radiative corrections on the way down to the EWSB scale, allowing for EWSB if
one also has an appropriate $\mu$ parameter. Whether in a constrained scenario
or not, this is known as radiative EWSB.
The soft Higgs masses will be of the correct order of magnitude for EWSB since
they will be of order the SSB scale.

\cleardoublepage
\newpage

\chapter{The $\E$SSM}
\label{chap:essm}

The main theoretical shortcoming of the MSSM as a model describing TeV scale
softly broken supersymmetry is the $\mu$ problem.
The model also predicts the tree level result that the lightest Higgs
boson must have a mass smaller than $m\dz{Z}|\cos(2\beta)|$. Large loop
corrections must then push the Higgs mass above the LEP limit in order for the model
not to be ruled out. In practice when this is done the model is
quite fine-tuned~\cite{Kane1999}.
In light of the shortcomings of the minimal model, it is worth considering
supersymmetric models that have a non-minimal structure at the TeV scale.

The $\E$SSM~\cite{King2006,King2006a,King2007}
is a string theory inspired supersymmetric model based on an $\E$
GUT group. The low energy gauge group contains
an extra $U(1)$, called $U(1)\dz{N}$, under which
the RH neutrinos that arise in the model are not charged.
This means that the RH neutrinos may acquire large
intermediate scale
Majorana masses. This choice, that the low energy gauge group
is $\GSM\otimes U(1)_N$, defines the model. The $U(1)\dz{N}$ gauge symmetry is
spontaneously broken at low energy by a SM-singlet field~--- charged under the
extra $U(1)\dz{N}$, but a singlet under $\GSM$.
This field radiatively acquires a VEV which is naturally
of order the SSB scale, meaning that there is $Z'$ boson with
an induced mass of order the TeV scale. This SM-singlet VEV also induces an
effective $\mu$ parameter, also naturally of order the SSB scale,
with the $\mu$ term
of the MSSM being forbidden by the enlarged gauge symmetry.

Although $\E$ is not a group without complex representations,
complete representations of $\E$ are
nonetheless free of gauge anomalies.
In the $\E$SSM automatic gauge anomaly cancellation is thereby ensured by allowing
three complete 27
representations of $\E$ to survive down to the low energy scale.
These three 27s contain the three generations of known matter, however they also
contain the VEV acquiring Higgs doublets and SM-singlet.
This means that there are two extra copies of the Higgs doublets and SM-singlet
in the low energy particle spectrum. Whereas in the MSSM the Higgs doublets
do not form a complete representation of the potential $SU(5)$ GUT group,
in the $\E$SSM
supermultiplets with the quantum numbers of Higgs doublets are contained within
each of the fundamental 27 representations of the GUT group that also each
contain one generation of SM matter.

In the $\E$SSM only one generation of Higgs doublets and SM-singlets,
defined to be the third, acquires the
required VEVs and is known as `active'.
The other two generations, the first and second, of Higgs doublets and SM-singlets
do not acquire VEVs and these are known as `inert'.
Furthermore, in the $\E$SSM it is assumed that the inert generations have
suppressed Yukawa couplings to SM matter, suppressed due to some
flavour symmetry. This means that new FCNCs from the enlarged Higgs sector are
suppressed and also explains why the inert generations do not radiatively
acquire VEVs.

\section{Gauge Symmetry and Matter Content}
\label{ref:e6g&m}

The subgroups of the $\E$ GUT group may be written
\be
E_6 &\supset& SO(10) \otimes U(1)\dz{\psi} \nn\\
&\supset& SU(5) \otimes U(1)\dz{\chi} \otimes U(1)\dz{\psi} \nn\\
&\supset& SU(3)\dz{c} \otimes SU(2)\dz{L} \otimes U(1)\dz{Y}
\otimes U(1)\dz{\chi} \otimes U(1)\dz{\psi}.
\ee
In the $\E$SSM $\E$ is spontaneously broken at the GUT scale directly to
$SU(3)\dz{c} \otimes SU(2)\dz{L} \otimes U(1)\dz{Y} \otimes U(1)\dz{N}$, where
\be
U(1)\dz{N} &=& \cos(\vartheta) U(1)\dz{\chi} + \sin(\vartheta) U(1)\dz{\psi}
\ee
and $\tan(\vartheta) = \sqrt{15}$. This is such that the RH neutrinos
that appear in the theory are completely uncharged. Three complete 27 representations of $\E$ then
survive down to low energy in order to ensure gauge anomaly cancellation.
They decompose under the $SU(5) \otimes U(1)\dz{N}$ subgroup as~\cite{Keith1997}
\be
27 &\rightarrow& \left(10,\f{1}{\sqrt{40}}\right) \oplus \left(\overline{5},\f{2}{\sqrt{40}}\right)
\nn\\
&& \vphantom{A} \oplus \left(\overline{5},-\f{3}{\sqrt{40}}\right) \oplus \left(5,-\f{2}{\sqrt{40}}\right)
\oplus \left(1,\f{5}{\sqrt{40}}\right) \oplus \left(1,0\vphantom{\f{0}{\sqrt{40}}}\right).
\ee
The first two terms contain normal matter, whereas the final term, which is
a singlet under the entire low energy gauge group, contains the RH
neutrino, or technically the LH antineutrino $\kk{N}\uz{c}$.
The second-to-last term, which is charged only under $U(1)\dz{N}$,
contains the SM-singlet $S$. The third generation SM-singlet acquires
a VEV
\be
\langle S\dz{3}\rangle &=& \f{s}{\sqrt{2}}
\ee
which, as we shall see,
generates the effective $\mu$ term and spontaneously breaks
$U(1)\dz{N}$ leading to a mass for the
$Z'$ boson.
The remaining two terms contain the down- and up-type
Higgs doublets $H\dz{d}$ and $H\dz{u}$, but also contain $SU(3)\dz{c}$ triplets.
These exotic coloured states are known as
$\bar{D}$ and $D$~--- the antitriplet from $\overline{5}$ and the triplet from $5$
respectively. Only the third generation of Higgs doublets
acquires VEVs
\be
\langle H\ud{0}{d3}\rangle &=& \f{v\dz{d}}{\sqrt{2}} = \f{v}{\sqrt{2}}\cos(\beta)\quad\mbox{and}\nn\\\nn\\
\langle H\ud{0}{u3}\rangle &=& \f{v\dz{u}}{\sqrt{2}} = \f{v}{\sqrt{2}}\sin(\beta).
\ee
The charge assignments of the matter of the supermultiplets of the $\E$SSM are
summarised in table~\ref{tab:eSSMCharges}.

\begin{table}
\begin{center}
\begin{tabular}{ccc|cccc|}
Supermultiplet & Boson & Fermion & $\!r\uz{c}\!$ & $\!r\uz{L}\!$ & $\!\sqrt{5/3}Q\uz{Y}\!$ & $\!\sqrt{40}Q\uz{N}\!$ \\\hline
LH quark doublet chiral
& $\tilde{Q}\dvp{c}{L}$ & $Q\dvp{c}{L}$ & $3$ & $2$ & $+\nf{1}{6}$ & $+1$ \\
LH down-type antiquark chiral
& $\tilde{d}\ud{c}{R}$ & $d\ud{c}{R}$ & $\overline{3}$ & $1$ & $+\nf{1}{3}$ & $+2$ \\
LH up-type antiquark chiral
& $\tilde{u}\ud{c}{R}$ & $u\ud{c}{R}$ & $\overline{3}$ & $1$ & $-\nf{2}{3}$ & $+1$ \\
LH lepton doublet chiral
& $\tilde{L}\dvp{c}{L}$ & $L\dvp{c}{L}$ & $1$ & $2$ & $-\nf{1}{2}$ & $+2$ \\
LH charged antilepton chiral
& $\tilde{e}\ud{c}{R}$ & $e\ud{c}{R}$ & $1$ & $1$ & $+1$ & $+1$ \\
LH antineutrino chiral
& $\tilde{N}\uvp{c}{R}$ & $N\uvp{c}{R}$ & $1$ & $1$ & $0$ & $0$ \\\hline
Down-type Higgs doublet chiral
& $H\dz{d}$ & $\tilde{H}\dz{d}$ & $1$ & $2$ & $-\nf{1}{2}$ & $-3$ \\
Up-type Higgs doublet chiral
& $H\dz{u}$ & $\tilde{H}\dz{u}$ & $1$ & $2$ & $+\nf{1}{2}$ & $-2$ \\
SM-singlet chiral
& $S$ & Singlino $\tilde{S}$ & $1$ & $1$ & $0$ & $+5$ \\\hline
Exotic colour antitriplet chiral
& $\bar{D}$ & $\tilde{\bar{D}}$ & $\overline{3}$ & $1$ & $+\nf{1}{3}$ & $-3$ \\
Exotic colour triplet chiral
& $D$ & $\tilde{D}$ & $3$ & $1$ & $-\nf{1}{3}$ & $-2$ \\\hline
Gluon vector & $G\uz{\mu}$ & Gluino $\tilde{G}$ & $8$ & $1$ & $0$ & $0$ \\
$SU(2)\dz{L}$ gauge vector & $W\uz{\mu}$ & Wino $\tilde{W}$ & $1$ & $3$ & $0$ & $0$ \\
$U(1)\dz{Y}$ gauge vector & $B\uz{\mu}$ & Bino $\tilde{B}$ & $1$ & $1$ & $0$ & $0$ \\
$U(1)\dz{N}$ gauge vector & ${B}\uz{\prime\mu}$ & Bino$'$ $\tilde{B}\uz{\prime}$ & $1$ & $1$ & $0$ & $0$ \\\hline
\end{tabular}
\caption{The $SU(3)\dz{c}$ and $SU(2)\dz{L}$ representations
and the $\E$ GUT normalised $U(1)\dz{Y}$ and $U(1)\dz{N}$ charges of the
supermultiplets of the
$\E$SSM.\label{tab:eSSMCharges}}
\end{center}
\end{table}

The low energy gauge invariant superpotential
\be
\cc{W} &=& \cc{W}\dz{0} + \cc{W}\dz{1} + \cc{W}\dz{2},
\label{eq:w}\ee
where
\be
\cc{W}\dz{0} &=& \lambda\dvp{c}{ijk}S\dvp{c}{i}H\dvp{c}{dj}.H\dvp{c}{uk}
                + \kappa\dvp{c}{ijk}S\dvp{c}{i}\bar{D}\dvp{c}{j}D\dvp{c}{k}
                     + h\ud{N}{ijk}\tilde{\kk{N}}\ud{c}{i}H\dvp{c}{uj}.\tilde{L}\dvp{c}{k}\nn\\
&& \vphantom{A}      + h\ud{U}{ijk}\tilde{u}\ud{c}{Ri}H\dvp{c}{uj}.\tilde{Q}\dvp{c}{Lk}
                     + h\ud{D}{ijk}\tilde{d}\ud{c}{Ri}H\dvp{c}{dj}.\tilde{Q}\dvp{c}{Lk}
                     + h\ud{E}{ijk}\tilde{e}\ud{c}{Ri}H\dvp{c}{dj}.\tilde{L}\dvp{c}{Lk},\label{eq:W0}\\\nn\\
\cc{W}_1 &=&           g\ud{Q}{ijk}D\dvp{c}{i}\tilde{Q}\dvp{c}{Lj}.\tilde{Q}\dvp{c}{Lk}
                     + g\ud{q}{ijk}\bar{D}\dvp{c}{i}\tilde{d}\ud{c}{Rj}\tilde{u}\ud{c}{Rk},\mbox{\quad and}\\\nn\\
\cc{W}_2 &=&           g\ud{N}{ijk}\tilde{\kk{N}}\ud{c}{i}D\dvp{c}{j}\tilde{d}\ud{c}{Rk}
                     + g\ud{E}{ijk}\tilde{e}\ud{c}{Ri}D\dvp{c}{j}\tilde{u}\ud{c}{Rk}
                     + g\ud{D}{ijk}\tilde{Q}\dvp{c}{Li}.\tilde{L}\dvp{c}{Lj}\bar{D}\dvp{c}{k}.
\ee
It is now clear that the effective $\mu$ parameter is given by
\be
\mu &=& \f{\lambda\dz{333}s}{\sqrt{2}},
\ee
generating the term $\mu H\dz{d3}.H\dz{u3}$ in the superpotential.
The $\mu$ problem is solved since $s$ is of order the SSB scale and $\lambda_{333}$
is perturbative.

\subsection{Discrete symmetries of the superpotential}
\label{essm:discrete}

It should be noted that simply due to gauge invariance the superpotential of the
$\E$SSM is already invariant
under the $\dd{Z}\ud{M}{2}$ imposed on the MSSM provided that the
exotic $\bar{D}$ and $D$ bosons and the SM-singlet bosons
are interpreted as being $\dd{Z}\ud{M}{2}$-even along with the Higgs doublets.
The squarks and sleptons, including the RH sneutrinos,
are $\dd{Z}\ud{M}{2}$-odd.
The $U(1)\dz{B-L}$-violating terms of the MSSM superpotential that
matter parity is invoked to forbid are never present in the renormalisable
$\E$SSM superpotential since they would
violate the extra surviving $U(1)\dz{N}$ gauge symmetry. Importantly, all of the
$U(1)\dz{B-L}$-preserving MSSM terms are gauge invariant with the exception of
the $\mu$ term. Again the $\dd{Z}\ud{M}{2}$-odd states are known as
the supersymmetric particles and in the $\E$SSM the LSP is automatically stable.

In order for non-diagonal flavour transitions arising
from the Higgs sector to be suppressed, the superpotential is assumed to obey
an approximate $\dd{Z}\dz{2}$
symmetry known as $\dd{Z}\ud{H}{2}$. Under this symmetry all of the fields in
the superpotential other
than $S\dz{3}$, $H\dz{d3}$, and $H\dz{u3}$ are odd. It is this approximate
symmetry that distinguishes between the active and inert generations of
Higgs doublets and SM-singlets, with the inert generations having suppressed
couplings to matter and not radiatively acquiring VEVs.
This approximate symmetry suppresses $\lambda\dz{ijk}$ couplings of
the forms $\lambda\dz{\alpha 33}$, $\lambda\dz{3\alpha 3}$,
$\lambda\dz{33\alpha}$, and $\lambda\dz{\alpha\beta\gamma}$,
where $\alpha,\beta,\gamma \in \{1,2\}$,
indexing the inert generations only.
Such an approximate
$\dd{Z}\ud{H}{2}$ symmetry, with a stable hierarchy of couplings, can
be realised in $\E$SSM flavour theories
such as the one proposed in ref.~\cite{Howl2010}.
The symmetry cannot be exact or else the lightest of the exotic coloured
states would be absolutely stable. The existence of such stable coloured exotics
contradicts observation~\cite{Smith1988}.

Although the $U(1)\dz{B-L}$-violating terms of the MSSM are forbidden by gauge
symmetry, since the $\dd{Z}\ud{H}{2}$ cannot be exact
another exact discrete symmetry must be imposed on the superpotential
in order to avoid rapid proton decay caused by the terms in the $\cc{W}\dz{1}$
and $\cc{W}\dz{2}$ that involve the exotic coloured states.
There are two ways to impose an appropriate $\dd{Z}\dz{2}$ symmetry on
$\cc{W}$ that lead to baryon and lepton number conservation.
The first option is to impose a symmetry called $\dd{Z}\ud{L}{2}$ under which
only the sleptons, including the RH sneutrinos, are odd.
In this case the superpotential is equal to $\cc{W}\dz{0} + \cc{W}\dz{1}$ and
the model is known as the $\E$SSM-I.
$U(1)\dz{B}$ and $U(1)\dz{L}$ are symmetries of the renormalisable
superpotential if the exotic coloured states $\bar{D}$ and $D$ are,
respectively, diquarks and antidiquarks, with $B = \pm\nf{2}{3}$ and $L = 0$.
The second option is to impose a symmetry called $\dd{Z}\ud{B}{2}$ under which
both the sleptons and the exotic $\bar{D}$ and $D$ bosons are odd.
In this case the superpotential is equal to $\cc{W}\dz{0} + \cc{W}\dz{2}$ and
the model is is known as the $\E$SSM-II.
$U(1)\dz{B}$ and $U(1)\dz{L}$ are symmetries of the superpotential
if the exotic coloured states $\bar{D}$ and $D$ are,
respectively, antileptoquarks and leptoquarks, with $B = \mp1$ and $L = \mp1$.

All of these potential exact and approximate discrete symmetries of the
superpotential (\ref{eq:w}) are summarised in table~\ref{tab:Z2}.

\begin{table}\begin{center}
\begin{tabular}{r|cccc|}
& $\dd{Z}\ud{M}{2}$ & $\dd{Z}\ud{L}{2}$ & $\dd{Z}\ud{B}{2}$ & $\dd{Z}\ud{H}{2}$ \\\hline
$S\dvp{c}{\alpha},H\dvp{c}{d\alpha},H\dvp{c}{u\alpha}$ & $+$ & $+$ & $+$ & $-$ \\&&&&\\
$S\dvp{c}{3},H\dvp{c}{d3},H\dvp{c}{u3}$                & $+$ & $+$ & $+$ & $+$ \\&&&&\\
$\tilde{Q}\dvp{c}{Li},\tilde{d}\ud{c}{Ri},\tilde{u}\ud{c}{Ri}$
                                                       & $-$ & $+$ & $+$ & $-$ \\&&&&\\
$\tilde{L}\dvp{c}{Li},\tilde{e}\ud{c}{Ri},\tilde{N}\ud{c}{i}$
                                                       & $-$ & $-$ & $-$ & $-$ \\&&&&\\
$\bar{D}\dvp{c}{i},D\dvp{c}{i}$                        & $+$ & $+$ & $-$ & $-$ \\\hline
\end{tabular}
\caption{The charges of the fields of the $\E$SSM superpotential under
various exact and approximate $\dd{Z}\dz{2}$ symmetries that the superpotential may or may
not obey. $\dd{Z}\ud{M}{2}$ is already a symmetry due to gauge invariance.
Either $\dd{Z}\ud{L}{2}$ or $\dd{Z}\ud{B}{2}$ is imposed in order to avoid rapid
proton decay. $\dd{Z}\ud{H}{2}$ is an approximate flavour symmetry.
$i\in\{1,2,3\}$ and $\alpha\in\{1,2\}$.\label{tab:Z2}}
\end{center}\end{table}

It should be noted that, although the matter that survives down to low energy
form three complete $27$ representations of the broken $\E$, with the exception of
the uncharged RH neutrinos, to ensure anomaly cancellation, the imposed
exact discrete symmetries and approximate flavour symmetries do not commute with $\E$.

\subsection{Non-Higgs supermultiplets and RH neutrinos}

It is known that in the model as presented thus far the gauge couplings, though
on course to unify, do not unify below the Plank scale. The beta functions above
the SSB scale are modified compared to those of the MSSM by the existence of the
extra matter. For example, above the SSB scale the QCD beta
function is in fact zero at one-loop order.
This issue can be solved by
having the $\E$ GUT group be broken to an intermediate group before being
broken finally to $\GSM \otimes U(1)\dz{N}$ as shown in ref.~\cite{Howl2008}.

The canonical solution~\cite{King2006,King2006a,King2007}, however, is to introduce into the
superpotential a bilinear term
involving extra fields, known as non-Higgs fields, from extra incomplete
$27$ and $\overline{27}$ representations
known as $27'$ and $\overline{27}\vphantom{27}'$
\be
\cc{W}' = \mu' H' \bar{H}\vphantom{H}',
\ee
where $H'$ is the $H\dz{d}$ field from $27'$ and $\bar{H}\vphantom{H}'$
is the corresponding field from $\overline{27}\vphantom{H}'$.
These supermultiplets taken together do not spoil gauge anomaly cancellation.
To some extent this solution reintroduces the
$\mu$ problem, but $\mu'$ is not required to be related to the EWSB scale and in
order to observe satisfactory
gauge coupling unification it is only required that $\mu' \lesssim 100$~TeV.
The unification of the gauge
couplings in the $\E$SSM can then be achieved for any phenomenologically
acceptable value of $\alpha\dz{3}$ at the EWSB scale consistent with the
measured low energy central value. This is unlike in the MSSM where
significantly higher values of $\alpha\dz{3}$ are required at the EWSB scale,
well above the central measured value~\cite{King2007}.

Since RH neutrinos are completely uncharged
they can acquire very heavy Majorana masses, allowing for a type-I see-saw mechanism.
Furthermore, in the early universe
the heavy RH neutrinos, which can each decay into final states with lepton number
either $+1$ or $-1$, can create a lepton asymmetry, leading to
successful leptogenesis~\cite{King2008}.

\section{$U(1)$ Gauge Boson and Gaugino Mixing, EWSB Scale Gaugino Mass Relations, and $Z$-$Z'$ Mixing}
\label{ref:GaugeMixing}

In the low energy Lagrangian of the $\E$SSM
as well as the $U(1)$ gauge boson kinetic terms contained in (\ref{eq:superLagrangian})
\be
-\f{1}{4}\cc{B}\uvp{\mu\nu}{\mu\nu}\cc{B}\dvp{\prime}{\mu\nu}
- \f{1}{4}\cc{B}\uvp{\prime\mu\nu}{\mu\nu}\cc{B}\ud{\prime}{\mu\nu},
\ee
where the first term is the kinetic term for the $U(1)_Y$ gauge boson $B$, with
$\cc{B}_{\mu\nu} = \partial_\mu B_\nu - \partial_\nu B_\mu$, and
the second term is the kinetic term for the $U(1)_N$ gauge boson $B\uz{\prime}$,
the term
\be
-\f{\sin(\chi)}{2}\cc{B}\uvp{\mu\nu}{\mu\nu}\cc{B}\ud{\prime}{\mu\nu}\label{eq:bosonMixing}
\ee
is also gauge invariant. At the GUT scale the coefficient $\sin(\chi)$
must be equal to zero since this kinetic mixing term violates the $\E$ gauge symmetry.
Furthermore, this $\E$-breaking mixing term is not induced by radiative corrections
as long as only complete representations of $\E$ survive down to low energy.
If the non-Higgs supermultiplets are present, however, a non-zero $\sin(\chi)$
can be induced~\cite{King2006}.

Making the change of variables~\cite{Langacker1998}
\be
B\dvp{\prime}{\mu} &\rightarrow& B\dvp{\prime}{\mu} - B\ud{\prime}{\mu}\tan(\chi),\nn\\
B\ud{\prime}{\mu} &\rightarrow& \f{B\ud{\prime}{\mu}}{\cos(\chi)}
\ee
the mixing term (\ref{eq:bosonMixing}) is eliminated from the Lagrangian, but
in the covariant derivative one must make the substitution
\be
g\ud{\prime}{1}Q\uz{N}B\uz{\prime} &\rightarrow&
\left(g\ud{\prime}{1}\f{Q\uz{N}}{\cos(\chi)} - g\dvp{\prime}{1}Q\uz{Y}\tan(\chi)\right)B\uz{\prime}
= g\ud{\eff}{1}Q\uz{\eff}B\uz{\prime},
\ee
using an effective $g\ud{\prime}{1}$ coupling and effective
$Q\uz{N}$ charges
\be
g\ud{\eff}{1} &=& \f{g\uz{\prime}}{\cos(\chi)}\quad\mbox{and}\nn\\
Q\uz{\eff} &=& Q\uz{N} - \f{g\dvp{\prime}{1}}{g\ud{\prime}{1}}Q\uz{Y}\sin(\chi).
\ee

However, even in the presence of non-Higgs doublets the EWSB scale relations
$g\ud{\eff}{1} = g\ud{\prime}{1} = g\dvp{\prime}{1}$ and $Q\uz{\eff} = Q\uz{N}$
are expected to be satisfied at one-loop level to within one-loop accuracy~\cite{King2006}.

\subsection{Soft Gaugino Masses}
\label{sub:smallGauginoMixing}

In the SSB breaking part of the Lagrangian the $\E$-violating soft mass term
\be
M\dz{11}\tilde{B}\uz{ac\dagger}\tilde{B}\uz{\prime a} + \mbox{c.c.}
\ee
can also be induced at low energy, even though it is forbidden at the GUT scale.
Along with the gauge kinetic mixing, however, this soft gaugino mass maxing
is also expected to be small~\cite{Kalinowski2009}.

If at the GUT scale the soft gaugino masses
$M\dvp{\prime}{3} = M\dvp{\prime}{2} = M\dvp{\prime}{1} = M\ud{\prime}{1} = M\dvp{\prime}{1/2}$
and $M\dvp{\prime}{11}=0$, as required by $\E$ gauge invariance, then, due to the RGEs,
at the EWSB scale one expects
$M\ud{\prime}{1} \approx M\dvp{\prime}{1} \approx \nf{1}{2}M\dvp{\prime}{2} \gg M\dvp{\prime}{11}$~\cite{King2006}.

\subsection{$Z$-$Z'$ mixing}
\label{sub:ZZ}

The three VEVs $v\dz{d}$, $v\dz{u}$, and $s$ do not just induce diagonal masses for the
$Z$ and $Z'$ bosons, but also induce a mixing term. The induced $Z$-$Z'$ mass-squared
matrix is
\be
\left(\ba{cc}m\ud{2}{Z} & m\ud{2}{ZZ'}\\m\ud{2}{ZZ'} & m\ud{2}{Z'}\ea\right),
\ee
where
\be
m\ud{2}{Z} &=& \f{\bar{g}\uz{2}}{4}v\uz{2},\nn\\\nn\\
m\ud{2}{ZZ'} &=& \f{\bar{g}g\ud{\prime}{1}}{2}v\uz{2}\bigg(Q\ud{N}{d}\cos^2(\beta)
- Q\ud{N}{u}\sin^2(\beta)\bigg),\quad\mbox{and}\nn\\\nn\\
m\ud{2}{Z'} &=& g\ud{\prime 2}{1}v\uz{2}\bigg(Q\ud{N2}{d}\cos^2(\beta)
+ Q\ud{N2}{u}\sin^2(\beta)\bigg) + g\ud{\prime 2}{1}Q\ud{N2}{S}s\uz{2},\label{eq:ZZ}
\ee
with $Q\ud{N}{d,u,S}$ the $U(1)\dz{N}$ charges a down-type Higgsinos,
up-type Higgsinos and singlinos respectively, given in table~\ref{tab:eSSMCharges}.
The mass eigenstates are then
\be
Z\dz{1} &=& Z\cos(\alpha\dz{ZZ'}) + Z'\sin(\alpha\dz{ZZ'})\quad\mbox{and}\nn\\
Z\dz{2} &=& -Z\sin(\alpha\dz{ZZ'}) + Z'\cos(\alpha\dz{ZZ'}),
\ee
where
\be
\alpha\dz{ZZ'} &=& \f{1}{2}\arctan\left(\f{2m\ud{2}{ZZ'}}{m\ud{2}{Z'} - m\ud{2}{Z}}\right).\label{eq:alphaZZ}
\ee

Experimental limits on the $Z\dz{2}$ boson mass and on the mixing angle $\alpha\dz{ZZ'}$
are model dependant since in different models that involve a $Z'$ boson the couplings
of that $Z'$ will depend on the model. In the $\E$SSM the most recent limit on the $Z\dz{2}$ boson,
set by the ATLAS collaboration~\cite{Collaboration2011}, searching for dilepton
resonances, is $m_{Z_2} > 1520$~GeV at a confidence level of 95\%.
This analysis is for a $Z'$ boson associated with the extra $U(1)_N$ of the $\E$SSM,
but neglects any other matter beyond that of the SM.
When decays of the $Z_2$ boson into inert neutralinos (inert Higgsino and singlino dominated mass eigenstates) are considered the $Z_2$ width
tends to increase by a factor of about 2~\cite{Athron2011}. This then means that the branching ratio
into leptons is decreased by a factor of about 2. Estimating the effect of halving this expected branching ratio
on the analysis in ref.~\cite{Collaboration2011} one can read off a 95\% confidence level
lower bound of around 1350~GeV.
At the times of the publications of \textbf{papers~I, II, and~III} the most recent available limits were
$m_{Z_2} > 861$~GeV~\cite{Aaltonen2009}, $m_{Z_2} > 865$~GeV~\cite{Erler2010}, and $m_{Z_2} > 892$~GeV~\cite{Accomando2011}
respectively, all at confidence levels of 95\%. Limits on the angle $\alpha\dz{ZZ'}$
typically require it be less than order $10\uz{-3}$~\cite{Erler2009}. This means that
neglecting $\alpha\dz{ZZ'}$ and setting $\mass{Z\dz{1}} = m\dz{Z}$ and
$\mass{Z\dz{2}} = m\dz{Z'} \approx g\ud{\prime}{1}Q\ud{N}{S}s$ is
in most cases an excellent approximation.

\section{EWSB and the Active Higgs Boson Mass Eigenstates}
\label{ref:Higgs}

The EWSB active Higgs potential of the two active Higgs doublets $H\dz{d} \equiv H\dz{d3}$
and $H\dz{u} \equiv H\dz{u3}$ and
the active SM-singlet $S \equiv S\dz{3}$ is
\be
V(H\dz{d},H\dz{\vphantom{d}u},S) &=& \lambda^2|S|^2\Big(|H_d^{\vphantom{\dagger}}|^2+|H_{\vphantom{d}u}^{\vphantom{\dagger}}|^2\Big)
+ \lambda^2|H_d^{\vphantom{\dagger}}.H_{\vphantom{d}u}^{\vphantom{\dagger}}|^2\nn\\
&&\vphantom{A} + \f{g_2^2}{2}\Big(H_d^\dagger \tau^a H_d^{\vphantom{\dagger}}+H_{\vphantom{d}u}^\dagger \tau^a H_{\vphantom{d}u}^{\vphantom{\dagger}}\Big)
\Big(H_d^\dagger \tau^a H_d^{\vphantom{\dagger}}+H_{\vphantom{d}u}^\dagger \tau^a H_{\vphantom{d}u}^{\vphantom{\dagger}}\Big)\nn\\
&&\vphantom{A} + \frac{g\uz{\prime 2}}{8}\Big(|H_d^{\vphantom{\dagger}}|^2-|H_{\vphantom{d}u}^{\vphantom{\dagger}}|^2\Big)^2  + \frac{g\ud{\prime 2}{1}}{2}\Big(Q\ud{N}{d}|H_d^{\vphantom{\dagger}}|^2
+ Q\ud{N}{{\vphantom{d}u}}|H_{\vphantom{d}u}^{\vphantom{\dagger}}|^2+Q\ud{N}{S}|S|^2\Big)^2\nn\\
&&\vphantom{A} + m_{S}^2|S|^2+m_d^2|H_d^{\vphantom{\dagger}}|^2+m_{\vphantom{d}u}^2|H_{\vphantom{d}u}^{\vphantom{\dagger}}|^2\nn\\
&&\vphantom{A} + \Big[\lambda A_{\lambda}SH_d^{\vphantom{\dagger}}.H_{\vphantom{d}u}^{\vphantom{\dagger}} + \mbox{c.c.}\Big] + \Delta,
\label{eq:higgs1}\ee
where $m_S$, $m_d$, and $m_{\vphantom{d}u}$ are the soft scalar masses for $S$, $H_d$, and $H_{\vphantom{d}u}$ respectively
and $\Delta$ represents the contributions from loop corrections.
Once again $\tau^a = \sigma^a/2$ in $SU(2)_L$ doublet space
and $H\dvp{0}{d}.H\dvp{0}{{\vphantom{d}u}} = H\ud{-}{d}H\ud{+}{{\vphantom{d}u}} - H\ud{0}{d}H\ud{0}{{\vphantom{d}u}}$.
We define $\lambda \equiv \lambda_{333}$ and $A_\lambda$ is then the corresponding SSB parameter.

Initially this EWSB sector involves ten degrees of freedom. Four of these, however, are
massless Goldstone modes which provide the longitudinal polarisations of the massive $W^{\pm}$, $Z_1$,
and $Z_2$ bosons.
When $CP$ invariance is preserved the other six degrees of freedom form
one charged complex scalar, one $CP$-odd pseudoscalar, and three $CP$-even real Higgs states.
The masses of the charged and pseudoscalar Higgs bosons are
\be
m^2_{H^{\pm}} &=& \f{\sqrt{2}\lambda A_{\lambda}}{\sin(2\beta)}s-\frac{\lambda^2}{2}v^2+m_W^2+\Delta_{\pm}\quad\mbox{and}\\\nn\\
m^2_{A} &=& \f{\sqrt{2}\lambda A_{\lambda}}{\sin(2\varphi)}v + \Delta_A,
\label{eq:higgs2}
\ee
where $\Delta_{\pm}$ and $\Delta_A$ are loop corrections and
\be
\tan(\varphi) &=& \f{v}{2s}\sin(2\beta).
\ee

The $CP$-even active Higgs sector comprises $\kk{Re}\,H_d^0$, $\kk{Re}\,H_u^0$ and $\kk{Re}\,S$.
In the field-space basis
\be
\left(\ba{ccc}h & H & N\ea\right)^\rr{T},\nn
\ee
rotated by $\beta$ with respect to the standard interaction basis such that
\be
\kk{Re}\,H_d^0 &=& \f{1}{\sqrt{2}}\Big(h\cos(\beta) - H\sin(\beta) + v_d\Big),\nn\\\nn\\
\kk{Re}\,H_u^0 &=& \f{1}{\sqrt{2}}\Big(h\sin(\beta) + H\cos(\beta) + v_u\Big),\quad\mbox{and}\nn\\\nn\\
\kk{Re}\,S     &=& \f{1}{\sqrt{2}}\Big(N+s\Big),
\label{eq:higgs3}
\ee
the mass matrix for the $CP$-even Higgs sector is~\cite{Athron2009}
\be
\left(
\ba{ccc}
\f{\partial^2 V}{\partial v^2}&
\f{1}{v}\f{\partial^2 V}{\partial v \partial\beta}&
\f{\partial^2 V}{\partial v \partial s}\\[0.3cm]
\f{1}{v}\f{\partial^2 V}{\partial v \partial\beta}&
\f{1}{v^2}\f{\partial^2 V}{\partial^2\beta}&
\f{1}{v}\f{\partial^2 V}{\partial s \partial\beta}\\[0.3cm]
\f{\partial^2 V}{\partial v \partial s}&
\f{1}{v}\f{\partial^2 V}{\partial s \partial\beta}&
\f{\partial^2 V}{\partial^2 s}
\ea
\right) &=& \left(
\ba{ccc}
\vphantom{\f{\partial^2 V}{\partial^2 s}}M_{11}^2 & M_{12}^2 & M_{13}^2\\
\vphantom{\f{\partial^2 V}{\partial^2 s}}M_{12}^2 & M_{22}^2 & M_{23}^2\\
\vphantom{\f{\partial^2 V}{\partial^2 s}}M_{13}^2 & M_{23}^2 & M_{33}^2
\ea
\right),
\label{eq:higgs4}
\ee
where
\be
M_{11}^2&=&\f{\lambda^2}{2}v^2\sin^2(2\beta)+\f{\bar{g}^2}{4}v^2\cos^2(2\beta)+g^{\prime 2}_1 v^2\Big(Q^N_d\cos^2(\beta)+
Q^N_u\sin^2(\beta)\Big)^2+\Delta_{11},\nn\\
M_{12}^2 &=& \left(\frac{\lambda^2}{4}-\f{\bar{g}^2}{8}\right)v^2
\sin (4\beta)\nn\\
&&\vphantom{A} + \f{g^{\prime 2}_1}{2}v^2\Big(Q^N_u-Q^N_d\Big)\Big(Q^N_d\cos^2(\beta)
+Q^N_u\sin^2(\beta)\Big)\sin(2\beta)+\Delta_{12},\nn\\
M_{22}^2&=&\f{\sqrt{2}\lambda A_{\lambda}}{\sin(2\beta)}s+\left(\frac{\bar{g}^2}{4}-\f{\lambda^2}{2}\right)v^2
\sin^2(2\beta)+\f{g^{\prime 2}_1}{4}\Big(Q^N_u-Q^N_d\Big)^2 v^2 \sin^2(2\beta)+\Delta_{22},\nn\\
M_{23}^2 &=& -\f{\lambda A_{\lambda}}{\sqrt{2}}v\cos(2\beta)+\f{g^{\prime 2}_1}{2}\Big(Q^N_u-Q^N_d\Big)Q^N_S
v s\sin(2\beta)+\Delta_{23},\nn\\
M_{13}^2 &=& -\f{\lambda A_{\lambda}}{\sqrt{2}}v\sin(2\beta)+\lambda^2 v s+g^{\prime 2}_1\Big(Q^N_d\cos^2(\beta)+
Q^N_u\sin^2(\beta)\Big)Q^N_S v s+\Delta_{13},\quad\mbox{and}\nn\\
M_{33}^2 &=& \f{\lambda A_{\lambda}}{2\sqrt{2}}\f{v^2}{s}\sin(2\beta)+m_{Z'}^2+\Delta_{33}.
\label{eq:higgs5}
\ee

In (\ref{eq:higgs5}) $\Delta_{ij}$ are the contributions from loop corrections
which in the leading one-loop
approximation are rather similar to the ones calculated in the NMSSM.
Explicit expressions for
$\Delta_{ij}$ in the leading one-loop approximation
are given in ref.~\cite{Athron2009}.
Since the smallest eigenvalue of the mass-squared
matrix (\ref{eq:higgs4}) is always less than its smallest diagonal element,
at least one Higgs scalar in the $CP$-even sector, approximately $h$, always remains light,
\ie~$m^2_{h_1}\lesssim M_{11}^2$.
In the leading two-loop approximation the mass of the lightest Higgs boson in the
E$_6$SSM does not exceed about $150$--$155$~GeV.
The field-space state $h$ has couplings to SM matter identical to those of the SM
Higgs boson for all values of $\tan(\beta)$.

When the visible sector SSB mass scale and
the active SM-singlet VEV $s$ are considerably larger
than the EWSB scale, the mass-squared matrix (\ref{eq:higgs4}) has a hierarchical structure and
the masses of the heaviest Higgs bosons are closely approximated by
the diagonal entries $M_{22}^2$ and
$M_{33}^2$~\cite{King2006}. As a result the mass of one of the two heavier $CP$-even Higgs bosons,
predominantly $H$, is approximately $m_A$ while the mass of the other, predominantly $N$,
is approximately $m_{Z'}$. When $\lambda\gtrsim g'_1$ vacuum stability requires $m_A$
to be considerably larger than $m_{Z'}$ and
the EWSB scale so that the qualitative pattern of the Higgs spectrum is rather
similar to the one that arises in the PQ-symmetric NMSSM~\cite{Miller2004}.
In this limit the heaviest $CP$-even, the $CP$-odd, and the charged states are
almost degenerate with masses around $m_A$~\cite{King2006}.

\cleardoublepage

\newpage
\chapter{Thermal Relic Dark Matter}
\label{ref:dm}

Non-baryonic dark matter is an unknown form of matter that is believed to make
up the majority of the matter energy density of the universe. It interacts
either very weakly or not at all electromagnetically. The existence of
dark matter was first proposed when analysis of orbital motion within the Coma galaxy cluster
using the virial theorem
implied that there was more mass present than just that
of the visible baryonic matter~\cite{Zwicky1937}.

Currently the evidence for the existence of cosmological dark matter is very
strong. Its existence is inferred from galactic rotation curves~\cite{Salucci1997} and from
various measurements of galaxy clusters (see for example ref.~\cite{Vikhlinin2006}). There is also evidence from
observations of mass inferred from gravitational lensing (see for example refs.~\cite{Taylor1998,Markevitch2004}), but our best
measurements of the amount of cosmological dark matter come from fits to
CMB data in the context of the standard cosmological model $\Lambda$CDM (CDM plus dark energy). Such
fits to WMAP data~\cite{Dunkley2009} in particular give
\be
\Omega\dz{B}h\uz{2} &=& 0.0227 \pm 0.0006\mbox{\quad(1-sigma)}
\ee
for the present baryon energy density and
\be
\Omega\dz{\DM} h\uz{2} &=& 0.110 \pm 0.006\mbox{\quad(1-sigma)}\label{eq:omegaDM}
\ee
for dark, non-baryonic matter, where $h \approx 0.73$ is the reduced Hubble
parameter and $\Omega$ is the energy density divided by the critical
density $\rhoCrit$. It is thought that the majority of dark matter must be
non-relativistic in order for the observed large structure formation to be
explained.

A standard assumption for pre-BBN cosmology is that the DMP
was at some time prior to BBN in thermal and chemical equilibrium with the
photon and other species
still themselves in equilibrium with the photon.  At some time in the past it
would have then decoupled from equilibrium and under this assumption one can
predict the relic density today of the DMP in some model if one knows all of the
model parameters.
The chemical decoupling happens roughly when the particle's
inelastic interaction rate (maintaining chemical equilibrium) becomes less than
the expansion rate of the universe $H = \dot{a}/a$.  When this freeze-out
occurs the number density of the frozen-out species typically remains much
larger than it would have been if the species had remained in
equilibrium with the photon as the universe cooled.
If such a thermal relic particle has a
freeze-out temperature $T\uz{F}$
that is much less than the mass of the
particle such that the particle was non-relativistic at freeze-out
then it is known as CDM.

\section{The Boltzmann Equation}
\label{dm:be}

Let us assume that in some model some number of particle species, labelled with
$i$ in order of ascending mass $m\dz{i}$, are odd under
some symmetry $\ZZ{D}$ such that the lightest one is stable and the DMP.
The evolution of the cosmological
number density $n\dz{i}$ of a $\ZZ{D}$-odd particle species in the early
universe can be expressed as
\be
\dot{n}\dz{i} &=& -3Hn\dz{i}
- \sum_j \langle \sigma\dvp{\eq}{ij}v\dvp{\eq}{ij}\rangle \bigl(n\dvp{\eq}{i}n\dvp{\eq}{j} - n\ud{\eq}{i}n\ud{\eq}{j}\bigr) \nn\\
&& \quad\vphantom{A} - \sum_{j\ne i} \Bigl[\Gamma\dvp{\eq}{ij}\bigl(n\dvp{\eq}{i}-n\ud{\eq}{i}\bigr)
- \Gamma\dvp{\eq}{ji}\bigl(n\dvp{\eq}{j} - n\ud{\eq}{j}\bigr)\Bigr] \nn\\
&& \quad\vphantom{A} - \sum_{j\ne i}\sum_X \Bigl[\langle\sigma'\dz{Xij}v\dvp{\eq}{iX}\rangle\bigl(n\dvp{\eq}{i}n\dvp{\eq}{X}
- n\ud{\eq}{i}n\ud{\eq}{X}\bigr)\nn\\
&& \qquad\qquad\qquad\quad\vphantom{A} - \langle\sigma'\dz{Xji}v\dvp{\eq}{jX}\rangle\bigl(n\dvp{\eq}{j}n\dvp{\eq}{X} - n\ud{\eq}{j}n\ud{\eq}{X}\bigr)\Bigr].
\label{eq:dotni}\ee
The first term accounts for Hubble expansion and the second term accounts for
annihilations with other $\ZZ{D}$-odd particles,
including self-annihilations. The third term represents the decays
of $\ZZ{D}$-odd particles $i$ into other $\ZZ{D}$-odd species $j$
as well as decays of other $\ZZ{D}$-odd species into species $i$. The final
term represents the inelastic scattering of supersymmetric particles $i$ off of
$\ZZ{D}$-even particles $X$ into other $\ZZ{D}$-odd species $j$
and vice versa~\cite{Griest1991,Schelke2004}.

Summing up these equations yields the somewhat simpler expression
\be
\dot{n} \equiv \sum_i\dot{n}\dz{i} &=&
-3Hn - \sum_i\sum_j\langle\sigma\dvp{\eq}{ij}v\dvp{\eq}{ij}\rangle\bigl(n\dvp{\eq}{i}n\dvp{\eq}{j}
- n\ud{\eq}{i}n\ud{\eq}{j}\bigr).
\label{eq:nsum}
\label{eq:dotn}
\ee
It should be noted that, assuming that all heavier $\ZZ{D}$-odd particles
decay into the DMP with not too long a lifetime, after thermal freeze-out the
relic DMP number density will subsequently becomes equal to $n$.

During thermal freeze-out the annihilation rates of
the $\ZZ{D}$-odd particles become small compared to
the expansion rate of the universe and their number densities
become larger than their (non-relativistic) equilibrium values.
The universe expands too
fast for the number densities to track their equilibrium values.
Let us assume, however, that these states inelastically scatter off of
SM states $X$
frequently enough that the ratios of the number densities of the $\ZZ{D}$-odd
particles do maintain their equilibrium values during the time of thermal freeze-out.
We shall refer to this as \textbf{condition~A}
and assuming that it is satisfied we have
\be
\frac{n\dvp{\eq}{j}}{n\dvp{\eq}{i}} = \frac{n\ud{\eq}{j}}{n\ud{\eq}{i}} &\Rightarrow&
\frac{\vphantom{n\ud{\eq}{j}}n\dvp{\eq}{i}}{\vphantom{n\ud{\eq}{i}}n} = \frac{\vphantom{n\ud{\eq}{j}}n\ud{\eq}{i}}{\vphantom{n\ud{\eq}{i}}n\dz{\eq}},
\ee
which allows us to rewrite (\ref{eq:nsum}) as
\be
\dot{n} &=& -3Hn - \langle\sigma v\rangle\bigl(n\uvp{2}{\eq} - n\ud{2}{\eq}\bigr),
\label{eq:dotn2}
\ee
where $n\dz{\eq} \equiv \sum_in\ud{\eq}{i}$ and the effective cross-section
\be
\langle\sigma v\rangle &=& \sum_i\sum_j\langle\sigma\dz{ij}v\dz{ij}\rangle
\frac{n\ud{\eq}{i}n\ud{\eq}{j}}{n\ud{2}{\eq}}.
\ee
Here we see that much heavier $\ZZ{D}$-odd states, with correspondingly
smaller non-relativistic
equilibrium number densities, would be present in smaller numbers
during the DMP's thermal freeze-out and
annihilation cross-sections involving them would be less important.

\section{The Freeze-Out Temperature}

The energy density of one relativistic species of \{boson, fermion\} is
\be
\rho\dvp{4}{i} &=& g\dvp{4}{i}\{1,\nf{7}{8}\}\frac{\pi\uz{2}}{30}T\ud{4}{i},
\ee
where $T\dz{i}$ is the temperature of that species and $g\dz{i}$ is
the number of degrees of freedom.  We define an effective number of relativistic
degrees of freedom $g\dz{\eff}$ for the whole system by writing
\be
\rho &=& g\dz{\eff}\frac{\pi\uz{2}}{30}T\uz{4},\label{eq:geff}
\ee
where $\rho$ is the total
density of relativistic matter and $T \equiv T\dz{\gamma}$ is the photon
temperature.  The effective number of degrees of freedom $g\dz{\eff}$
takes into account the factor of $\nicefrac{7}{8}$ for fermions and also takes
into account the fact that some species no longer in equilibrium
with the photon may have a different temperature.
The entropy density of a single species of \{boson, fermion\} with temperature
$T\dz{i}$ is
\be
s\dvp{3}{i} &=& g\dvp{3}{i}\{1,\nf{7}{8}\}\f{2\pi\uz{2}}{45}T\ud{3}{i}\label{eq:sden}
\ee
and similarly an effective number of relativistic degrees of freedom $h\dz{\eff}$ for the
whole system is defined by
\be
s &=& h\dz{\eff}\f{2\pi\uz{2}}{45}T\uz{3}.
\ee
The numbers $h\dz{\eff}$ as $g\dz{\eff}$ will differ when any species $i$
has a different temperature to the photon, with $h\dz{\eff}$ containing factors
of $(T\dz{i}/T)\uz{3}$ and $g\dz{\eff}$ containing factors of
$(T\dz{i}/T)\uz{4}$.  The number density of a non-relativistic species,
which the cold DMP should be at freeze-out, is
\be
n\dz{i} &=& g\dz{i}\left(\f{m\dz{i} T\dz{i}}{2\pi}\right)\uz{3/2}\exp\left(\frac{-m\dz{i}}{T\dz{i}}\right)
\ee
and the energy density is simply
\be
\rho\dz{i} &=& m\dz{i} n\dz{i}.
\ee

In a radiation dominated universe the expansion rate is then given by
\be
H\uz{2} &=& \f{8\pi G}{3}\rho = \f{1}{M\ud{2}{\rr{P}}}g\dz{\eff}\f{4\pi\uz{3}}{45}T\uz{4}
\equiv k\ud{2}{1}g\dz{\eff} T\uz{4},\label{eq:k1}
\ee
where we define the constant $k\dz{1}$ for future convenience.
We can approximate the freeze-out temperature $T\uz{F}$ by equating an effective
DMP interaction rate with the radiation dominated expansion rate
\be
n\ud{\eq\,F}{1}\langle\sigma v\rangle\uvp{\vphantom{\eq}F}{1} &=& \sqrt{g\dz{\eff}}\f{(T\uz{F})\uz{2}}{M\dz{\rr{P}}}
\sqrt{\frac{4\pi\uz{3}}{45}}.
\ee
It is useful to scale the temperature by the DMP mass and define
\be
x &=& \f{T}{m\dz{1}}.
\ee
One can then use the expression for the non-relativistic DMP number density to
derive the transcendental equation
\be
x\uz{F} &=& \f{1}{\ln(\xi Mm\dz{1}\langle\sigma v\rangle\uz{F}) - \nf{1}{2}\ln(x\uz{F})},
\ee
where
\begin{equation}
\xi = \frac{1}{4\pi\uz{3}}\sqrt{\f{45}{2g\dz{\eff}\uz{F}}}.
\label{eq:logxF}
\end{equation}

To see how $n$ evolves after freeze-out we first note that for isentropic
expansion the total
entropy density of the system $s \propto a\uz{-3}$, where $a$ is the scale factor of the universe.  This means that
$\dot{s}/s = -3H$.  Defining
\be
y &=& \f{n}{s}
\ee
we find
\be
\f{\rr{d}y}{\rr{d}x} &=& \f{1}{3H}\f{\rr{d}s}{\rr{d}x}\langle\sigma v\rangle\left(y\uvp{2}{\eq} - y\ud{2}{\eq}\right)\nn\\\nn\\
&=& \sqrt{\f{\pi}{45}}g\dz{*}Mm\dz{1}\langle\sigma v\rangle\left(y\uvp{2}{\eq} - y\ud{2}{\eq}\right),
\label{eq:yxODE}\ee
where
\be
g\dz{*} &=& \frac{h\dz{\eff}}{\sqrt{g\dz{\eff}}}\left[1 + \f{T}{3h\dz{\eff}}\f{\rr{d}h\dz{\eff}}{\rr{d}T}\right].
\ee
This equation can be used to find the relic density today numerically.

By integrating from $x=x\uz{0}$ today to $x=x\uz{F}$ at freeze-out
one can determine the value of $y$ today $y\uz{0}$ and
the current DMP relic density is
\be
\Omega &=& \f{m\dz{1} y\uz{0}s\uz{0}}{\rhoCrit}.
\ee
The entropy density today $s\uz{0}$ is
dominated by the cosmic microwave and neutrino backgrounds. The CMB
temperature is measured and the neutrino temperature can then be calculated as
in section~\ref{ezssm:Neff}.

The freeze-out temperature $x\uz{F}$ depends only logarithmically on the
effective cross-section, as in (\ref{eq:logxF}), but $\langle\sigma v\rangle\uz{F}$
is critical to determining how small $y$ is driven during the time around
freeze-out, before interactions become negligible. Subsequently, after the
period of thermal freeze-out, $y$ approximately remains constant as the universe
expands.

\section{Supersymmetric Dark Matter}

In supersymmetric theories with $R$-parity $\ZZ{M}$ this $\ZZ{M}$ plays the
role of $\ZZ{D}$ and the DMP is the LSP~\cite{Bertone2005}.
In such theories the LSP is typically either the lightest neutralino or the gravitino,
depending on the nature of the SSB mechanism which determines the typical scale
of the gravitino mass relative to the visible sector SSB masses. In gravity mediated supersymmetry breaking
the LSP is typically the lightest neutralino.

A sub-weak-strength interacting neutralino is generally
considered a good candidate for LSP dark matter~\cite{Ellis1984,Jungman1996}.
Neutralinos do not typically form Dirac states and as such a neutralino DMP's relic abundance in standard
cosmology is determined by thermal freeze-out and not by matter-antimatter asymmetry
as in the case of baryons.

Thermal relic neutralino dark matter has been widely studied in the
MSSM~\cite{Cotta2009,King2006b,King2007a,Ellis2008}
and cMSSM~\cite{Kane1994,Feng2000,Ellis2000,Ellis2003,Baer2002}.
A successful dark matter scenario may be realised if the LSP is the
lightest neutralino and there are various successful regions of parameter space
that have different dominant annihilation mechanisms. For example there is the bulk region,
which involves annihilation via t-channel slepton exchange; the focus point region,
which involves annihilation via t-channel chargino exchange; and the funnel region,
which involves annihilation via s-channel Higgs boson exchange.
There are also regions corresponding to coannihilation with staus or stops.

Typically in these scenarios $x\uz{F} \sim 1/20$, meaning that the dark matter is
indeed cold. Since, again, the freeze-out
temperature is only logarithmically dependent on the effective cross-section,
as in (\ref{eq:logxF}), this approximate value does not vary significantly for a wide variety
of sub-weak-strength interacting neutralino dark matter scenarios, including the
$\E$SSM scenarios described in chapters~\ref{chap:ndmin}, \ref{chap:nhde}, and~\ref{chap:ezssm}.

\cleardoublepage

\newpage
\chapter{Dark Matter in the $\E$SSM}
\label{chap:ndmin}

In this chapter, which contains work that was first published in \textbf{paper I},
we present a
study of neutralino dark matter in the presence of inert
Higgsinos and singlinos, using the
extended neutralino sector of the $\E$SSM
as an example.
The study here should be compared to the study of dark matter in the USSM
in ref.~\cite{Kalinowski2009}. The particle content of the USSM, in addition to the
states of the MSSM, also contains a SM-singlet $S$ and
a $Z'$ boson together with their fermionic superpartners
the singlino $\tilde{S}$ and the gaugino $\tilde{B}'$.
The existence of these interaction states can modify the nature of the neutralino LSP.
In this study we include the above states of the USSM and also
the extra inert doublet Higgsinos and singlinos predicted by the $\E$SSM,
but not included in the USSM~---
$\tilde{H}_{d2}$, $\tilde{H}_{d1}$, $\tilde{H}_{u2}$, $\tilde{H}_{u1}$,
$\tilde{S}_2$, and $\tilde{S}_1$. We do not, however, include the
corresponding inert scalars which do not play a role in the heavy inert scalar limit.
We also do not include any of the exotic coloured $\bar{D}$ and $D$ states
since in general we would not expect them to play a significant role in the
calculation of the dark matter relic abundance.

We study neutralino dark matter in the $\E$SSM, as defined above,
both analytically and numerically, using {\tt micrOMEGAs}~\cite{Belanger2010}.
We find that results for the relic abundance
in the $\E$SSM are radically different from those of both the MSSM
and the USSM. This is because the two inert generations of doublet Higgsinos
and singlinos predicted by the $\E$SSM provide an almost
decoupled neutralino sector with a naturally light LSP that can account for
the CDM relic abundance somewhat independently of the
rest of the model.
In plausible scenarios the LSP
annihilates predominantly through an s-channel $Z$ boson.

Imposing the conditions that the LSP has a mass greater than half of the $Z$ boson mass,
so that the LSP does not contribute at all to the $Z$ boson invisible decay width,
and accounts for all of the observed dark matter implies that $\tan(\beta)$ must be less than about 2.
Apart from this requirement on $\tan(\beta)$,
the very stringent constraints on MSSM or USSM parameter space
that come from requiring that the model explains the
observed dark matter relic density become completely relaxed since
in the $\E$SSM the neutralino dark matter depends almost
exclusively on the parameters of the almost decoupled inert neutralino sector.
We expect similar results to apply to any singlet extended
supersymmetric model with an almost decoupled inert neutralino sector comprising
extra generations of inert Higgsinos and singlinos.

In section~\ref{ndmin:inert} we discuss the inert neutralino sector
of the E$_6$SSM, introduce the effective model that we study, and highlight the most important
couplings for our analysis of the LSP dark matter relic density.
In section~\ref{ndmin:matrices} we display the complete neutralino and chargino mass
matrices of the $\E$SSM. In section~\ref{sec:analytical} we present some
analytical results that provide useful insights into the new inert
sector physics. These results are subsequently used to understand and interpret the
results of the full numerical
dark matter relic density calculation using {\tt micrOMEGAs} which are
presented in section~\ref{ndmin:results}.
The conclusions are summarised in section~\ref{ndmin:conclusions}.

\section{The Trilinear Higgs Yukawa Couplings}
\label{ndmin:inert}

The most important couplings in our analysis
are the trilinear couplings between
the three generations of down- and up-type Higgs doublets and SM-singlets
contained in the superpotential of the $\E$SSM (\ref{eq:W0})
\be
\lambda\dvp{0}{ijk}S\dvp{0}{i}H\dvp{0}{dj}.H\dvp{0}{uk} &=& \lambda\dvp{0}{ijk}(S\dvp{0}{i}H\ud{-}{dj}H\ud{+}{uk} -
S\dvp{0}{i}H\ud{0}{dj}H\ud{0}{uk}).\label{eq:lambda}
\ee
The trilinear coupling tensor $\lambda_{ijk}$
consists of 27 numbers which play various roles.
The purely third family coupling
$\lambda \equiv \lambda_{333}$ is very important, because it is the
combination $\mu = \lambda s/\sqrt{2}$ that plays the role of
an effective $\mu$ term in this theory. Some
other neutralino mass terms, such as those involving $\tilde{S}$, are
also proportional to $\lambda$. The couplings of the inert Higgs doublets to
the third generation SM-singlet
$\lambda_{\alpha\beta} \equiv \lambda_{3\alpha\beta}$
directly contribute to neutralino and chargino mass terms for the
inert Higgsino doublets. $f_{d\alpha\beta} \equiv
\lambda_{\alpha 3\beta}$ and $f_{u\alpha\beta} \equiv \lambda_{\alpha\beta 3}$
directly contribute to neutralino mass terms coupling an inert doublet Higgsino
to an inert singlino.

The 13 Higgs trilinear couplings mentioned thus far are the only
couplings that obey the proposed $\ZZ{H}$ symmetry. This
approximate flavour symmetry is proposed in order to
prevent FCNCs in the SM matter sector
by eliminating non-diagonal flavour transitions originating from the Higgs sector.
The $\ZZ{H}$ cannot be exact as discussed in subsection~\ref{essm:discrete}.
If $\lambda_{ijk}$
obeyed $\ZZ{H}$ exactly then, as we will see below, the
neutralino mass matrix (and also the chargino mass matrix) would be
decoupled into two independent systems and the lightest from each sector
would be absolutely stable. We
shall refer to the $\ZZ{H}$-breaking couplings involving two
third generation fields as $x_{d\alpha} \equiv
\lambda_{3\alpha 3}$, $x_{u\alpha} \equiv \lambda_{33\alpha}$, and
$z_\alpha \equiv \lambda_{\alpha 33}$. The notation for the
$\lambda_{ijk}$ couplings used is compiled in table~\ref{couplings}.

\begin{table}[ht]
  \centering
\begin{tabular}{c|ccccccc|}
$ijk$ & $333$ & $3\alpha\beta$ & $\alpha 3\beta$
& $\alpha\beta 3$ & $33\alpha$ & $3\alpha 3$ & $\alpha 33$ \\
\hline
$\lambda_{ijk}$ & $\lambda$ & $\lambda_{\alpha\beta}$ & $f_{d\alpha\beta}$
& $f_{u\alpha\beta}$ & $x_{d\alpha}$ & $x_{u\alpha}$ & $z_\alpha$ \\
\hline
\end{tabular}
\caption{The abbreviated notation for the
$\lambda_{ijk}$ couplings.\label{couplings}}
\end{table}

The 8 remaining $\ZZ{H}$-breaking couplings
$\lambda_{\alpha\beta\gamma}$ are of less importance for our study. As long as
only the third generation Higgs doublets and SM-singlet acquire VEVs
then these couplings do not appear in the neutralino or chargino
mass matrices. Additionally, they only appear in Feynman
rules that involve the inert Higgs scalars and we assume
that these are given SSB
masses that are heavy enough such that these particles do
not contribute to any processes relevant for this study.
Similarly we neglect the exotic coloured $\bar{D}$ and $D$ states
since we expect them to be too heavy to play a significant role in the dark
matter relic density calculation.

\section{The Neutralino and Chargino Mass Matrices}
\label{ndmin:matrices}

In the MSSM there are four neutralino interaction states~--- the
neutral wino, the bino, and the two neutral Higgsinos. In the USSM~\cite{Kalinowski2009} two
extra states are added~--- the singlino and the bino$'$. In
the conventional USSM interaction basis
\be
\tilde{N}\ud{\rr{int}}{\rr{USSM}} &=& \left(\ba{cccccc} \tilde{B} & \tilde{W}\uvp{3}{d} & \tilde{H}\ud{0}{d}&
\tilde{H}\ud{0}{u} & \tilde{S} & \tilde{B}' \ea\right)\uz{\rr{T}}
\ee
and neglecting bino-bino$'$ mixing,
as justified in ref.~\cite{Kalinowski2009}
(see also subsection~\ref{sub:smallGauginoMixing}),
the USSM neutralino mass matrix
\be\small
M\ud{N}{\rr{USSM}} &=& \left(\ba{cccccc} M\dvp{\prime}{1} & 0 & -\frac{1}{2}g'v_d & \frac{1}{2}g'v_u & 0 & 0 \\
0 & M\dvp{\prime}{2} & \frac{1}{2}gv_d & -\frac{1}{2}gv_u & 0 & 0 \\
-\frac{1}{2}g'v_d & \frac{1}{2}gv_d & 0 & -\mu & -\frac{\lambda v_u}{\sqrt{2}} & Q\ud{N}{d}g\ud{\prime}{1}v_d \\
\frac{1}{2}g'v_u & -\frac{1}{2}gv_u & -\mu & 0 & -\frac{\lambda v_d}{\sqrt{2}} & Q\ud{N}{u}g\ud{\prime}{1}v_u \\
0 & 0 & -\frac{\lambda v_u}{\sqrt{2}} & -\frac{\lambda v_d}{\sqrt{2}} & 0 & Q\ud{N}{S}g\ud{\prime}{1}s \\
0 & 0 & Q\ud{N}{d}g\ud{\prime}{1}v_d & Q\ud{N}{u}g\ud{\prime}{1}v_u & Q\ud{N}{S}g\ud{\prime}{1}s & M\ud{\prime}{1}\ea\right),
\label{eq:USSM}
\ee
where $M\dvp{\prime}{1}$, $M\dvp{\prime}{2}$, and $M\ud{\prime}{1}$ are the soft gaugino masses
and $Q\ud{N}{d,u,S}$ are the $U(1)\dz{N}$ charges of down-type Higgsinos,
up-type Higgsinos, and singlinos respectively, given in table~\ref{tab:eSSMCharges}.
In the $\E$SSM this is
extended. We take the full basis of neutralino interaction states to be
\be
\tilde{N}\uvp{\rr{int}}{\rr{USSM}} &=& \left(\ba{ccccccc}
\tilde{N}\ud{\rr{int\,T}}{\rr{USSM}} & \tilde{H}\ud{0}{d2} & \tilde{H}\ud{0}{u2}
& \tilde{S}\dvp{0}{2} & \tilde{H}\ud{0}{d1} & \tilde{H}\ud{0}{u1} & \tilde{S}\dvp{0}{1}
\ea\right)\uz{\rr{T}}.
\ee
The final six states are the extra inert doublet Higgsinos and singlinos
that appear in the full $\E$SSM.
Under the assumption that only the third generation Higgs doublets
and singlet acquire VEVs the full Majorana mass matrix is then
\be
M\uvp{N}{\rr{USSM}} &=& \left(\ba{ccc} M\ud{N}{\rr{USSM}} & B\dvp{\rr{T}}{2} & B\dvp{\rr{T}}{1}\\
B\ud{\rr{T}}{2} & A\dvp{\rr{T}}{22} & A\dvp{\rr{T}}{21}\\
B\dvp{\rr{T}}{1} & A\ud{\rr{T}}{21} & A\dvp{\rr{T}}{11}\ea\right),\label{eq:nmm}
\ee
where the submatrices involving the inert interaction states are given by
\be
A\dvp{\rr{T}}{\alpha\beta} = A\ud{\rr{T}}{\beta\alpha}
&=& -\f{1}{\sqrt{2}} \left(\ba{ccc}
0 & \lambda_{\alpha\beta} s & f_{u\beta\alpha}v\sin(\beta)\\
\lambda_{\beta\alpha} s & 0 & f_{d\beta\alpha}v\cos(\beta)\\
f_{u\alpha\beta}v\sin(\beta) & f_{d\alpha\beta}v\cos(\beta) & 0 \ea\right)\label{eq:AA}
\ee
and the $\ZZ{H}$-breaking submatrices by
\be
B\dz{\alpha} &=& -\f{1}{\sqrt{2}}
\left( \ba{ccc} 0 & 0 & 0 \\ 0 & 0 & 0 \\
0 & x_{d\alpha}s & z_\alpha v\sin(\beta) \\
x_{u\alpha}s & 0 & z_\alpha v\cos(\beta) \\
x_{u\alpha}v\sin(\beta) &
x_{d\alpha}v\cos(\beta) & 0 \\ 0 & 0 & 0
\ea\right).
\ee

Similarly we take our basis of chargino interaction states to be
\be
\tilde{C}\dvp{+}{\rr{int}} &=& \left(\ba{c}
\tilde{C}\ud{+}{\rr{int}} \\ \tilde{C}\ud{-}{\rr{int}} \ea\right),
\ee
where
\be
\tilde{C}\ud{+}{\rr{int}} = \left(\ba{c}\tilde{W}\uvp{+}{d} \\
\tilde{H}\ud{+}{u} \\ \tilde{H}\ud{u2}{+} \\ \tilde{H}\ud{+}{u1} \ea\right)
& \mbox{\qquad and\qquad} & \tilde{C}\ud{-}{\rr{int}} = \left(\ba{c}
\tilde{W}\uvp{-}{d} \\ \tilde{H}\ud{-}{d} \\ \tilde{H}\ud{-}{d2}
\\ \tilde{H}\ud{-}{d1} \ea\right).
\ee
The corresponding mass matrix is then
\be
M\uz{C} = \left(\ba{cc} & P^\rr{T} \\ P
\ea\right),
\ee
where
\be
P = \left(\ba{cccc} M_2 & \sqrt{2}m_W\sin(\beta) & 0 & 0\\
\sqrt{2}m_W\cos(\beta) & \mu & \frac{1}{\sqrt{2}}x_{d2}s &
\frac{1}{\sqrt{2}}x_{d1}s \\
0 & \frac{1}{\sqrt{2}}x_{u2}s & \frac{1}{\sqrt{2}}\lambda_{22}s & \frac{1}{\sqrt{2}}\lambda_{21}s \\
0 & \frac{1}{\sqrt{2}}x_{u1}s & \frac{1}{\sqrt{2}}\lambda_{12}s &
\frac{1}{\sqrt{2}}\lambda_{11}s \ea\right).
\label{eq:cmm}
\ee

One can already see from (\ref{eq:AA}) from that a typical feature of the
$\E$SSM is that the LSP is
composed mainly of inert singlino and ends
up being typically very light. One can see this by inspecting the
submatrices $A_{\alpha\beta}$
and assuming a hierarchy of the form
$\lambda_{\alpha\beta} s \gg f_{(u,d)\alpha\beta}v$.
This is a natural assumption since we
already require that $s \gg v$ in order to satisfy the experimental limits
on the $Z_2$ boson mass. At the time of the publication of \textbf{paper~I} the
experimental lower limit was 861~GeV, from ref.~\cite{Aaltonen2009}. The
current limit is around 1350~GeV as discussed in subsection~\ref{sub:ZZ}.

For both the neutralinos and the charginos we see that if the
$\ZZ{H}$-breaking couplings are exactly zero then the inert parts
of the neutralino and chargino mass matrices becomes decoupled from the USSM parts.
However, as previously discussed, although approximate decoupling is expected,
exact decoupling is not and will therefore not be considered.

\section{Analytical Discussion}
\label{sec:analytical}

It will be useful to get some analytical understanding of the calculation of the
dark matter relic abundance coming from the new
neutralino/chargino physics of the $\E$SSM before looking at
the results of the full numerical simulation. To this end,
in this section, we consider just
one inert generation consisting of two inert Higgs doublets and one inert SM-singlet.
We label this generation as the first generation. We shall assume that
the $\ZZ{H}$-breaking Yukawa couplings of the first Higgs generation to the
third conventional Higgs generation are large enough to allow the neutralino/chargino
states of the USSM to decay into the LSP, formed mostly from inert neutralino interaction states,
but also small enough such that we can consider the
inert neutralinos to be approximately decoupled from the rest of the
neutralino mass matrix for the purposes of obtaining an
analytical estimate of the mass eigenstates.
This all amounts to considering the single block $A_{11}$
of the extended neutralino mass matrix (\ref{eq:nmm}).

\subsection{The neutralino masses and mixing for one inert generation}

Within the first generation we use the basis
\be
\tilde{N}\uz{\rr{int}} &=& \left(\begin{array}{ccc} \tilde{H}\ud{0}{d1} &
\tilde{H}\ud{0}{u1} & \tilde{S}\dvp{0}{1} \end{array}\right)\uz{\rr{T}}
\ee
and the neutralino mass matrix is then, from (\ref{eq:AA}),
\be
A = A_{11} &=& -\frac{1}{\sqrt{2}} \left( \begin{array}{ccc} 0 &
\lambda^\prime s & f_uv\sin \beta \\ \lambda^\prime s & 0 &
f_dv\cos \beta \\ f_uv\sin \beta & f_dv\cos \beta & 0 \end{array}\right),
\ee
where $\lambda^\prime = \lambda_{11} \equiv \lambda_{311}$, $f_d =
f_{d11} \equiv \lambda_{131}$, and $f_u = f_{u11} \equiv \lambda_{113}$. As
discussed earlier, it is natural to assume that $\lambda^\prime s
\gg fv$ and this will lead to a light, mostly
first generation singlino lightest neutralino.

Finding the eigenvalues of the matrix $A$ amounts to solving a
reduced cubic equation. Expanding in $fv/\lambda^\prime s$
the three neutralino masses from the first generation are
\be
m_1 &=& \frac{1}{\sqrt{2}} \frac{f_df_u}{\lambda^\prime} \frac{v^2}{s} \sin(2\beta)
+ \cdots,\label{eq:sin2beta}\label{eq:m1}\\\nn\\
m_2 &=& \frac{\lambda^\prime s}{\sqrt{2}} - \frac{m_1}{2} + \cdots,\mbox{\quad and}\\\nn\\
m_3 &=& -\frac{\lambda^\prime s}{\sqrt{2}} - \frac{m_1}{2} + \cdots.
\ee
The lightest state $\tilde{N}\dz{1}$, with mass $m\dz{1}$,
is mostly singlino (as we will confirm below)
and the two heavier states have nearly degenerate masses, split by
$m\dz{1}$. At $\beta = 0$ or $\pi/2$ the lightest neutralino
becomes massless. This is when only one of the third generation
active Higgs doublets has a VEV.

We shall define the neutralino mixing matrix $N$ by
\be
N\ud{a}{i} M\uvp{ab}{ij} N\ud{b}{j} &=& m\dvp{a}{i}\delta\dvp{ab}{ij}\mbox{\quad(no sum on $i$)}
\ee
with superscripts indexing the interaction states and subscripts indexing the
mass eigenstates.
The lightest state is then made up of the following superposition
of interaction states:
\be
\tilde{N}_1^0 &=& N_1^1 \tilde{H}_{d1}^0 + N_1^2 \tilde{H}_{u1}^0
+ N_1^3 \tilde{S}^{\vphantom{0}}_1.
\ee
Again expanding in $fv/\lambda^\prime s$
\be
N_1 &=& \left( \begin{array}{c} -\f{f_dv}{\lambda^\prime s} \cos(\beta) + \cdots\\\\
-\f{f_uv}{\lambda^\prime s} \sin \beta + \cdots\\\\
1 - \f{1}{2} \left(\f{v}{\lambda^\prime s}\right)^2
\Big[f_d^2\cos^2(\beta) + f_u^2\sin^2(\beta)\Big] + \cdots\label{eq:N1}
\end{array}\right),
\ee
confirming that the LSP is mostly singlino in this limit. The
other eigenvectors, which determine the composition of neutralinos 2 and 3, are
\be
N_i &=& \sqrt{\frac{1}{a_i^2 + b_i^2 + \cdots}} \left(
\begin{array}{c} a_i \\ b_i \\ 1 \end{array} \right)\mbox{\quad(no sum on $i$)},
\ee
where
\begin{eqnarray}
-b_2 = a_2 &=& \frac{\lambda^\prime s}{v} \left[f_d\cos(\beta) - f_u\sin(\beta)\right]^{-1} + \cdots
\mbox{\quad and}\\ \nonumber \\
b_3 = a_3 &=& \frac{\lambda^\prime s}{v} \left[f_d\cos(\beta) + f_u\sin(\beta)\right]^{-1} + \cdots.
\end{eqnarray}
Note that $a,b \gg 1$ and that $a_2$ and $b_2$ flip sign at
$f_d\cos(\beta) = f_u\sin(\beta)$ whereas $a_3$ and $b_3$ are
always positive. Very approximately these eigenvectors are then
\begin{eqnarray}
N_2 &=& \frac{1}{\sqrt{2}} \left( \begin{array}{c} -1 \\ 1 \\ 0 \end{array} \right) \rr{sign}(f_us_\beta - f_dc_\beta) +\cdots
\mbox{\quad and}\\\nn\\\nn\\
N_3 &=& \frac{1}{\sqrt{2}} \left( \begin{array}{c} 1 \\ 1 \\ 0
\end{array} \right) +\cdots.
\end{eqnarray}

Under the assumptions of this section the chargino from the first generation is
the first generation charged Higgsino with mass $\lambda^\prime s/\sqrt{2}$.

\subsection{Annihilation Channels}

From (\ref{eq:m1}) it can be seen that the LSP mass $m_1$ is proportional to $v^2/s$
and so is naturally small since $v\ll s$. To understand this, recall that
$Z$-$Z^\prime$ mixing leads to two mass eigenstates~--- $Z_2 \approx Z'$ and $Z_1 \approx Z$~---
and limits on $Z$-$Z^\prime$ mixing and on the $Z_2$ mass place lower limits on $s$
that imply that $v\ll s$ must be satisfied. For example, when $s=3000$ GeV
the $Z_2$ mass is about 1100~GeV and $v^2/s \approx 20$ GeV. The LSP mass further decreases
as $s$ becomes larger in the considered limit.
In practice it is quite difficult to arrange the parameters such that
the LSP mass exceeds about 100 GeV, although this depends on the sizes of
Yukawa couplings that one is willing to accept
(an issue explored more thoroughly in chapter~\ref{chap:nhde}).

\begin{figure}
\begin{center}
\begin{picture}(260,100)(-20,-50)
\SetWidth{1}
\Text(-5,40)[r]{$\tilde{N}_1$}
\Text(-5,-40)[r]{$\tilde{N}_1$}
\Line(0,40)(20,0)
\Line(0,-40)(20,0)
\Photon(20,0)(60,0){2}{4}
\Text(40,5)[b]{$Z$}
\Line(60,0)(80,40)
\Line(60,0)(80,-40)
\Text(155,40)[r]{$\tilde{N}_1$}
\Text(155,-40)[r]{$\tilde{N}_1$}
\Line(160,40)(180,0)
\Line(160,-40)(180,0)
\DashLine(180,0)(220,0){4}
\Text(200,5)[b]{$H_1$}
\Line(220,0)(240,40)
\Line(220,0)(240,-40)
\end{picture}
\end{center}
\caption{S-channel LSP annihilation diagrams.\label{fig:sChan}}
\end{figure}
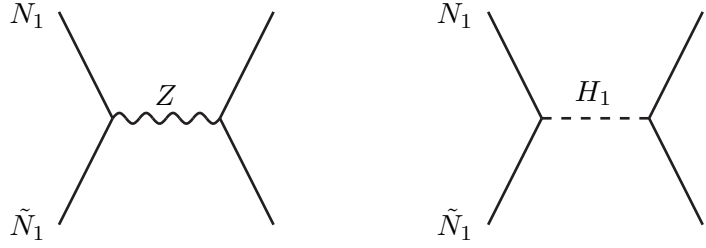

In view of the above discussion the LSP is expected to be relatively light.
When determining the important early universe annihilation channels we therefore
begin by looking at s-channel annihilation,
which can result in lighter mass final states. The most important
diagrams are shown in figure~\ref{fig:sChan} and it will turn out that the
most important of these annihilations are those with a $Z$ boson in the s-channel.
The $Z$-$\tilde{N}_1$-$\tilde{N}_1$ coupling in this diagram
is suppressed by a factor
\be
\frac{1}{2} \left(\frac{v}{\lambda^\prime s}\right)^2 \left[f_u^2\sin^2(\beta) -
f_d^2\cos^2(\beta)\right] + \cdots \nonumber
\ee
relative to the $Z$-neutrino-neutrino coupling under the assumptions of this section
since the LSP only couples through its small Higgsino components. This coupling vanishes
completely at $f_d\cos(\beta) = f_u\sin(\beta)$, which is when the
LSP contains a completely symmetric combination of
$\tilde{H}_{d1}^0$ and $\tilde{H}_{u1}^0$.
While in the MSSM a Higgsino dominated LSP would be expected to
be such a symmetric combination of down-type and up-type (active) Higgsino, with
mass around $\mu$, an inert neutralino LSP in the $\E$SSM a priori has no
reason to be close to such a symmetric combination.

Full gauge coupling strength s-channel $Z$ boson
annihilations tend to leave a relic density that is too low to
account for the observed amount of dark matter,
but in this model the coupling of the
mostly singlino LSP to the $Z$ boson is typically suppressed,
as it only couples through its doublet Higgsino admixture, leading to
an increased relic density.
As $\lambda' s$ decreases
the proportion of the LSP that is made up of inert doublet Higgsino,
rather than inert singlino, increases. This can be seen in (\ref{eq:N1}).
This then increases the strength of the
overall $Z$-$\tilde{N}_1$-$\tilde{N}_1$ coupling. The inclusive cross-section for
s-channel annihilation through a $Z$-boson is therefore highly dependent
on $\lambda' s$, which affects both the coupling and the LSP mass $m_1$.
The effect of independently increasing the coupling is always to increase the cross-section,
but the effect of independently increasing the LSP mass can be to either increase or decrease the
cross-section, depending on which side of the $Z$ boson resonance the centre-of-mass energy is
on in typical collisions during the period of thermal freeze-out.
S-channel annihilation through the
lightest Higgs boson will also become important if typical LSP collisions are on resonance.

\begin{figure}
\begin{center}
\begin{picture}(280,100)(-20,-50)
\SetWidth{1}
\Text(-5,40)[r]{$\tilde{N}_1$}
\Text(-5,-40)[r]{$\tilde{N}_1$}
\Line(0,40)(40,20)
\Line(0,-40)(40,-20)
\Line(40,20)(40,-20)
\DashLine(40,20)(80,40){4}
\DashLine(40,-20)(80,-40){4}
\Text(85,40)[l]{$H_1$}
\Text(85,-40)[l]{$H_1$}
\Text(155,40)[r]{$\tilde{N}_1$}
\Text(155,-40)[r]{$\tilde{N}_1$}
\Line(160,40)(200,20)
\Line(160,-40)(200,-20)
\Line(200,20)(200,-20)
\Photon(200,20)(240,40){2}{4}
\Photon(200,-20)(240,-40){-2}{4}
\Text(245,40)[l]{$Z,W$}
\Text(245,-40)[l]{$Z,W$}
\end{picture}
\end{center}
\caption{T-channel LSP annihilation diagrams.\label{fig:tChan}}
\end{figure}
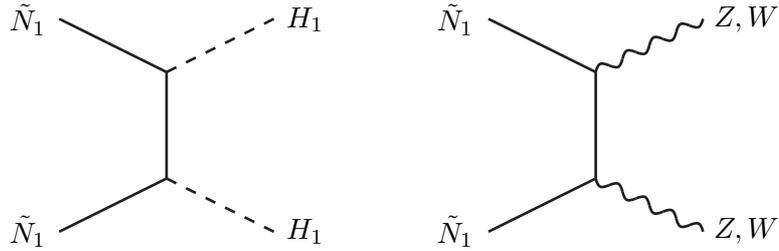

The most important of the potential t-channel processes are shown in figure~\ref{fig:tChan}.
In practice these channels will not play a significant role compared to the
s-channel annihilations considered previously, but we discuss them for completeness.
The t-channel particle for these
processes is one of the neutralinos or the chargino of the first generation.
In the first
diagram~--- t-channel annihilation to active third generation
Higgs scalars~--- the couplings are approximately just $f$ couplings of the first
generation and appropriate mixing matrix elements.  With the inert
chargino or with inert neutralino 2 or 3 in the t-channel the diagram is
approximately inert singlinos annihilating with an inert doublet
Higgsino in the t-channel and the couplings are approximately just
$f_d$ and $f_u$ for producing $H_d$ and $H_u$ interaction states respectively.
The LSP mass is smaller than the other masses by a
factor of order $v^2/s^2$. With the LSP itself in the t-channel the
first diagram therefore receives an enhancement of order $s^2/v^2$
for the t-channel propagator at low momentum, but has a
suppression of order $v^2/s^2$ in the couplings due to the LSP
only containing doublet type first generation Higgsinos with
amplitudes of order $v/s$.

The second diagram in figure~\ref{fig:tChan} represents annihilation to massive gauge bosons. To
very good approximation these bosons only couple to weak isospin
doublets and not to SM-singlets (since $Z$-$Z'$ mixing must be very small).
These diagrams therefore have a
suppression of order $v^2/s^2$ relative to the full gauge interaction strength
due to the couplings even with an inert
chargino or with inert neutralino 2 or 3 in the t-channel. On top
of this suppression these diagrams also receive an additional
suppression of order $v^2/s^2$ in the couplings, but an enhancement
of order $s^2/v^2$ in the propagator when the LSP is in the t-channel.
This second type of diagram
has a greater
chance of being kinematically allowed
than the first.

As previously stated,
inert Higgs scalars are assumed heavy and annihilation to
and/or through these particles is not considered. It should be noted though
that these particles have suppressed couplings to SM matter due to the
approximate $\ZZ{H}$ symmetry and diagrams for the annihilation of LSPs into SM matter
that involve these particles would be suppressed by these couplings.

\section{Numerical Analysis}
\label{ndmin:results}

We now turn to the full model, in which the LSP is determined from the neutralino mass matrix in
(\ref{eq:nmm}). There are two copies of the inert generation considered in
the previous section as well as six unknown mixing parameters between the
two generations. In general, after rotation to the mass
eigenstate basis, we expect that two states are much lighter than the
rest~--- both inert-singlino-like in the $\lambda^\prime s \gg fv$
limit\footnote{An exception to this would be the large $M'_1$ limit
in which the LSP could originate from the lower-right block of the USSM neutralino mass
matrix (\ref{eq:USSM}) due to a mini see-saw mechanism as discussed in ref.~\cite{Kalinowski2009}.}.

In this section we use numerical methods to
predict the relic density. We first diagonalise
the neutralino, chargino, and Higgs scalar mass
matrices numerically.
Having done this \texttt{micrOMEGAs 2.2}~\cite{Belanger2010} is then used to numerically compute the
present day relic density, including the relevant annihilation and coannihilation channel
cross-sections and the LSP freeze-out temperature $x\uz{F}$. {\tt micrOMEGAs}
achieves this by calculating all of the relevant tree level
Feynman diagrams using {\tt CalcHEP}. The {\tt CalcHEP} model files for the
considered model are generated using
{\tt LanHEP}~\cite{Semenov2010}.
The {\tt micrOMEGAs} relic density calculation
assumes standard cosmology in which the LSP dark matter was in equilibrium
with the photon at some time in the past, numerically solving (\ref{eq:yxODE}).

\subsection{The parameter space of the model}

As justified in section~\ref{ref:GaugeMixing}
we assume that differences between the GUT normalised couplings of the two $U(1)$
gauge groups $U(1)_Y$ and $U(1)_N$, as well as the mixing
between the two groups, is negligible, giving $g_1^\prime \approx
0.46$.
The free parameters are then
the trilinear Higgs couplings $\lambda_{ijk}$, the singlet VEV
$s$, $\tan(\beta)$, the soft $\lambda_{333}$ coupling $A_\lambda$,
and the soft gaugino masses. It will turn out that the soft
gaugino masses usually have little effect on the dark matter
physics. One can see this by observing the neutralino mass
matrix (\ref{eq:nmm}) where
the USSM terms coming from the soft gaugino masses do not directly
mix with terms from the new $\E$SSM inert sector. The active scalar
Higgs doublet and SM-singlet SSB masses
are determined from the minimisation conditions of the scalar potential
(\ref{eq:higgs1}) given $s$, $v$, $\tan(\beta)$, and $A_\lambda$.

In the following analysis we shall choose $s=3000$~GeV and
$\mu=400$~GeV which gives $\lambda=2\sqrt{2}\,/15 \approx 0.19$
and makes the $Z_2$ mass about 1100~GeV. Although much of the physics
is highly dependent on $s$, this specific choice of $s$ does not
limit the generality of the results obtained since $s$ always appears multiplied
by a Yukawa coupling. This is explained in more detail below.
We also choose $M\dvp{\prime}{1} = M\ud{\prime}{1} = M\dvp{\prime}{2}/2 = 250$~GeV.
These relations between the
SSB gaugino masses are motivated by their RG running from the GUT
scale (see subsection~\ref{sub:smallGauginoMixing}), but the value is not. In this analysis the squarks and sleptons will not
play a significant role in the calculation of dark matter relic abundance
since the LSP will always be much lighter.
We choose equal SSB sfermion masses $M_s = 800$~GeV and
set the stop mixing parameter $X_t$, defined by
\be
X_t = A_t - \f{\mu}{\tan(\beta)},\label{eq:stopMixing}
\ee
where $A_t$ is the SSB parameter associated with the top Yukawa coupling,
to be equal to $\sqrt{6}M_s$, resulting in large loop corrections to the lightest $CP$-even (SM-like) Higgs mass
as in ref.~\cite{King2006}. This is known as the maximal mixing scenario and results in a lightest
$CP$-even Higgs mass in excess of 114~GeV for all parameter
space considered. The SSB $\lambda$ coupling $A_\lambda$
is set by choosing the pseudo-scalar Higgs mass $m_A$, from (\ref{eq:higgs2}). We choose
$m_A = 500$~GeV.

We initially assume the $\ZZ{H}$-breaking $\lambda_{ijk}$
couplings to be small (0.01) for the following analysis. The main
properties of the physics can then be seen by varying the three
parameters $\lambda^\prime=\lambda_{22}=\lambda_{11}$,
$f=f_{d22}=f_{u22}=f_{d11}=f_{u11}$, and $\tan(\beta)$. The first
and second generation mixing couplings are set such that
$\lambda_{21,12}=\epsilon\lambda^\prime$ and
$f_{(d,u)(21,12)}=\epsilon f$. Assuming this parameter choice the
sub-matrices of the neutralino mass matrix (\ref{eq:AA}) become
\be
A_{22} &=& A_{11} = -\frac{1}{\sqrt{2}} \left( \begin{array}{ccc}
0 & \lambda^\prime s & fv\sin(\beta)\\
\lambda^\prime s & 0 & fv\cos(\beta)\\
fv\sin(\beta) & fv\cos(\beta) & 0 \end{array} \right)\quad\mbox{and}\label{eq:param1} \\\nn\\
A_{21} &=& \epsilon A_{22}. \label{eq:param2}
\ee
This simple parametrisation is sufficient for
illustrating the generic properties of the physics. Deviations
from this parametrisation are discussed afterwards.

With the above parametrisation, the two generations
are approximately degenerate when the mixing terms are not very large. In this
case the LSP and the NLSP will each contain approximately
equal contributions from each interaction basis generation.

Finally, it is worth remarking that, assuming the above parametrisation, the effect
on the neutralino and chargino inert sectors of changing $s$ is
simply equivalent to that of changing $\lambda^\prime$
(although the  $Z_2$ mass will depend on $s$). This means that the
following results are applicable for any experimentally
consistent values of $s$ as long as one accordingly scales $\lambda^\prime$.

\subsection{The neutralino and chargino spectra}

\begin{figure}
\begin{center}
\includegraphics[width=\linewidth]{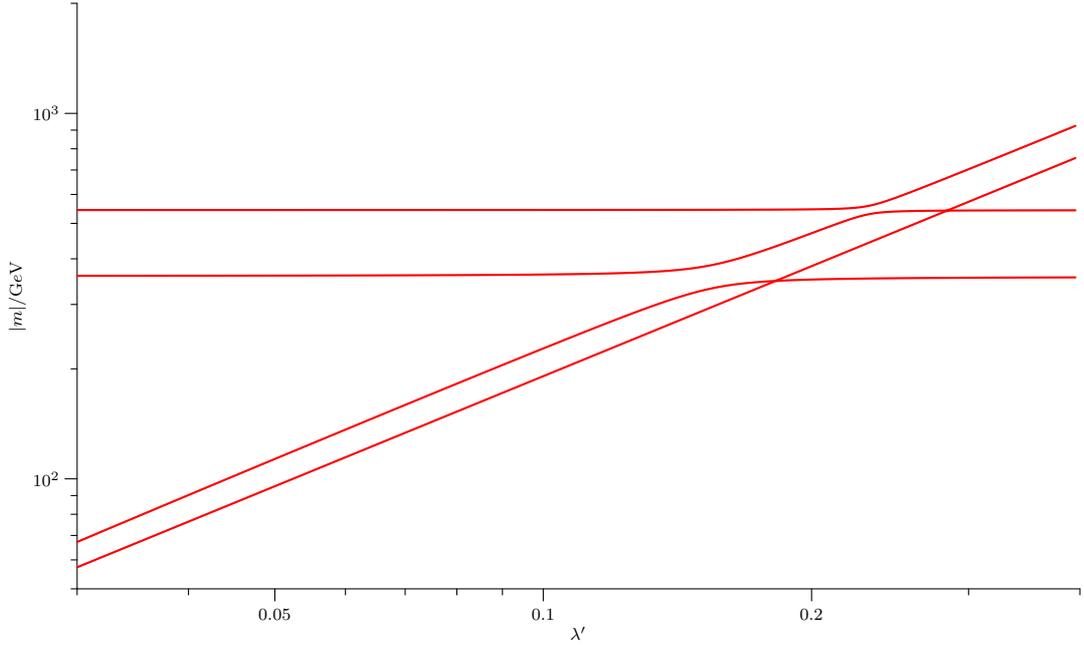}
\end{center}
\caption{Inert chargino masses (magnitude only) against
$\lambda^\prime$ with $f=1$, $\epsilon = 0.1$, $\tan(\beta) = 1.5$,
$s=3000$~GeV, and $\ZZ{H}$-breaking $\lambda_{ijk}$ couplings set to
0.01.}\label{fig:specC}
\end{figure}

Figure~\ref{fig:specC} shows how the spectrum of chargino masses
varies with $\lambda^\prime$. Although the plot is for $\tan(\beta) =
1.5$, as one can see from (\ref{eq:cmm}) the inert sector of the chargino mass
matrix has no dependence on $\tan(\beta)$, with the mass terms just being
proportional to the SM-singlet VEV. The almost constant masses are
those mass eigenvalues coming mostly from the USSM sector~--- the
third generation charged Higgsino and the wino. The charginos coming
mostly from the inert sector vary with $\lambda^\prime$ as
expected and drop below the LEP lower limit around 100~GeV~\cite{Kraan2005}
at some value of $\lambda^\prime$, depending on the value of $s$.
The effect of the $\epsilon=0.1$ mixing between generations can be seen
in the splitting between the two inert sector charginos. Where lines
cross in figure~\ref{fig:specC} the chargino masses are of opposite sign.
When chargino mass lines of the same sign approach each other, they veer away from each other
at the would-be crossing point due to interference.

\begin{figure}
\begin{center}
\includegraphics[width=\linewidth]{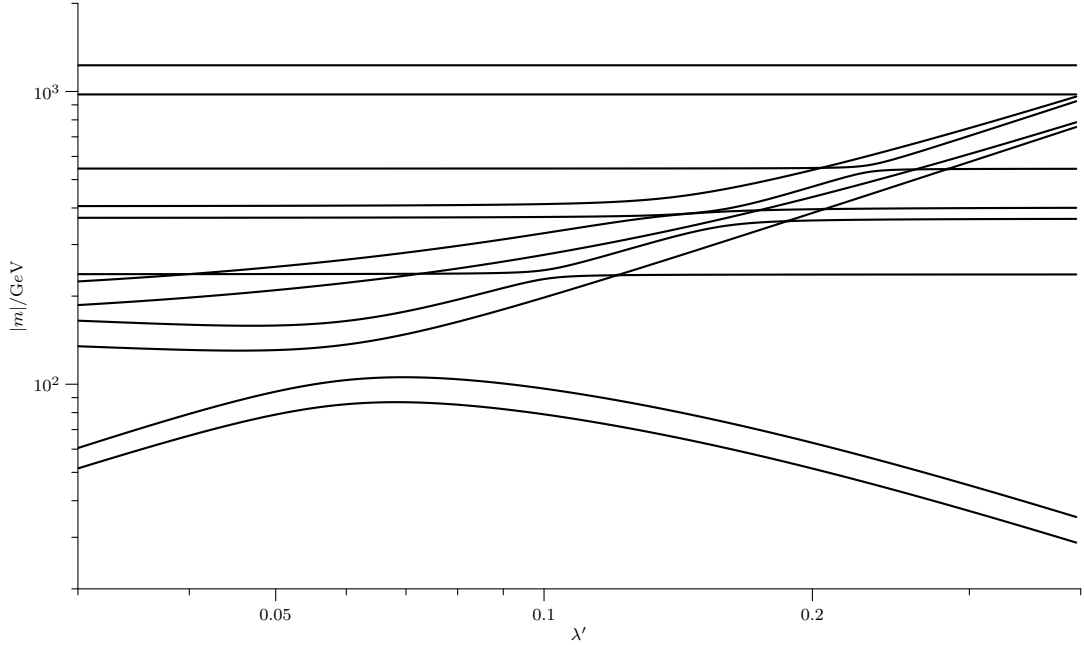}
\end{center}
\caption{Inert neutralino masses (magnitude only) against
$\lambda^\prime$ with $f=1$, $\epsilon = 0.1$, $\tan(\beta) = 1.5$,
$s=3000$~GeV, and $\ZZ{H}$-breaking $\lambda_{ijk}$ couplings set to
0.01.}\label{fig:specN}
\end{figure}

Figure~\ref{fig:specN} shows how the spectrum of neutralino
masses varies with $\lambda^\prime$. The inert neutralino spectrum
is dependent on $\tan(\beta)$, but each of the qualitative
features can be understood. We see the two light neutralino
states that become heavier as $\lambda^\prime$ decreases from
unity until the approximation $\lambda^\prime s \gg fv$ breaks down. At this
point $fv\sin(\beta)$ begins to dominate and the LSP mass
decreases with decreasing $\lambda^\prime$ as the dominance of
$fv\sin(\beta)$ becomes greater. In this low $\lambda^\prime$
region the LSP is no longer mostly inert singlino, but is mostly
inert up-type Higgsino. The six almost unvarying neutralino masses are
those mostly from the USSM sector, which is not mixing very much
with the inert sector. We have already seen that
the inert sector chargino
masses continue to be set by $\lambda^\prime$ as we go down into
the low $\lambda^\prime$ region, resulting in light charginos in
this region. By contrast, the four heavier inert sector
neutralinos begin to be governed by the $fv$ terms rather than
the $\lambda^\prime s$ terms in the low $\lambda^\prime$ region and
therefore approach a constant value in this region.

As in the case of the charginos,
the effect of the $\epsilon=0.1$ mixing can be seen in the splitting
between the two light neutralinos and the four heavier inert
neutralinos which are both split by this mixing and further split
by the light neutralino mass as predicted in the previous section.

\begin{figure}
\begin{center}
\includegraphics[width=\linewidth]{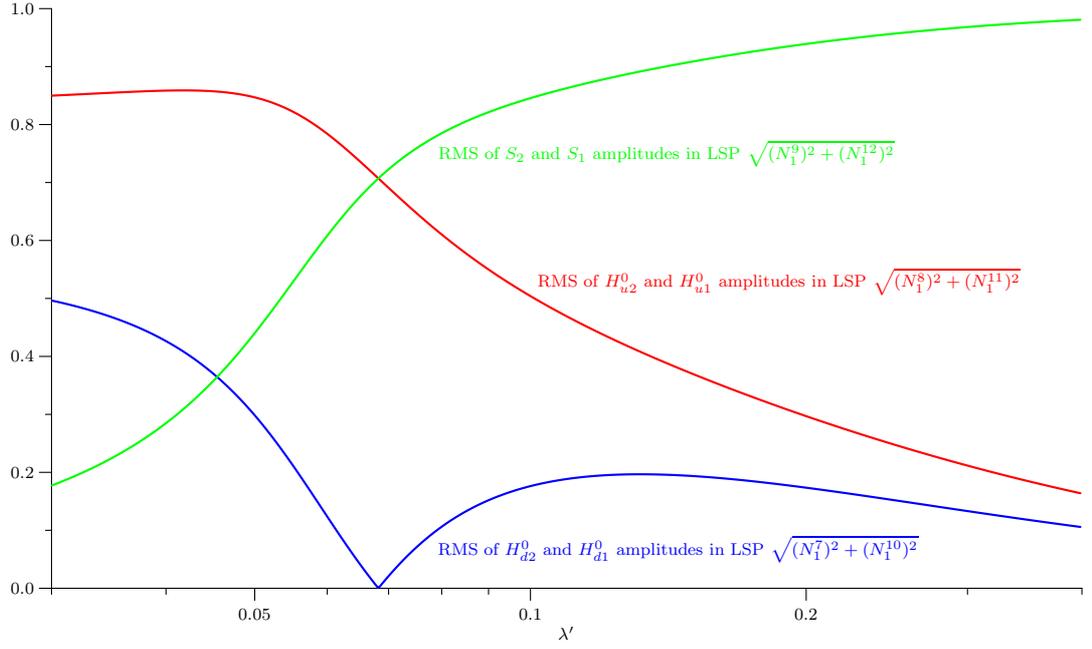}
\end{center}
\caption{The component structure of the LSP
in terms of the inert interaction states
against $\lambda^\prime$ with $f=1$, $\epsilon = 0.1$, $\tan(\beta) = 1.5$,
$s=3000$~GeV, and $\ZZ{H}$-breaking $\lambda_{ijk}$ couplings set to
0.01.}\label{fig:compLSP}
\end{figure}

Figure~\ref{fig:compLSP} shows how the composition of the LSP in terms of
the inert interaction states varies with $\lambda'$.
The behaviour in the $\lambda' s \gg fv$ limit is as predicted in
(\ref{eq:N1}). We also see how the dominant component of the
LSP changes from inert singlino to inert up-type Higgsino
in the low $\lambda'$ region.

\subsection{The dark matter relic density}

\begin{figure}
\begin{center}
\includegraphics[width=\linewidth]{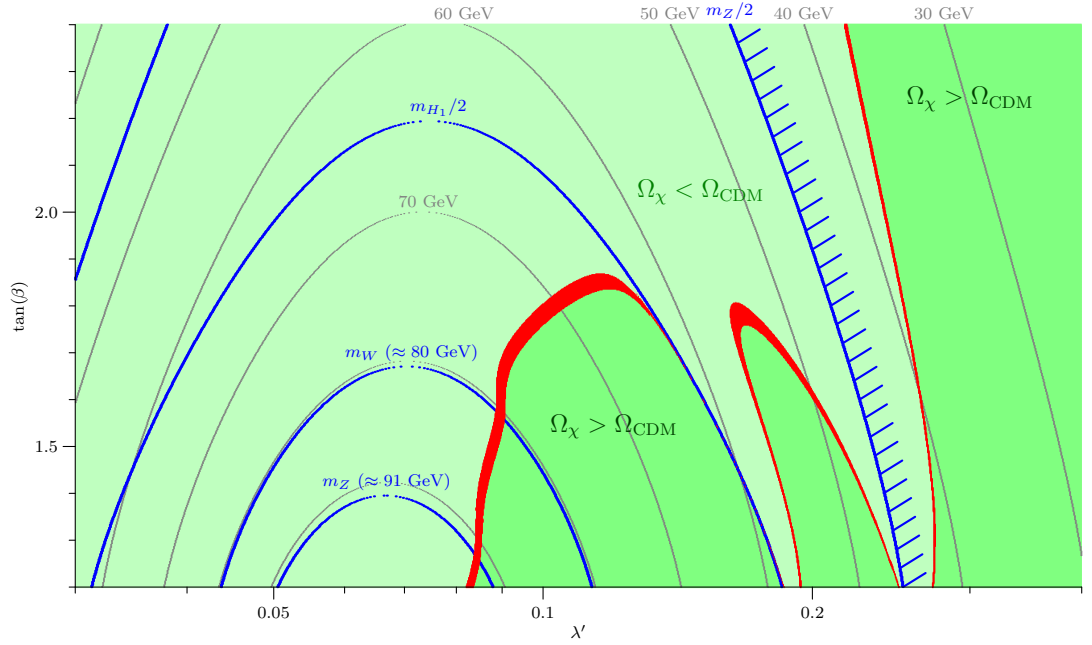}
\end{center}
\caption{Contour plot of the LSP mass
and relic density $\Omega\dz{{\chi}}h\uz{2}$ regions in the
$(\lambda^\prime,\tan(\beta))$-plane with $s=3000$~GeV, $\epsilon=0.1$, and
$f=1$. The red region is where the prediction for $\Omega\dz{{\chi}}h\uz{2}$
is consistent with the measured 1-sigma range of
$\Omega\dz{\rr{DM}}h\uz{2}$. In the region to the right of the hatched line the
LSP mass is less than half of the $Z$ boson mass.\label{fig:mBeta}}
\end{figure}

\begin{figure}
\begin{center}
\includegraphics[width=\linewidth]{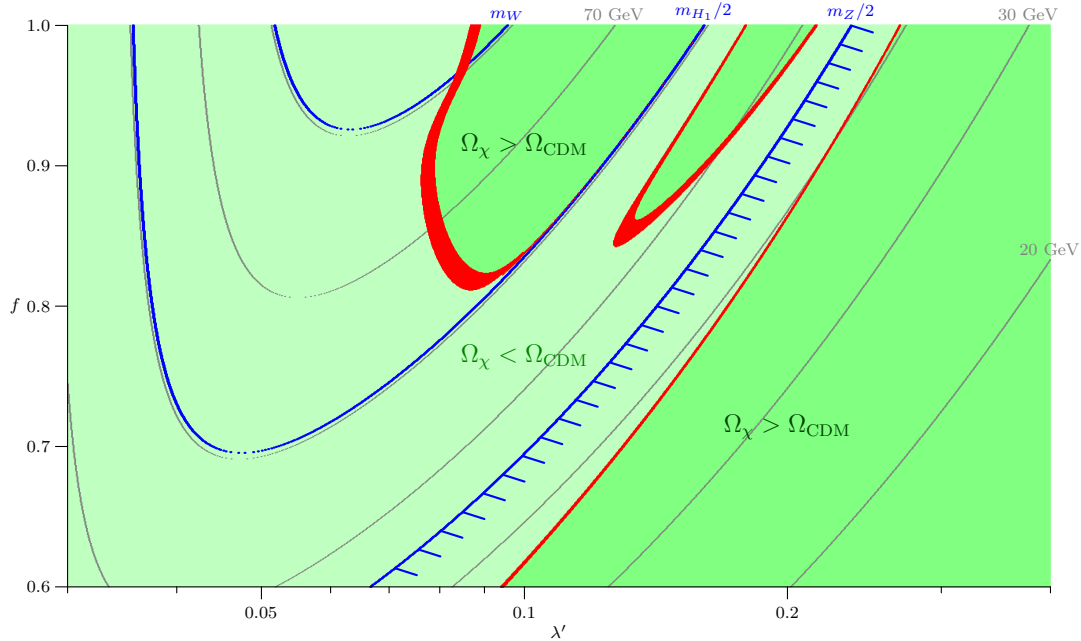}
\end{center}
\caption{Contour plot of the LSP mass
and relic density $\Omega\dz{{\chi}}h\uz{2}$ regions in the
$(\lambda^\prime,f)$-plane with $s=3000$~GeV, $\epsilon=0.1$, and
$\tan(\beta)=1.5$. The red region is where the prediction for
$\Omega\dz{{\chi}}h\uz{2}$ is consistent with the measured 1-sigma range of
$\Omega\dz{\rr{DM}}h^2$. In the region to the right of the hatched line the
LSP mass is less than half of the $Z$ boson mass.\label{fig:m}}
\end{figure}

Using the parametrisation in (\ref{eq:param1}) and (\ref{eq:param2}) we
use {\tt micrOMEGAs 2.2} to numerically compute the
present day relic density.
Figure~\ref{fig:mBeta} shows a contour plot of the LSP mass
and predicted relic density $\Omega\dz{{\chi}}h\uz{2}$ regions in the
$(\lambda^\prime,\tan(\beta))$-plane, with $s=3000$~GeV, $\epsilon=0.1$, and
$f=1$. We focus on small values of $\lambda'<0.4$ since for large $\lambda'$
the LSP is a very light, predominantly inert singlino state which does not annihilate
very efficiently through any channel, leading to a too high relic density
$\Omega\dz{{\chi}}h\uz{2} > \Omega\dz{\rr{DM}}h\uz{2}$. (Such regions are shaded dark green.)
As $\lambda'$ is decreased below 0.3 the LSP mass increases and approaches about half of the $Z$ boson mass
and there is a region where the prediction for $\Omega\dz{{\chi}}h\uz{2}$
is consistent with the measured 1-sigma range of $\Omega\dz{\rr{DM}}h\uz{2}$.
(Such regions are shaded red.)
When the LSP mass is around 40~GeV it contains enough inert doublet Higgsino
such that s-channel annihilation via the $Z$ boson
becomes strong enough to account for the observed relic density.
As the LSP mass is increased further from 40~GeV
and approaches 45~GeV, the annihilations before freeze-out
become on resonance for annihilation with a $Z$ boson in the s-channel and the
predicted relic density becomes lower than that observed.
(Such regions are shaded light green.)

In the regions where the LSP mass is less than half of the $Z$ boson mass
the LSP will contribute to the effective number of neutrinos as
inferred from the invisible $Z$ boson width at LEP defined in
(\ref{eq:NeffZ}). The same couplings that lead to a successful relic density,
via annihilations through an s-channel $Z$ boson, also mean that there may be
a significant contribution to the invisible $Z$ boson width. For the present
discussion it is assumed that such points, with an LSP mass lower that half of the
$Z$ mass, are unsafe from an experimental point
of view. A detailed discussion of the validity of such points is postponed until
chapter~\ref{chap:nhde}.
Note that in the MSSM this issue does not arise since either the LSP is bino-like,
and so does not couple to the $Z$, or is Higgsino- or Wino-like in which case it would
have accompanying almost degenerate charginos and therefore must have a mass greater
than about 100~GeV in any case. Here we can have an inert Higgsino/singlino LSP with a mass
lower than half of the $Z$ boson mass while still having experimentally
consistent inert-doublet-Higgsino-like charginos.

We note at this point that the requirement that the LSP mass exceeds 45~GeV implies
low $\tan(\beta)$ and this is the reason for the restricted range of $\tan(\beta)$
in figure~\ref{fig:mBeta}. This can be seen from (\ref{eq:sin2beta}) where we found that the LSP
mass should be approximately proportional to $\sin(2\beta)$,
\ie~to the product of the two doublet Higgs VEVs, which is maximized at
$\sin(2\beta) = 1$, corresponding to $\tan(\beta)=1$.
In the $\E$SSM an experimentally acceptable
lightest Higgs mass can be achieved even with $\tan(\beta)$ as low
as about 1.2~\cite{King2006}, so having low $\tan(\beta)$ is not a problem in such models.

Decreasing $\lambda'$ further results in LSP masses above 45 GeV and
to the left of the hatched line in figure~\ref{fig:mBeta}
other successful relic density regions (shaded in red) appear.
These regions are punctuated by the light Higgs resonance, leading to
the interesting double loop shape of the successful red regions to the left of the hatched
line in figure~\ref{fig:mBeta}.
In these regions the LSP can have a mass significantly larger than half of the $Z$ boson
mass, moving far enough off the Higgs and $Z$ boson resonances that annihilation is weakened
just enough to give the observed relic density.

However, another effect is observed as $\lambda'$ decreases. The composition
of the LSP changes from being singlino dominated to being Higgsino
dominated.
For low $\lambda'$ the cross-section begins to increase with decreasing $\lambda'$,
even where this still corresponds to increasing LSP mass,
leading to a lower relic density. This is because the inert doublet Higgsino components in the
LSP rapidly grow, as can be seen in figure~\ref{fig:compLSP}.
At low $\lambda'$, when the LSP is largely inert doublet Higgsino,
annihilation is too strong, leading to the predicted relic
density being lower than that observed (as indicated by the light green shading in figure~\ref{fig:mBeta}).
The effects of the t-channel
$W$ and then $Z$ pair production channels can also be seen as they
each become relevant.

Furthermore, for the entire successful region to the left of the hatched line in figure~\ref{fig:mBeta}
the lightest chargino is
heavy enough to be consistent with experiment, as can be seen on figure~\ref{fig:specC}.
This result will be recreated
for all high enough values of $s$. For larger values of $s$ the
successful regions and corresponding inert chargino masses are shifted down by
the corresponding amount in $\lambda'$.

When $\lambda^\prime s \gg fv$ lowering $f$
results in a lower LSP mass, as in (\ref{eq:m1}). It also extends the range of
$\lambda^\prime$ in which this approximation is valid, \ie~it moves
the boundary of the previously discussed
low $\lambda'$ region to be further down in $\lambda'$.
Figure~\ref{fig:m} shows the LSP mass and predicted present day relic
density for different values of $\lambda^\prime$ and $f$ with
$\epsilon = 0.1$ and $\tan(\beta)=1.5$. The shifting of the
successful region, where the
LSP mass is above $m_Z/2$, down in $\lambda'$ at lower values of $f$ is apparent.
At lower values of $\tan(\beta)$ this successful region
extends further down in $f$. It should be
noted that in order to predict the correct dark matter relic
density, $\lambda^\prime$ should be much smaller than $f$ and that this
disparity becomes greater if $s$ is increased.
Increasing $s$ effectively just shifts all of the features on
figures~\ref{fig:mBeta} and~\ref{fig:m} to the left.

\subsection{Deviations from the considered parametrisation}

Breaking the relation $f_{u(22,11)}=f_{d(22,11)}$ can have similar effects
to those of changing $\tan(\beta)$. However, because these
parameters cannot be too high (in order for the theory to be perturbative up to
the GUT scale) and because lowering
them to much less than unity makes the LSP too light,
$\tan(\beta)$ can be varied much more freely than the $f_u/f_d$ ratio.

The effect of increasing the inert generation mixing parameter $\epsilon$
is to increase the various mass splittings between similar inert
mass eigenstates. Increasing the mixing between the first and
second generations thus results in a lighter LSP, shrinking the
successful region, and
a lighter lightest chargino, potentially inconsistent with
current chargino non-observation.

The physics of the inert sector when deviating from the currently considered
parametrisation is studied much more carefully in chapter~\ref{chap:nhde}.

As long as the LSP is still mostly from the
inert sector, as considered here,
other parameters do not greatly affect the dark matter physics and are effectively free.
Squark and slepton parameters do not affect the dark matter
physics of the considered model. Top and stop loops can have a
significant effect on the lightest Higgs mass, but as long as this
mass is experimentally allowed then these parameters are also
not constrained by the requirement that the model produce successful dark matter.

\section{Summary and Conclusions}
\label{ndmin:conclusions}

In this work we studied inert
neutralino dark matter arising in supersymmetric models with extra inert
Higgsinos and singlinos. As an example we considered the
extended neutralino sector of the $\E$SSM.
This work represents a first study of the inert neutralino sector of the
$\E$SSM and it is found that in this model the LSP does typically arise
from this sector.
We studied this novel dark matter scenario both analytically and numerically,
using \texttt{micrOMEGAs}.

The dark matter scenario differs greatly from those of the MSSM and USSM since
the two inert neutralino generations provide an almost
decoupled neutralino sector with a naturally light LSP that can account for
the CDM relic abundance somewhat independently of the
rest of the model.
Although the $\E$SSM has two inert generations, the presence of the second
inert generation is not crucial to the dark matter scenario.

In the successful regions where the observed dark matter relic density is reproduced
the neutralino mass spectrum is well described by
the analytical results of section~\ref{sec:analytical}.
In this region the LSP is mostly inert singlino and has a mass
approximately proportional to $v^2/s$, as in (\ref{eq:sin2beta}),
and as $\lambda's$ is decreased
the LSP becomes heavier and also less inert singlino dominated, picking up
significant inert doublet Higgsino contributions.

To avoid potential conflict with high precision LEP data we considered the
case where the LSP mass is above half of the $Z$ boson mass.
Since the LSP mass in (\ref{eq:sin2beta}) is proportional to $f_d f_u \sin(2\beta)$,
we found that such regions of parameter space in which the dark matter relic density
prediction is consistent with observation require
low values of $\tan(\beta)$~--- less than about 2. Depending on the
value of the singlet VEV $s$,
the $f_{(u,d)\alpha\beta}$ trilinear Higgs coupling parameters should also be
reasonably large compared to the
$\lambda_{\alpha\beta}$ ones.

One of the main messages arising from this work is that neutralino dark
matter could arise from an almost decoupled sector of inert
Higgsinos and singlinos and that if it does then the parameter space
of the rest of the model is completely opened up. For example, if
such a model is regarded as an extension of the MSSM then
the lightest MSSM-like supersymmetric particle is not even required to be a
neutralino and could even be a sfermion.
In the $\E$SSM the lightest MSSM-like supersymmetric particle can decay into the inert LSP via
$\ZZ{H}$-breaking $\lambda_{ijk}$ couplings that need not be extremely small.

Similar results should apply to any singlet extended SSM
with one or more extra, inert generations of Higgsinos and singlinos
with a trilinear Higgs coupling tensor equivalent to that in (\ref{eq:lambda}).

\cleardoublepage

\newpage
\chapter{Novel Higgs Decays in the $\E$SSM}
\label{chap:nhde}

The discovery of the Higgs boson, the last remaining undiscovered particle of the
SM, is one of the main goals
of the LHC.
The strategy for Higgs searches depends on the decay branching fractions
of the Higgs boson into different channels. Physics beyond the SM may affect
the Higgs decay rates to SM particles and give
rise to new Higgs decay channels necessitating a drastic change in
the strategy for Higgs boson searches. (For recent reviews of
non-standard Higgs boson decays see refs.~\cite{Chang2008,Djouadi2009,Dermisek2009}.)
In particular there exist several extensions to the SM in which the
Higgs boson can decay with a substantial branching fraction into
particles that cannot be directly detected.

These invisible Higgs boson decays can occur in supersymetry, with the lighest
Higgs boson decaying into neutralino LSPs.
In some regions
of MSSM parameter space the lightest Higgs boson
decays into the lightest neutralino with a
relatively large branching ratio, giving rise to invisible
final states if $R$-parity is conserved~\cite{Baer1987}.
LEP and Tevatron data allow the neutralino LSP to be sufficiently light such that
the decays of the lightest Higgs boson into these neutralinos is kinematically allowed
and such light neutralinos can annihilate efficiently
through a $Z$ boson pole resulting in a reasonable density of dark matter.

The presence of invisible decays considerably modifies
Higgs boson searches and makes discovery much more difficult. If
the Higgs boson is mainly invisible then the usual visible branching ratios will be
dramatically reduced, preventing detection in the much studied
channels at the LHC and Tevatron. In the case where invisible
Higgs boson decays dominate it is impossible to fully reconstruct
a resonance and it is very challenging to identify the Higgs boson at collider
experiments, \ie~the quantum numbers remain unknown. At $e^{+} e^{-}$
colliders, the problems relating to the observation of an invisible
Higgs boson are less severe~\cite{Eboli1994,DeCampos1997,Djouadi1994} since it
can be tagged through the recoiling $Z$.
The LEP exclusion of Higgs boson masses up to 114.4~GeV applies even in the case
of invisibly decaying Higgs bosons~\cite{TheLEPHiggs2001}
and similar limits could apply to Higgs bosons decaying
into soft lepton pairs some fraction of the time, as happens for some of
the novel Higgs decay scenarios discussed in this chapter.

Higgs boson searches at hadron colliders, however, are
more difficult in the presence of such invisible decays. Previous
studies have analysed $Zh$ and $Wh$ associated
production~\cite{Choudhury1994,Godbole2003,Davoudiasl2005}
as well as $t\bar{t} h$ production~\cite{Gunion1994,Kersevan2003} and $t\bar{t} VV$ 
and $b\bar{b} VV$ production~\cite{Boos2011} as promising channels, where $h$ is
the Higgs boson and the $V$s stand for vector bosons.
The possibility of observing an invisible Higgs boson in central
exclusive diffractive production at the LHC was studied in~\cite{Belotsky2004}.
Another proposal is to observe such an invisible Higgs in inelastic events
with large missing transverse energy and two high $E_T$ jets. In this case
the Higgs boson is produced by $VV$ fusion and has a large transverse momentum
resulting in a signal with two quark jets with distinctive
kinematic distributions compared to $Zjj$ and $Wjj$
backgrounds~\cite{Battaglia2004,Davoudiasl2005,Eboli2000}.

In this chapter, which contains work that was first published in \textbf{paper II},
we consider novel decays
of the lightest Higgs boson and associated collider signatures
within the $\E$SSM.
If the Yukawa couplings of the inert neutralino sector are required to be small
enough such that perturbation theory remains valid up to the GUT scale then the masses
of the two lightest inert neutralino states are expected to be smaller than about 60--65~GeV.
As a result the lightest inert
neutralino tends to be the LSP.
As noted in the previous chapter
such an inert neutralino can give an appropriate contribution to the dark
matter density, consistent with recent observations, if it has
a mass around 35--50~GeV.
In this case the lightest Higgs
boson decays predominantly into inert neutralino states and the usual Higgs
boson branching ratios to SM particles are less than a few percent.

In section~\ref{ref:novelInert} we look in more detail at the inert sector of the
model and the couplings of the inert neutralinos, inert charginos, and active Higgs bosons are
specified. Novel decays of the lightest $CP$-even Higgs state and dark matter constraints are
discussed in section~\ref{novel:hd}. In section~\ref{novel:bm} we specify
some benchmark points and discuss the experimental constraints
and predictions. The conclusions are summarised in section~\ref{novel:conclusions}.

\section{Inert Charginos and Neutralinos}
\label{ref:novelInert}

In our analysis we will assume that $\ZZ{H}$-violating couplings are small
and can be neglected. This assumption can be justified if
one takes into account that the $\ZZ{H}$-violating operators can give an
appreciable contribution to the amplitude of $K^0$-$\bar{K}^0$ oscillations and
give rise to new muon decay channels such as $\mu\rightarrow e^{-}e^{+}e^{-}$. In order to suppress
processes with non-diagonal flavour transitions the Yukawa couplings of the exotic
particles to the quarks and leptons of the first two generations should be smaller
than $10^{-3}$--$10^{-4}$. Such small $\ZZ{H}$-violating couplings can be neglected
in the first approximation.

In this approximation, and given the assumption that only $H_u$, $H_d$,
and $S$ acquire non-zero VEVs, the charged components of the inert Higgsinos
do not mix with the MSSM-like chargino states. The neutral components of the inert Higgsinos
and inert singlinos also do not mix with
the USSM-like neutralino states. If $\ZZ{H}$ symmetry was exact then both the lightest state in the
ordinary neutralino sector and the lightest inert neutralino would be absolutely stable. Therefore
although $\ZZ{H}$-violating couplings are expected to be rather small we shall
assume that they are large enough to allow either the lightest USSM-like neutralino state or the lightest
inert neutralino state to decay within a reasonable time~--- the lighter of the two
being the stable LSP and the dark matter candidate.

In the basis
\be
\tilde{N}\ud{\rr{int}}{\rr{inert}} &=& \left(\ba{cccccc}
\tilde{H}\ud{0}{d2} & \tilde{H}\ud{0}{u2}
& \tilde{S}\dvp{0}{2} & \tilde{H}\ud{0}{d1} & \tilde{H}\ud{0}{u1} & \tilde{S}\dvp{0}{1}
\ea\right)\uz{\rr{T}}\label{eq:inertBasis}
\ee
the inert part of the neutralino mass matrix is given by
\be
M\ud{N}{\rr{inert}} &=& \left(\ba{cc}
A\dvp{\rr{T}}{22} & A\dvp{\rr{T}}{21}\\
A\ud{\rr{T}}{21} & A\dvp{\rr{T}}{11}\ea\right),\label{eq:icn1}
\ee
with the submatrices given in (\ref{eq:AA}). The inert part of the
chargino mass matrix $P$, given in (\ref{eq:cmm}), may be written
\be
P\ud{\rr{inert}}{\alpha\beta} &=& \dfrac{1}{\sqrt{2}}\lambda\dvp{\rr{inert}}{\alpha\beta}.\label{eq:icn3}
\ee

From (\ref{eq:icn1}) and (\ref{eq:icn3}) one can see that
in the exact $\ZZ{H}$ symmetry limit the spectrum of the inert neutralinos and
charginos in the $\E$SSM can
be parametrised in terms of $\lambda_{\alpha\beta}$, $f_{d\alpha\beta}$,
$f_{u\alpha\beta}$, $\tan(\beta)$, and $s$.
In other words the masses and couplings of the inert neutralinos are determined by 12 Yukawa
couplings, which can in principle be complex, $\tan(\beta)$, and $s$. Four of the Yukawa couplings mentioned above~---
$\lambda_{\alpha\beta}$~--- as well as the VEV of the SM singlet field $s$ set the masses and couplings
of the inert chargino states. Six off-diagonal Yukawa couplings define the mixing between the two generations of
inert Higgsinos and singlinos.

In the following analysis the VEV of the active SM-singlet field is chosen to be large enough
($s\gtrsim 2400$~GeV) so that experimental constraints from ref.~\cite{Erler2010} on the $Z\dz{2}$ boson mass
($\mass{\dz{Z\dz{2}}} > 892$~GeV) and $Z$-$Z'$ mixing are satisfied. Since the publication of \textbf{paper~II}
the limit on the $Z_2$ mass in the $\E$SSM has increased as discussed in subsection~\ref{sub:ZZ}.
In order to avoid the LEP lower
limit on the masses of inert charginos~\cite{Kraan2005} the Yukawa couplings $\lambda_{\alpha\beta}$ are chosen
such that all inert chargino states are heavier than $100$~GeV. In addition, we also require
the validity of perturbation theory up to the GUT scale and this constrains the allowed range of
all Yukawa couplings.

The theoretical and experimental restrictions specified above set very strong limits on the masses
and couplings of the lightest inert neutralinos. In particular, our numerical analysis indicates that
the lightest and second lightest inert neutralinos are always light. They typically have masses
below 60--65~GeV. These neutralinos are predominantly inert singlino in nature.
From our numerical analysis it follows that the lightest and second lightest inert neutralinos
might have rather small couplings to the $Z$ boson so that any possible signal that these neutralinos
could give rise to at LEP would be extremely suppressed. As a consequence such inert neutralinos
would remain undetected. At the same time four other inert neutralinos, which are approximately linear
superpositions of neutral components of inert doublet Higgsinos, are normally heavier than 100~GeV.

\subsection{The diagonal inert Yukawa coupling approximation}
\label{sub:diagonalInertYukawas}

In order to clarify the results of our numerical analysis it is useful to consider a few simple
cases that give some analytical understanding of our calculations.
The simplest case is when all of the Yukawa coupling from the off-diagonal blocks
of (\ref{eq:icn1}) are zero such that
\be
\lambda_{\alpha\beta} &=& \lambda_{\alpha}\delta_{\alpha\beta},\nn\\
f_{d\alpha\beta} &=& f_{d\alpha}\delta_{\alpha\beta},\quad\mbox{and}\nn\\
f_{u\alpha\beta} &=& f_{\alpha}\delta_{\alpha\beta}\quad\mbox{(no sum on $\alpha$)}.
\label{eq:diagonalInertYukawas}
\ee
This leads to two decoupled generations with the
properties studied in section~\ref{sec:analytical}.
The mass matrix of inert neutralinos (\ref{eq:icn1}) reduces to block diagonal form while the masses of
the inert charginos are given by
\be
\mass{\tilde{C}\dz{\alpha}} &=& \f{\lambda_{\alpha}}{\sqrt{2}}s.
\label{eq:icn5}
\ee

When $f\dz{\alpha} = f\dz{d\alpha} = f\dz{u\alpha}$ one can prove using the method
proposed in ref.~\cite{Hesselbach2008} that there are theoretical upper bounds on the masses
of the lightest and second lightest inert neutralino states. The theoretical
restrictions are $|\mass{\tilde{N}\dz{\alpha}}|\uz{2} \lesssim \mu_{\alpha}^2$, where
\be
\mu_{\alpha}^2 &=& \f{1}{2}\left[|\mass{\tilde{C}\dz{\alpha}}|\uz{2}
+ \f{f_{\alpha}^2 v^2}{2}
\Big(1+\sin^2(2\beta)\Big)
\vphantom{\sqrt{\left(|\mass{\tilde{C}\dz{\alpha}}|\uz{2} + \f{f_{\alpha}^2 v^2}{2}(1+\sin^2(2\beta))
\right)\uz{2}
- f_{\alpha}^4 v^4 \sin^2(2\beta)}}\right.\nn\\
&&\quad\vphantom{A} - \left.\sqrt{\left(|\mass{\tilde{C}\dz{\alpha}}|\uz{2} + \f{f_{\alpha}^2 v^2}{2}\Big(1+\sin^2(2\beta)\Big)
\right)\uz{2}
- f_{\alpha}^4 v^4 \sin^2(2\beta)}\right].\label{eq:icn6}
\ee

The value of $\mu_{\alpha}$ decreases with increasing $|\mass{\tilde{C}\dz{\alpha}}|$ and $\tan(\beta)$,
approaching its maximum value
\be
\mu_{\alpha} &\rightarrow& \f{f\dz{\alpha}}{\sqrt{2}}v
\ee
as $\mass{\tilde{C}\dz{\alpha}} \rightarrow 0$ and $\tan(\beta) \rightarrow  1$.

The upper bound on the mass of the
lightest inert neutralino also depends on the values of the Yukawa couplings $f_{d\alpha}$ and $f_{u\alpha}$.
The theoretical restrictions on these couplings due to the requirement that
the theory should remain perturbative up to the GUT scale become weaker with increasing $\tan(\beta)$.
At large values of $\tan(\beta)$ the upper bounds on $|\mass{\tilde{C}\dz{\alpha}}|$
from (\ref{eq:icn6}) becomes rather small and as $\tan(\beta)$ tends to unity
the upper bounds on $|\mass{\tilde{C}\dz{\alpha}}|$ again become rasther small, because
theoretical constraints on $f_{d\alpha}$ and $f_{u\alpha}$ become rather stringent.
Taking both of these effects in account the upper bounds on $|\mass{\tilde{C}\dz{\alpha}}|$
achieve their maximum values around $\tan(\beta) \approx 1.5$. For this value of $\tan(\beta)$
the requirement of the validity of perturbation theory up to the GUT scale implies that
for $f = f_{d1}=f_{u1}=f_{d2}=f_{u2}$ $f$ must be less than about $0.6$. As a consequence the
lightest inert neutralinos are
lighter than around $60$--$65$~GeV for $|\mass{\tilde{C}\dz{\alpha}}| > 100$~GeV.

Using the results from section~\ref{sec:analytical} for the compositions of
the light neutralinos from each inert generation one can derive the couplings of
these states to the $Z$ boson.
We define $R_{Z\alpha\beta}$ couplings such that the
$Z$-$\tilde{N}_\alpha$-$\tilde{N}_\beta$ coupling is equal to $R_{Z\alpha\beta}$ times
the $Z$-$\nu$-$\nu$ coupling
\be
R_{Z\alpha\beta} &=& N_{\alpha}^1 N_{\beta}^1 - N_{\alpha}^2 N_{\beta}^2
+ N_{\alpha}^4 N_{\beta}^4 - N_{\alpha}^5 N_{\beta}^5,
\ee
where $N_i^a$ is the neutralino mixing matrix element corresponding to
mass eigenstate $i$ and inert interaction state $a$ in the basis
(\ref{eq:inertBasis}).

In the case where the off-diagonal inert Yukawa coupling blocks vanish while
$\lambda_{\alpha} s\gg f_{(u,d)\alpha} v$
the relative couplings of the lightest and second lightest inert
neutralino states to the $Z$ boson are given by
\be
R_{Z\alpha\beta} &=& R_{Z\alpha\alpha}\delta_{\alpha\beta}\quad\mbox{(no sum on $\alpha$)},
\ee
where
\be
R_{Z\alpha\alpha} &=& \f{v^2}{2 \masss{2}{\tilde{C}_{\alpha}}}
\biggl(f_{d\alpha}^2\cos^2(\beta)-f_{u\alpha}^2\sin^2(\beta)\biggr).
\label{eq:icn14}
\ee
This demonstrates that the couplings of $\tilde{N}_1$ and $\tilde{N}_2$ to the
$Z$ boson can be very strongly suppressed. It becomes zero when
$|f_{d\alpha}|\cos(\beta)=|f_{u\alpha}|\sin(\beta)$,
which is when $\tilde{N}_{\alpha}$ contains a completely symmetric combination
of $\tilde{H}^{0}_{d\alpha}$ and
$\tilde{H}^{0}_{u\alpha}$. (\ref{eq:icn14}) also indicates that the couplings
of $\tilde{N}_1$ and $\tilde{N}_2$ to the $Z$ boson
are always small if the inert charginos are rather heavy or if
$f_{d\alpha}$ and $f_{u\alpha}$ are small,
\ie~the masses of $\tilde{N}_1$ and $\tilde{N}_2$ are small.

\subsection{$\Delta_{27}$ and pseudo-Dirac lightest inert neutralino states}
\label{sub:Delta27}

In order to provide an explanation for the origin of the approximate
$\ZZ{H}$ symmetry that singles out the third generation of Higgs doublets and SM-singlets,
and to account for tribimaximal mixing and other features of the quark and lepton
spectra, a $\Delta_{27}$ flavour symmetry may be applied to the $\E$SSM~\cite{Howl2010}.
The addition of the $\Delta_{27}$ flavour symmetry implies
an inert neutralino mass matrix with $A_{11}\approx A_{22} \approx 0$,
leading to approximately degenerate lightest neutralino states with a pseudo-Dirac
(see appendix~\ref{ap:spinors}) structure.

When all flavour diagonal Yukawa couplings $\lambda_{\alpha\alpha}$, $f_{d\alpha\alpha}$,
and $f_{u\alpha\alpha}$ exactly vanish, \ie~$A_{11}=A_{22}=0$,
the inert neutralinos form Dirac states.
In this limit the Lagrangian of the $\E$SSM is invariant under
an extra $U(1)$ global symmetry.
Under this symmetry the fermionic components of the inert supermultiplets transform
\be
\tilde{S}_2 &\rightarrow& e^{i\alpha} \tilde{S}_2,\nn\\
\tilde{H}_{d2} &\rightarrow& e^{i\alpha} \tilde{H}_{d2},\nn\\
\tilde{H}_{u2} &\rightarrow& e^{i\alpha} \tilde{H}_{u2},\nn\\
\tilde{S}_1 &\rightarrow& e^{-i\alpha} \tilde{S}_1,\nn\\
\tilde{H}_{d1} &\rightarrow& e^{-i\alpha} \tilde{H}_{d1},\nn\\
\tilde{H}_{u1} &\rightarrow& e^{-i\alpha} \tilde{H}_{u1}.
\label{eq:icn19}
\ee
In the above limiting case
the lightest inert neutralino is a Dirac state formed predominantly from $\tilde{S}_1$ and $\tilde{S}_2$.
In this case the LSP and its antiparticle have opposite charges with respect to the extra global
$U(1)$ and this could lead to the scenario known as asymmetric dark matter~\cite{Hooper2005,Kaplan2009,An2010,Frandsen2010}.
The ADM scenario supposes that there could be an asymmetry between the density of dark matter particles
and their antiparticles in the early universe similar to that for baryons. This could have a considerable
effect on the relic density calculations. In particular, if an asymmetry exists between
the number densities of dark matter particles and their antiparticles in the early universe then one can get an
appreciable dark matter density even if the dark matter particle-antiparticle annihilation cross section is very
large, like in the case of baryons. Furthermore, if most of the dark matter
antiparticles are eliminated by annihilation
with their particles then such an ADM scenario does not have the usual indirect
signatures associated with the
presence of dark matter. For example, there would be no high energy neutrino
signal from annihilations in the Sun.
At the same time, a relatively high concentration of dark matter particles can
build up in the Sun, altering heat transport in the solar interior and affecting
low energy neutrino fluxes~\cite{Frandsen2010}.

In practice the $\Delta_{27}$ scenario tells us that we are somewhat away from
the above limiting case, with a broken global $U(1)$ symmetry
leading to almost degenerate pseudo-Dirac lightest neutralinos, where the
relic density of the LSP can be calculated by standard methods.
It will turn out that the LSP cannot be too light (must be of order $m_Z/2$)
in order not to have a too high cosmological relic density. At the same time we
will see that the two lightest neutralinos
cannot be too heavy in order for perturbation theory to be valid up to the GUT scale.
In practice this means that in realistic scenarios the two lightest inert neutralino states
are rather close in mass. The $\Delta_{27}$ scenario provides an explanation
for this feature of the successful neutralino mass pattern.

It is worth noting
that the results from the previous
section can be reinterpreted in terms of this scenario. Specifically in the case
where $A_{11} = A_{22} = 0$ and $A_{21} = A_{12}$ a block diagonalisation of the
inert neutralino mass matrix
(\ref{eq:icn1}) results in
\bea{c}
A_{22} \rightarrow A_{22}' = -A_{21},\qquad\mbox{and}\qquad
A_{11} \rightarrow A_{11}' = A_{21},
\eea
with $A_{21} = A_{12} \rightarrow A_{21}' = A_{12}' = 0$.
This only corresponds to a redefinition of the generations 1 and 2 and does not mix
fields of different hypercharge. This provides a dictionary between
these two scenarios
\be
-\lambda_{22}' = \lambda_{11}' &=& \lambda_{21},\nn\\
-f_{d22}' = f_{d11}' &=& f_{d21},\\
-f_{u22}' = f_{u11}' &=& f_{u21}.
\ee
Rewriting the inert neutralino mass matrix in this block diagonal form also
makes it clear that the $R_{Z12}$ coupling vanishes in this limit in the same
way that it did for the diagonal case in subsection~\ref{sub:diagonalInertYukawas}.

\subsection{The couplings of Higgs bosons to inert neutralinos}

The presence of light inert neutralinos in the particle spectrum of the $\E$SSM
makes possible the decays of the Higgs bosons into these final states.
Now and in the next section we argue that
such decays may result in the modification of the SM-like Higgs signal at
current and future colliders. Since our main concern in this work is the
decays of the SM-like lightest Higgs boson, we shall
ignore the effects of the inert Higgs scalars and pseudoscalars which do not mix
appreciably with the
active scalar sector responsible for EWSB.
We also assume that all of the inert bosons are heavier than the lightest $CP$-even Higgs boson.

If all other Higgs boson states are much heavier than the lightest $CP$-even Higgs
boson then the lightest Higgs state, approximately given by $h$, as defined in (\ref{eq:higgs3}),
manifests itself in interactions with SM gauge bosons and fermions as a
SM-like Higgs boson. Since within the $\E$SSM the mass of this state is predicted to be relatively
low, its production cross section at the LHC should be large enough so that it can be observed in the
near future. In this context it is particularly interesting and important to analyse the decay modes of
the lightest $CP$-even Higgs state. Furthermore, we concentrate on the decays of
the SM-like Higgs boson into the lightest and second lightest inert neutralinos.

The couplings of the Higgs states to the inert neutralinos originate from the interactions of $H_u$, $H_d$,
and $S$ with the inert Higgs fields in the superpotential. Using (\ref{eq:higgs3}) one can express
$\kk{Re}\,H_d^0$, $\kk{Re}\,H_u^0$, and $\kk{Re}\,S$ in terms of the field-space basis states
$h$, $H$, and $N$. The components of the field-space basis are related to the physical $CP$-even mass
eigenstates by a unitary transformation
\be
\left(
\begin{array}{c}
h_1 \\ h_2\\ h_3
\end{array}
\right)&=&
U
\left(
\begin{array}{c}
h\\ H\\ N
\end{array}
\right).
\label{higgs6}
\ee
Combining all these expressions together one obtains an effective Lagrangian term that describes
the interactions of the inert neutralinos with the $CP$-even Higgs mass eigenstates
\bea{c}
X\ud{h\dz{m}}{ij}h\dz{m}\tilde{N}\ud{c\dagger}{i}\tilde{N}\dvp{\dagger}{j} + \mbox{c.c.},\label{eq:higgs7}
\eea
where
\be
X\ud{h\dz{m}}{ij} &=& -\f{1}{\sqrt{2}} U\ud{N}{h\dz{m}}\Lambda\dvp{N}{ij}
- \f{1}{\sqrt{2}} \left(U\ud{h}{h\dz{m}}\cos(\beta) - U\ud{H}{h\dz{m}}\sin(\beta)\right)F\dvp{N}{dij}\nn\\
&&\qquad \vphantom{A}
- \f{1}{\sqrt{2}} \left(U\ud{h}{h\dz{m}}\sin(\beta) + U\ud{H}{h\dz{m}}\cos(\beta)\right)F\dvp{N}{uij},
\ee
with
\be
\Lambda\dvp{N}{ij} &=& \lambda\dvp{1}{11} N^4_i N^5_j + \lambda\dvp{1}{12} N^4_i N^2_j
+\lambda\dvp{1}{21} N^1_i N^5_j + \lambda\dvp{1}{22} N^1_i N^2_j,\\
F\dvp{N}{dij} &=& f\dvp{1}{d11} N^6_i N^5_j + f\dvp{1}{d12} N^6_i N^2_j
+ f\dvp{1}{d21} N^3_i N^5_j + f\dvp{1}{d22} N^3_i N^2_j,\quad\mbox{and}\\
F\dvp{N}{uij} &=& f\dvp{1}{u11} N^6_i N^5_j + f\dvp{1}{u12} N^6_i N^2_j
+ f\dvp{1}{u21} N^3_i N^5_j + f\dvp{1}{u22} N^3_i N^2_j.
\ee

The expressions for the couplings of the active $CP$-even Higgs scalars to the inert neutralinos become
much simpler in the case where the Higgs spectrum has the usual hierarchical structure. In this case
$U$ is almost the identity. As a consequence the couplings of the SM-like Higgs boson to
the lightest and second lightest inert neutralino states are approximately given by
\be
X\ud{h\dz{1}}{\alpha\beta} &=& -\f{1}{\sqrt{2}} \Big(F\dz{d\alpha\beta}\cos(\beta)
+ F\dz{u\alpha\beta}\sin(\beta)\Big).
\label{eq:higgs8}
\ee

In the case of the diagonal inert Yukawa coupling limit defined in
subsection~\ref{sub:diagonalInertYukawas} and if
$\lambda_{\alpha} s\gg f\dz{(d,u)\alpha} v$ one can use the expression (\ref{eq:N1})
to find values for $N^{a}_{1}$ and $N^{a}_{2}$ and derive approximate formulae
for $X\ud{h\dz{1}}{\alpha\beta}$.
Substituting
into (\ref{eq:higgs8}) one obtains
\be
X\ud{h\dz{1}}{\alpha\beta} &=& \f{\mass{\tilde{N}\dz{\alpha}}}{v}\delta\dz{\alpha\beta} + \cdots\quad\mbox{(no sum on $\alpha$)}.
\label{eq:higgs9}
\ee
This simple analytical expression for the couplings of the SM-like Higgs boson to the lightest and
second lightest inert neutralinos is not as surprising as it may first appear.
When the Higgs spectrum is hierarchical, with $s \gg v$, the VEV of the lightest $CP$-even state $v$ is responsible for
all light fermion masses in the $\E$SSM. As a result we expect that their couplings to the SM-like Higgs can be
written as usual as being proportional to the mass divided by the VEV. We see that this is exactly what is found
in the limit of $\mass{\tilde{N}\dz{\alpha}}$ being small.

\section{Novel Higgs Decays and Dark Matter}
\label{novel:hd}

The interaction Lagrangian (\ref{eq:higgs7}) gives rise to decays of the
lightest Higgs boson into inert neutralino pairs with partial widths given by
\be
\Gamma(h\dz{1}\rightarrow\tilde{N}\dz{\alpha}\tilde{N}\dz{\beta})
&=& \f{\Delta_{\alpha\beta}}{16\pi \mass{h\dz{1}}}
\biggl(X\ud{h\dz{1}}{\alpha\beta}+X\ud{h\dz{1}}{\beta\alpha}\biggr)\uz{2}
\biggl[\masss{2}{h\dz{1}}-(\mass{\tilde{N}\dz{\alpha}} - \mass{\tilde{N}\dz{\beta}})\uz{2}\biggr]\nn\\
&&\quad \sqrt{\left(1-\f{\masss{2}{\tilde{N}\dz{\alpha}}}{\masss{2}{h\dz{1}}}
- \f{\masss{2}{\tilde{N}\dz{\beta}}}{\masss{2}{h\dz{1}}}\right)\uz{2}
- 4\f{\masss{2}{\tilde{N}\dz{\alpha}}\masss{2}{\tilde{N}\dz{\beta}}}{\masss{4}{h\dz{1}}}},
\label{eq:higgs10}
\ee
where $\Delta_{\alpha\beta} = \{1,2\}$ for \{$\alpha=\beta$, $\alpha\neq\beta$\}.

The partial widths associated with these inert decays of the SM-like Higgs boson
(\ref{eq:higgs10}) have to be compared to decay rates into SM particles.
When the SM Higgs boson is relatively light (less than about 140~GeV)
it decays predominantly into $b$ quark and
$\tau$ lepton pairs. The partial decay width of the lightest $CP$-even
Higgs boson into Dirac fermion pairs is given by~\cite{Djouadi2008}
\be
\Gamma(h_1\rightarrow f\bar{f}) &=& N_c\dfrac{\vphantom{m^2_f}g_2^2}{32\pi}\dfrac{m^2_f}{m^2_W} g^2_{h_1ff} m\dvp{2}{h_1}
\biggl(1-4\dfrac{ m_f^2}{m_{h_1}^2}\biggr)^{3/2}.
\label{eq:higgs11}
\ee
For the case of the decays into $\tau$ leptons the coupling of the lightest
$CP$-even Higgs state to the $\tau$ lepton normalised to the
corresponding SM coupling
\be
g_{h_1\tau\tau} &=& \dfrac{1}{\cos(\beta)}\biggl(U\ud{h}{h_1}\cos(\beta)-U\ud{H}{h_1}\sin(\beta)\biggr).
\label{eq:higgs12}
\ee

For a final state that involves $b$ quarks one has to include the QCD corrections. In particular,
the fermion mass in (\ref{eq:higgs11}) should be associated with the running $b$ quark mass $\bar{m}_b(\mu)$.
The bulk of the QCD corrections are absorbed by using the running $b$ quark mass defined at the appropriate
renormalisation scale~--- the scale of the lightest Higgs boson mass $\mu=m_{h_1}$ in the
considered case. In addition to the corrections that are associated with the running $b$ quark mass
there are other QCD corrections to the Higgs coupling to $b$ quarks that should be taken into account~\cite{Gorishnii1990}.
As a consequence, the partial decay width of the lightest $CP$-even Higgs boson into
$b$ quark pairs can be calculated using (\ref{eq:higgs11}) if one sets
\be
N_c &=& 3,\nn\\
m_f &=& \bar{m}_b(m_{h_1}),\quad\mbox{and}\\
g^2_{h_1ff} &=& \dfrac{1}{\cos^2(\beta)}\biggl(U\ud{h}{h_1}\cos(\beta)-U\ud{H}{h_1}\sin(\beta)\biggr)^2
\biggl[1+\Delta_{bb}+\Delta_H\biggr],\nn
\ee
where
\be
\Delta_{bb} &\approx& 5.67\dfrac{\bar{\alpha}_s(m_{h_1})}{\pi}+(35.94-1.36 N_f)
\dfrac{\bar{\alpha}^2_s(m_{h_1})}{\pi^2}\quad\mbox{and}\nn\\\\
\Delta_H    &\approx& \dfrac{\bar{\alpha}^2_s(m_{h_1})}{\pi^2}
\left(1.57-\dfrac{2}{3}\ln\left(\dfrac{m_{h_1}^2}{m_t^2}\right)+
\dfrac{1}{9}\ln^2\left(\dfrac{\bar{m}_b^2(m_{h_1})}{m_{h_1}^2}\right)\right).\nn
\label{eq:higgs13}
\ee
Here we neglect radiative corrections that originate from
loop diagrams that contain non-SM particles\footnote{Radiative corrections that are induced
by supersymmetric particles can be very important, particularly in the case
of the bottom quark at high values of
$\tan(\beta)$. For a review see ref.~\cite{Heinemeyer2004}.}.

From (\ref{eq:higgs9}) one can see that in the $\E$SSM the branching ratios
of the SM-like Higgs state into the lightest and second lightest inert
neutralinos depend rather strongly on the masses of these particles. When the lightest inert
neutralino states are heavy relative to the $b$ quark the
lightest Higgs boson decays predominantly into $\tilde{N}\dz{\alpha}\tilde{N}\dz{\beta}$
while the branching ratios for decays into SM particles are suppressed.
On the other hand if the lightest inert
neutralinos have masses that are considerably smaller than the masses of the $b$ quark and $\tau$ lepton
then the branching ratios of the SM-like Higgs into inert neutralino final states are small.

Constraints on the mass of the lightest inert neutralino can be obtained if we require
that this particle accounts for all or some of the observed dark matter relic density.
In the limit where all non-SM fields other than the two lightest inert neutralinos
are heavy the lightest inert neutralino state in the $\E$SSM is responsible for too large
a thermal relic density of dark matter. The LSP $\tilde{N}_1$ is composed mainly of
inert singlino and has a mass inversely proportional to the charged inert Higgsino mass.
In this limit it is typically very light with $|\mass{\tilde{N}\dz{\alpha}}|\ll m_Z$. As a result
the couplings of the lightest inert neutralino to gauge bosons, the SM-like Higgs boson, quarks, and leptons
are quite small, leading to a relatively small dark matter annihilation cross-section
into SM particles and giving rise to a relic density that is typically much larger
than the measured value. Thus, in the limit considered, the bulk of the $\E$SSM parameter space that
leads to small inert neutralino masses is ruled out.

The situation changes dramatically when the mass of the lightest inert neutralino increases.
In this case the Higgsino components of $\tilde{N}_1$ become larger and as a consequence
the couplings of $\tilde{N}_1$ to the $Z$ boson grow. A reasonable density
of dark matter can be obtained for $|\mass{\tilde{N}\dz{\alpha}}| \sim m_Z/2$ when the lightest inert
neutralino states annihilate mainly through an s-channel $Z$ boson. It is worth noting that if
$\tilde{N}_1$ were pure inert Higgsino then the s-channel $Z$ boson annihilation would proceed
with the full gauge coupling strength leaving a relic density too low to account for the
observed dark matter. In the $\E$SSM the LSP is mostly inert singlino so that its coupling
to the $Z$ boson is typically suppressed, since it only couples through its inert
Higgsino admixture, leading to an increased relic density. In practice an appropriate value
of $\Omega_{\mathrm{DM}}h^2$ can be achieved even if the coupling of $\tilde{N}_1$ to
the $Z$ boson is relatively small. This happens when $\tilde{N}_1$ annihilation proceeds
through the $Z$ boson resonance.
Thus, scenarios that result in a reasonable inert neutralino dark matter relic density correspond to lightest inert
neutralino masses that are much larger than $\bar{m}_b(m_{h_1})$ and hence to the
SM-like Higgs boson having very small branching ratios into SM particles.

\section{Benchmarks, Constraints, and Predictions}
\label{novel:bm}

\begin{table}
\begin{center}
\begin{tabular}{r|cccc|}
\hline
\textbf{Benchmark}          & \textbf{i}	&	\textbf{ii}	    &	\textbf{iii}	    &	\textbf{iv}	\\\hline
$\tan(\beta)$	            &	1.5	    &	1.5	    &	1.7	    &	1.564	\\
$m_{H^{\pm}}\approx m_{A}\approx m_{h_3}$ [GeV]&1977& 1977&	2022	&	1990	\\
$m_{h_1}$ [GeV]	            &	135.4	&	135.4	&	133.1	&	134.8	\\\hline
$\lambda_{22}$	            &	0.001	&	0.001	&	0.094	&	0.0001	\\
$\lambda_{21}$	            &	0.077	&	0.062	&	0	    &	0.06	\\
$\lambda_{12}$	            &	0.077	&	0.062	&	0	    &	0.06	\\
$\lambda_{11}$	            &	0.001	&	0.001	&	0.059	&	0.0001	\\\hline
$f_{d22}$	                &	0.001	&	0.001	&	0.53	&	0.001	\\
$f_{d21}$	                &	0.61	&	0.61	&	0.05	&	0.476	\\
$f_{d12}$	                &	0.6	    &	0.6	    &	0.05	&	0.466	\\
$f_{d11}$	                &	0.001	&	0.001	&	0.53	&	0.001	\\\hline
$f_{u22}$	        &	0.001	&	0.001	&	0.53	&	0.001	\\
$f_{u21}$	        &	0.426	&	0.426	&	0.05	&	0.4	\\
$f_{u12}$	        &	0.436	&	0.436	&	0.05	&	0.408	\\
$f_{u11}$	        &	0.001	&	0.001	&	0.53	&	0.001	\\\hline
$\tilde{N}_1$ mass [GeV]	&	41.91	&	47.33	&	33.62	&	-36.69	\\
$\tilde{N}_2$ mass [GeV]	&	-42.31	&	-47.84	&	47.78	&	36.88	\\
$\tilde{N}_3$ mass [GeV]	&	-129.1	&	-103.6	&	108.0	&	-103.11	\\
$\tilde{N}_4$ mass [GeV]	&	132.4	&	107.0	&	-152.1	&	103.47	\\
$\tilde{N}_5$ mass [GeV]	&	171.4	&	151.5	&	163.5	&	139.80	\\
$\tilde{N}_6$ mass [GeV]	&	-174.4	&	-154.4	&	-200.8	&	-140.35	\\\hline
$\tilde{C}_1$ mass [GeV]&	129.0	&	103.5	&	100.1	&	101.65	\\
$\tilde{C}_2$ mass [GeV]&	132.4	&	106.9	&	159.5	&	101.99	\\\hline
$\Omega_\chi h^2$	        &	0.096	&	0.098	&	0.109	&	0.107	\\\hline
$R_{Z11}$	                &	-0.0250	&	-0.0407	&	-0.144	&	-0.132	\\
$R_{Z12}$	                &	0.0040	&	0.0048	&	0.051	&	0.0043	\\
$R_{Z22}$	                &	-0.0257	&	-0.0429	&	-0.331	&	-0.133	\\\hline
$\Delta \NeffZ$	    &  0.000090	&	   0	&	0.0068	&	0.0073	\\
$D$	                        &	2.011	&	2.000	&	2.85	&	2.91	\\\hline
$X^{h_1}_{11}$	            &	0.137	&	0.147	&	0.110	&	-0.114	\\
$X^{h_1}_{12}+X^{h_1}_{21}$	&$-1.9\times 10^{-6}$&$-3.4\times 10^{-6}$&	0.0136	&$1.15\times 10^{-6}$\\
$X^{h_1}_{22}$	            &	-0.138	         &	-0.148	          &	0.125	&0.115	\\\hline
$\sigma_{\rr{SI}}$ [$10^{-44}$~cm$^2$]& 2.6--10.5        & 3.0--12.1	        &1.7--7.1&2.0--8.2\\\hline
$\mathrm{Br}(h\rightarrow \tilde{N}_1 \tilde{N}_1)$& 49.5\%	           & 49.7\%              & 57.8\% & 49.1\%\\
$\mathrm{Br}(h\rightarrow \tilde{N}_1 \tilde{N}_2)$& $7.9\times 10^{-11}$& $2.5\times 10^{-10}$& 0.34\% & 49.2\%\\
$\mathrm{Br}(h\rightarrow \tilde{N}_2 \tilde{N}_2)$& 49.0\%              & 48.5\%              & 39.8\% &$3.5\times 10^{-11}$\\
$\mathrm{Br}(h\rightarrow b\bar{b})$                         & 1.36\%              & 1.58\%              & 1.87\% & 1.59\%\\
$\mathrm{Br}(h\rightarrow \tau\bar{\tau})$                   & 0.142\%             & 0.165\%             & 0.196\%& 0.166\%\\\hline
$\Gamma(h\rightarrow \tilde{N}_1 \tilde{N}_1)$ [MeV] & 98.3	               &85.1                 & 81.7   & 82.9\\
$\Gamma(h)$ [MeV]                                           &198.7                &171.1                & 141.2  & 169.0\\\hline
\end{tabular}\end{center}
\caption{Benchmark scenarios for $m_{h_1} \approx 133$--135~GeV. The branching
ratios and decay widths of the lightest Higgs boson;
the masses of the active Higgs bosons, inert neutralinos, and charginos;
and the couplings of the inert neutralinos $\tilde{N}_1$ and $\tilde{N}_2$ are
calculated for $s=2400$~GeV, $\lambda=0.6$, 
$A_{\lambda}=1600$~GeV, $m_Q=m_u=M_s=700$~GeV, and $X_t=\sqrt{6} M_s$,
corresponding to $m_{h_2}\approx m_{Z_2}\approx 890$~GeV.
$\Delta\NeffZ$ and $D$ are defined in (\ref{eq:higgs14}) and (\ref{eq:higgs16}) respectively.\label{tab:tab1}}
\end{table}

\begin{table}
\begin{center}
\begin{tabular}{r|ccc|}
\hline
\textbf{Benchmark}       &	\textbf{v}	    &	\textbf{vi}	    &	\textbf{vii}	\\\hline
$\tan(\beta)$	            &	1.5	    &	1.7	    &	1.5	\\
$m_{H^{\pm}}\approx m_{A}\approx m_{h_3}$ [GeV]&1145&1165&  1145	\\
$m_{h_1}$ [GeV]	            &	115.9   &	114.4	&	115.9	\\

$\lambda_{22}$	            &	0.004	&	0.104	&	0.094	\\
$\lambda_{21}$	            &	0.084	&	0	    &	0	    \\
$\lambda_{12}$	            &	0.084	&	0	    &	0	    \\
$\lambda_{11}$	            &	0.004	&	0.09	&	0.059	\\\hline
$f_{22}$	                &	0.025	&	0.72	&	0.53	\\
$f_{21}$	                &	0.51	&	0.001	&	0.053	\\
$f_{12}$	                &	0.5	    &	0.001	&	0.053	\\
$f_{11}$	                &	0.025	&	0.7	    &	0.53	\\\hline
$f_{u22}$	        &	0.025	&	0.472	&	0.53	\\
$f_{u21}$	        &	0.49	&	0.001	&	0.053	\\
$f_{u12}$	        &	0.5	    &	0.001	&	0.053	\\
$f_{u11}$	        &	0.025	&	0.472	&	0.53	\\\hline
$\tilde{N}_1$ mass [GeV]	&	-35.76	&	41.20	&	35.42	\\
$\tilde{N}_2$ mass [GeV]	&	39.63	&	44.21	&	51.77	\\
$\tilde{N}_3$ mass [GeV]	&	-137.8	&	153.1	&	105.3	\\
$\tilde{N}_4$ mass [GeV]	&	151.7	&	176.7	&	-152.7	\\
$\tilde{N}_5$ mass [GeV]	&	173.6	&	-197.3	&	162.0	\\
$\tilde{N}_6$ mass [GeV]	&	-191.3	&	-217.9	&	-201.7	\\\hline
$\tilde{C}_1$ mass [GeV]&	135.8	&	152.7	&	100.1	\\
$\tilde{C}_2$ mass [GeV]&	149.3	&	176.5	&	159.5	\\\hline
$\Omega_\chi h^2$	        &	0.102	&	0.108	&	0.107	\\\hline
$R_{Z11}$	                &	-0.116	&	-0.0278	&	-0.115	\\
$R_{Z12}$	                &	0.0037	&  -0.00039	&	-0.045	\\
$R_{Z22}$	                &	-0.118	&	-0.0455	&	-0.288	\\\hline
$\Delta \NeffZ$	    &	0.0049	&  0.00009	&	0.0034	\\
$D$	                        &	2.62	&	2.011	&	2.43	\\\hline
$X^{h_1}_{11}$	            &	-0.117	&	0.141	&	0.117	\\
$X^{h_1}_{12}+X^{h_1}_{21}$	& -0.000027	& -0.00025	&	-0.0127	\\
$X^{h_1}_{22}$	            &	0.130	&	0.147	&	0.141	\\\hline
$\sigma_{\rr{SI}}$ [$10^{-44}$~cm$^2$]&3.9--15.7&5.4--21.9	&3.5--14.2      \\\hline
$\mathrm{Br}(h\rightarrow \tilde{N}_1 \tilde{N}_1)$& 49.6\%            & 53.5\%            &76.3\%\\
$\mathrm{Br}(h\rightarrow \tilde{N}_1 \tilde{N}_2)$&$2.1\times 10^{-8}$&$7.2\times 10^{-7}$&0.26\%\\
$\mathrm{Br}(h\rightarrow \tilde{N}_2 \tilde{N}_2)$& 48.4\%            & 44.2\%            &20.3\%\\
$\mathrm{Br}(h\rightarrow b\bar{b})$                         & 1.87\%            & 2.04\%            &2.83\%\\
$\mathrm{Br}(h\rightarrow \tau\bar{\tau})$                   & 0.196\%           & 0.21\%            &0.30\%\\\hline
$\Gamma(h\rightarrow \tilde{N}_1 \tilde{N}_1)$ [MeV] & 61.5	             & 60.1              &62.6  \\
$\Gamma(h)$ [MeV]                                           &124.1              & 112.2             &82.0  \\\hline
\end{tabular}\end{center}
\caption{Benchmark scenarios for $m_{h_1} \approx 114$--116~GeV. The branching
ratios and decay widths of the lightest Higgs boson;
the masses of the active Higgs bosons, inert neutralinos, and charginos;
and the couplings of the inert neutralinos $\tilde{N}_1$ and $\tilde{N}_2$ are
calculated for $s=2400$~GeV, $\lambda=g'_1=0.468$,
$A_{\lambda}=600$~GeV, $m_Q=m_u=M_s=700$~GeV, and $X_t=\sqrt{6} M_s$,
corresponding to $m_{h_2}\approx m_{Z_2}\approx 890$~GeV.
$\Delta\NeffZ$ and $D$ are defined in (\ref{eq:higgs14}) and (\ref{eq:higgs16}) respectively.
Continued in table~\ref{tab:tab3}\label{tab:tab2}}
\end{table}

\begin{table}
\begin{center}
\begin{tabular}{r|cc|}
\hline
\textbf{Benchmark} 	    &  \textbf{viii}               & \textbf{ix}\\\hline
$\tan(\beta)$	            & 1.5                 & 1.5\\
$m_{H^{\pm}}\approx m_{A}\approx m_{h_3}$ [GeV]	& 1145                & 1145\\
$m_{h_1}$ [GeV]	        	& 115.9               & 115.9\\

$\lambda_{22}$	          	& 0.001               & 0.468\\
$\lambda_{21}$	             & 0.079               & 0.05\\
$\lambda_{12}$	            & 0.080               & 0.05\\
$\lambda_{11}$	        	& 0.001               & 0.08\\\hline
$f_{22}$	            & 0.04                & 0.05\\
$f_{21}$	            	& 0.68                & 0.9\\
$f_{12}$	            & 0.68                & 0.002\\
$f_{11}$	          & 0.04                & 0.002\\\hline
$f_{u22}$	       	& 0.04                & 0.002\\
$f_{u21}$	      	& 0.49                & 0.002\\
$f_{u12}$	        	& 0.49                & 0.05\\
$f_{u11}$	        & 0.04                & 0.65\\\hline
$\tilde{N}_1$ mass [GeV]		& -45.08              & -46.24\\
$\tilde{N}_2$ mass [GeV]		& 55.34               & 46.60\\
$\tilde{N}_3$ mass [GeV]		& -133.3              & 171.1\\
$\tilde{N}_4$ mass [GeV]		& 136.9               & -171.4\\
$\tilde{N}_5$ mass [GeV]	& 178.4               & 805.4\\
$\tilde{N}_6$ mass [GeV]		& -192.2              & -805.4\\\hline
$\tilde{C}_1$ mass [GeV]		& 133.0               & 125.0\\
$\tilde{C}_2$ mass [GeV]		& 136.8               & 805.0\\\hline
$\Omega_\chi h^2$	        	& 0.0324              & 0.00005\\\hline
$R_{Z11}$	          & -0.0217             & -0.0224\\
$R_{Z12}$	                	& -0.0020             & -0.213\\
$R_{Z22}$	               	& -0.0524             & -0.0226\\\hline
$\Delta \NeffZ$	 & $1.57\times 10^{-6}$& 0\\
$D$	                      	& 2.0002              & 2.0\\\hline
$X^{h_1}_{11}$	          	& -0.147              & -0.148\\
$X^{h_1}_{12}+X^{h_1}_{21}$		&-0.0000140           & -0.000031\\
$X^{h_1}_{22}$	          	& 0.174               & 0.149\\\hline
$\sigma_{\rr{SI}}$ [$10^{-44}$~cm$^2$]     & 6.0--24.4                    & 6.1--25.0\\\hline
$\mathrm{Br}(h\rightarrow \tilde{N}_1 \tilde{N}_1)$&83.4\%             &49.3\%\\
$\mathrm{Br}(h\rightarrow \tilde{N}_1 \tilde{N}_2)$&$7.6\times 10^{-9}$&$3.0\times 10^{-8}$\\
$\mathrm{Br}(h\rightarrow \tilde{N}_2 \tilde{N}_2)$&12.3\%             &47.9\%\\
$\mathrm{Br}(h\rightarrow b\bar{b})$       &3.95\%             &2.58\%\\
$\mathrm{Br}(h\rightarrow \tau\bar{\tau})$         &0.41\%             &0.27\%\\\hline
$\Gamma(h\rightarrow \tilde{N}_1 \tilde{N}_1)$ [MeV] &49.0               &44.4\\
$\Gamma(h)$ [MeV]                        &58.8               &90.1\\\hline
\end{tabular}\end{center}
\caption{Continued from table~\ref{tab:tab2},
more benchmark scenarios for $m_{h_1} \approx 114$--116~GeV.
Again, the branching ratios and decay widths of the lightest Higgs boson;
the masses of the active Higgs bosons, inert neutralinos, and charginos;
and the couplings of the inert neutralinos $\tilde{N}_1$ and $\tilde{N}_2$ are
calculated for $s=2400$~GeV, $\lambda=g'_1=0.468$,
$A_{\lambda}=600$~GeV, $m_Q=m_u=M_s=700$~GeV, and $X_t=\sqrt{6} M_s$,
corresponding to $m_{h_2}\approx m_{Z_2}\approx 890$~GeV.
$\Delta\NeffZ$ and $D$ are defined in (\ref{eq:higgs14}) and (\ref{eq:higgs16}) respectively.\label{tab:tab3}}
\end{table}

In order to illustrate the features of the $\E$SSM mentioned in the previous section
we specify the set of benchmark points in tables~\ref{tab:tab1}, \ref{tab:tab2}, and~\ref{tab:tab3}.
For each benchmark scenario we calculate the spectra of inert neutralinos, inert charginos,
and active Higgs bosons as well as their couplings, the decay branching ratios of the lightest $CP$-even
Higgs state, and the dark matter relic density.
{\tt micrOMEGAs 2.2} is used to numerically compute the
present day density of dark matter.

\subsection{Benchmark scenarios}

\begin{figure}
\begin{center}
\includegraphics[width=\linewidth]{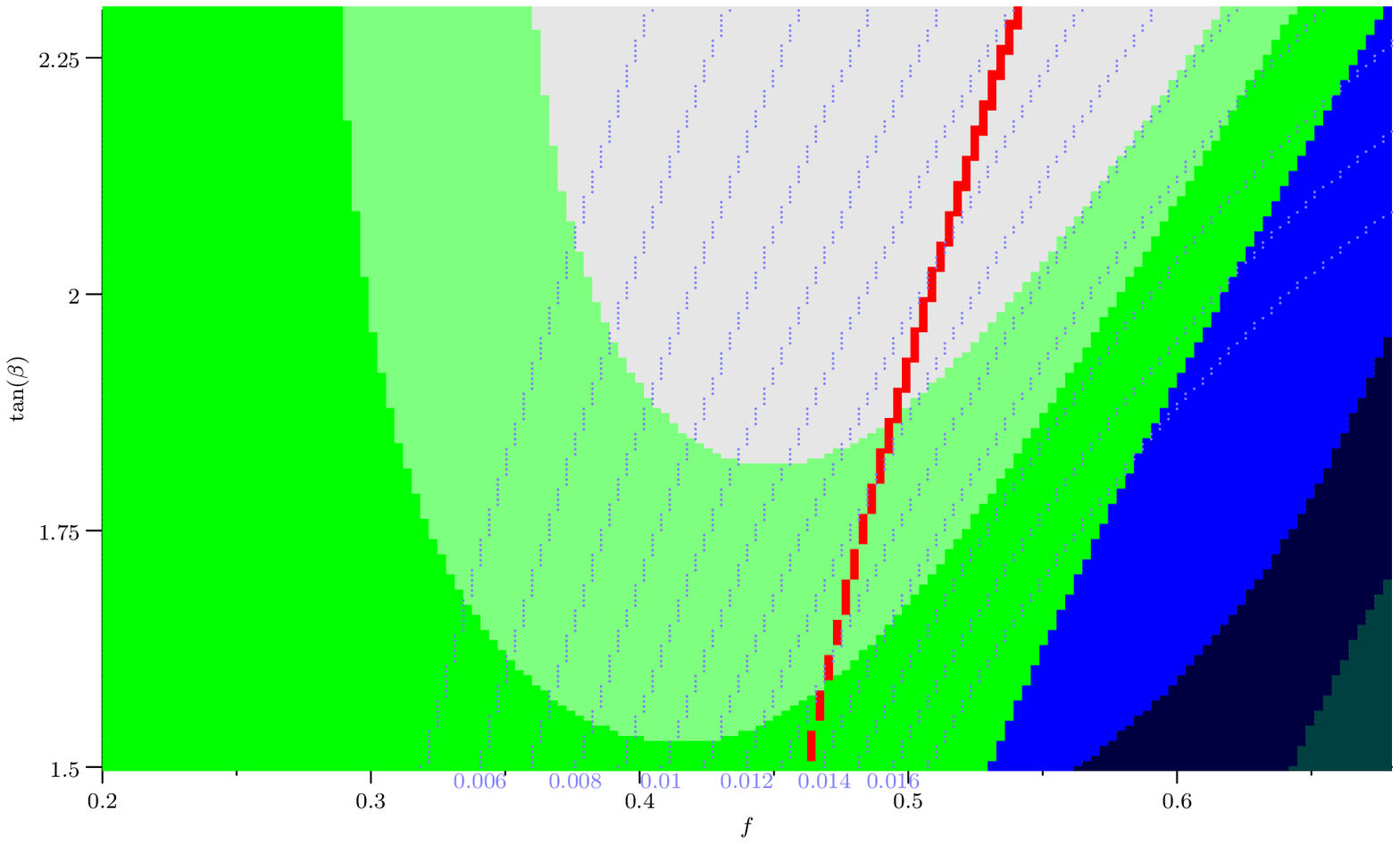}\\[5mm]
\includegraphics[width=\linewidth]{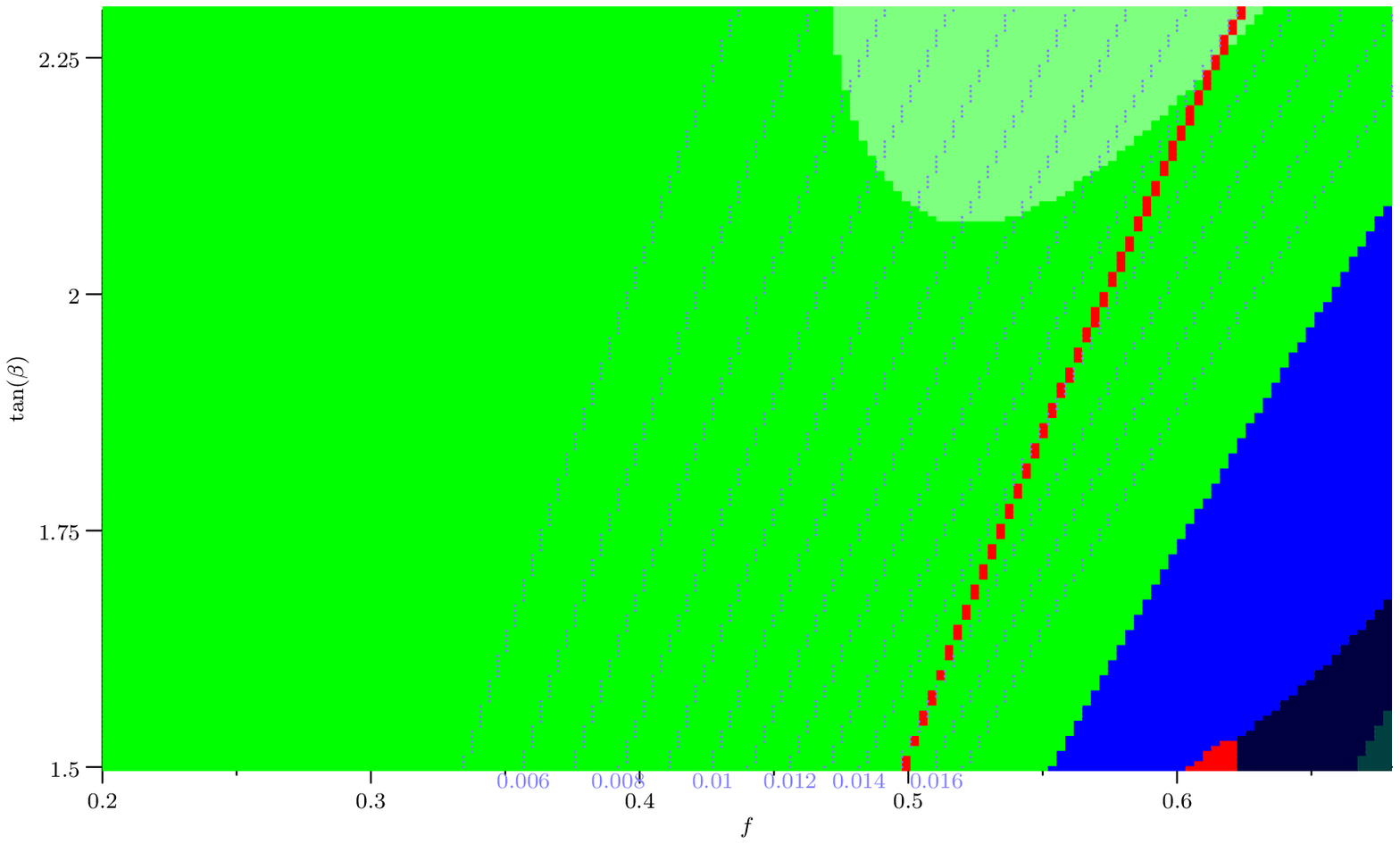}
\end{center}
\caption{Contour plots of $(X^{h_1}_{11})^2$ and various
regions in the $(f,\tan(\beta))$-plane with $s=2400$~GeV,
$f_{d\alpha\alpha}=f_{u\alpha\alpha}=\lambda_{\alpha\alpha}=0$~$\forall\alpha$,
$f_{d21}=f$, $f_{u21}=f_{d21}/a$, $f_{d12}=1.02f_{d21}$,
$f_{u12}=0.98f_{u21}$,
and $\lambda_{21}=\lambda_{12}=0.06$, implying that
$\mass{\tilde{C}_{1,2}}=101.8$~GeV. The upper plot is for $a=0.75+0.25\tan(\beta)$
and in the lower plot is for $a=0.5+0.5\tan(\beta)$.
The red region is where the prediction for $\Omega_{\chi} h^2$ is consistent
with the measured 1-sigma range of $\Omega_{\mathrm{DM}}h^2$ given in (\ref{eq:omegaDM}).
The dark green region corresponds to $D <3$ while
the pale green region represents the part of the parameter space in which $D$ is between
$3$ and $4$. The grey area indicates that $D > 4$. $D$ is defined in (\ref{eq:higgs16}).
The blue region corresponds to
$\mass{\tilde{N}_1} > m_Z/2$ while the dark blue region to the right is ruled out by the
requirement that perturbation theory remains valid up to the GUT scale.}\label{fig:fig2}
\end{figure}

In order to construct benchmark scenarios that are consistent with cosmological observations
and collider constraints we restrict our considerations to $\tan(\beta) \lesssim 2$.
The plots in figure~\ref{fig:fig2} show that in principle an appropriate
value of the dark matter density can be obtained
even when $\tan(\beta) > 2$. However, larger values of $\tan(\beta)$ lead to the lightest
and second lightest inert neutralinos having smaller masses as discussed in section~\ref{sec:analytical}.
As a result larger couplings of the lightest inert neutralinos to the $Z$ boson are required to
reproduce the measured value of $\Omega_{\mathrm{DM}}h^2$ and such light inert
neutralinos with substantial couplings to $Z$ boson
give a considerable contribution to its invisible width leading to a conflict with LEP
measurements. This is discussed in more detail in the following subsection.

Even for $\tan(\beta)\lesssim 2$ the lightest inert neutralino states can get
appreciable masses only if at least one of the inert chargino mass eigenstates
is light $\mass{\tilde{C}_1} \approx 100$--200~GeV. As clarified in
sections~\ref{ref:novelInert} and~\ref{sec:analytical}, the masses of the lightest
inert neutralino states decrease with increasing $\mass{\tilde{C}_{1,2}}$ and it is therefore
rather difficult to find benchmark scenarios consistent with cosmological observations for
$\mass{\tilde{C}_{1}} \gtrsim 200$~GeV. At the same time we demonstrate (with \textbf{benchmark~ix}
in table~\ref{tab:tab3}) that one light inert chargino mass eigenstate is enough to ensure
that the lightest inert neutralino state gains a mass of order $m_Z/2$.

To obtain the kind of inert neutralino and chargino spectra discussed above one has to
assume that the couplings $\lambda_{\alpha\beta}$ are rather small. They are expected
to be much smaller than the largest $f_{d\alpha\beta}$ and $f_{u\alpha\beta}$ couplings.
On the other hand, in order
to get $|\mass{\tilde{N}_1}|\sim |\mass{\tilde{N}_2}|\sim m_Z/2$ the Yukawa couplings $f_{d\alpha\beta}$ and
$f_{u\alpha\beta}$ need to be relatively close to their theoretical upper bounds
caused by the requirement of the validity of perturbation theory up to the GUT scale.
Since gauge coupling unification and RG flow
determine the low energy value of $g\ud{\prime}{1}$, the mass
of the $Z_2$ gauge boson is approximately set by the SM-singlet VEV $s$ only. In our study we choose
$s=2400$~GeV so that the $Z_2$ mass is about 890~GeV. This value of the $Z_2$ boson
mass is just above the
lower bound of 865~GeV found in ref.~\cite{Erler2010}~--- the most recent limit at the time of the publication of
\textbf{paper~II}~---
and allows satisfaction of stringent limits on the $Z_2$ mass and $Z$-$Z'$ mixing that come from
precision electroweak tests~\cite{Erler2009}.

Since we restrict our analysis to low values of $\tan(\beta)\lesssim 2$ the mass of the SM-like
Higgs boson is very sensitive to the choice of the coupling $\lambda$. Stringent LEP constraints
require $\lambda$ at the EWSB scale to be larger than the low energy value of $g'_1 \approx 0.47$ and if one
increases $\lambda$ much further then the theoretical upper bounds on $f_{d\alpha\beta}$ and
$f_{u\alpha\beta}$ from RG running become substantially stronger. As a consequence, it is rather difficult
to find solutions with $|\mass{\tilde{N}_1}|\sim |\mass{\tilde{N}_2}|\sim m_Z/2$. Therefore in our analysis we
concentrate on values of $\lambda$ at the EWSB scale less than about 0.6. In addition we set stop SSB masses
$m_Q=m_u=M_s=700$~GeV and restrict our consideration to the
maximal mixing scenario with the stop mixing parameter, defined in (\ref{eq:stopMixing}), $X_t=\sqrt{6} M_s$.
This choice of parameters limits the range of variations of
the lightest $CP$-even Higgs mass. In the leading two-loop approximation the mass of the
SM-like Higgs boson varies from 115~GeV for $\lambda=g'_1$ to 136~GeV for $\lambda=0.6$.
From tables~\ref{tab:tab1}, \ref{tab:tab2}, and~\ref{tab:tab3} one can see that
the large values of $\lambda\gtrsim g'_1$ that
we choose in our analysis result in an extremely hierarchical Higgs spectrum,
as pointed out in section~\ref{ref:Higgs}.
In tables~\ref{tab:tab1}, \ref{tab:tab2}, and~\ref{tab:tab3} the masses of the heavier
Higgs states are computed in the leading one-loop approximation. In the case of the lightest
Higgs boson mass the leading two-loop corrections are taken into account.

The set of benchmark points that we specify demonstrates that one can get a reasonable
dark matter density consistent with recent observations if
$|\mass{\tilde{N}_1}|\sim |\mass{\tilde{N}_2}|\sim m_Z/2$.
Our benchmark scenarios also indicate that in this case the SM-like Higgs boson decays predominantly
into the lightest inert neutralinos $\tilde{N}_1$ and $\tilde{N}_2$ while the total branching ratio
into SM particles varies from 2\% to 4\%.

\textbf{Benchmarks~i, ii, iv, v, and~viii} are motivated by the non-Abelian flavour
symmetry $\Delta_{27}$ which describes well the observed hierarchy in the quark and lepton sectors.
As was discussed in subsection~\ref{sub:Delta27}, these scenarios imply that all flavour diagonal Yukawa couplings
$\lambda_{\alpha\alpha}$, $f_{d\alpha\alpha}$, and $f_{u\alpha\alpha}$ are rather small.
Due to the approximate global $U(1)$ symmetry (\ref{eq:icn19}) that originates from the flavour symmetry
$\Delta_{27}$, the spectrum of inert neutralinos comprises a set of pseudo-Dirac states.
When the masses of the lightest and second lightest inert neutralinos are close, or they form an exact
Dirac state, then the decays $h_1 \rightarrow \tilde{N}_{\alpha}\tilde{N}_{\beta}$
lead to missing transverse energy in the final state.
These decay channels give rise to a large invisible branching ratio for the SM-like Higgs boson.

In tables~\ref{tab:tab1}, \ref{tab:tab2}, and~\ref{tab:tab3}
\textbf{benchmarks~i, ii, iv--vi, and~ix} have almost degenerate
lightest and second lightest inert neutralinos. In some of these benchmark points both lightest inert
neutralinos are lighter than $m_Z/2$ and as such the $Z$ boson can decay into
$\tilde{N}\dz{\alpha}\tilde{N}\dz{\beta}$ so that
the lightest and second lightest inert neutralino states contribute to the invisible $Z$ boson width.
In other benchmark scenarios both of the lightest inert neutralinos have masses above $m_Z/2$ and
the decays $Z \rightarrow \tilde{N}_{\alpha}\tilde{N}_{\beta}$ are kinematically forbidden.

When the LSP and NLSP are close in mass, LSP-NLSP coannihilation might be an important factor in 
determining the dark matter relic density. If this is the case then the LSP-NLSP mass splitting should 
be an important factor. Since annihilations of two identical neutralinos are p-wave suppressed, one should 
compare $\beta R_{Z11}$
with $R_{Z12}$ when trying to determine how important coannihilations are,
where $\beta$ is the relative speed of the incoming particles, approximately 1/6
around the time of thermal freeze-out.
It is useful to consider the following situations: With the LSP and NLSP almost degenerate
and with equal self-annihilation cross-sections, but a negligible coannihilation cross-section,
the relic density of dark matter would be twice what it would have have been if the NLSP
had not been present. If, alternatively, the coannihilation cross-section was
equal to the self-annihilation cross-sections then the existence of this extra channel would lead to a lower
present day relic density. In this case it would in fact be equal to the relic density
calculated in the absence of the NLSP. In this way, in such a scenario where coannihilations
and self-annihilations are about as important as each other, the relic density actually ends
up being largely independent of the LSP-NLSP mass splitting.

For the \textbf{benchmarks~i and~ii} this latter situation is approximately the case
and the LSP-NLSP mass splitting turns out not to be an important
factor. The mass splitting is in fact small~--- about half a GeV~--- but if it were larger
and the NLSPs were made to have frozen-out much earlier, the relic density would only
be decreased slightly (in this case by about a tenth).
In \textbf{benchmark~iv}, even though the LSP and NLSP are close in mass, coannihilations 
are unimportant due to the small value of $R_{Z12}$. In this case increasing the NLSP mass substantially 
while keeping everything else fixed would lead to an approximate halving of the predicted relic density, 
since the NLSP would have decoupled much earlier than, rather than at the same time as, the LSP. 
The only other benchmark scenario where the LSP and NLSP are close enough in mass for coannihilations 
to be potentially important is \textbf{benchmark~ix}. Here coannihilation is in fact the dominant process and 
changing the LSP-NLSP mass splitting would have a large effect on the predicted relic density. In fact, 
in this scenario if the NLSP were not present the predicted relic density would be within 
the measured range.

If the mass difference between the second lightest and the lightest inert neutralino is around
10~GeV or more then some of the decay products of a $\tilde{N}_2$ that originates from a
SM-like Higgs boson decay might be observed at the LHC. In our analysis we assume that
all scalar particles, except for the lightest Higgs boson, are heavy and that the couplings of the
inert neutralino states to quarks, leptons, and their superpartners are relatively small.
As a result the second lightest inert neutralino decays into the lightest one and a
fermion-antifermion pair mainly via a virtual $Z$. In our numerical analysis we did
not manage to find any scenario with $|\mass{\tilde{N}_2}|-|\mass{\tilde{N}_1}|\gtrsim 20$~GeV
leading to reasonable values of $\Omega_\chi h^2$. Hence we do not expect
any observable jets at the LHC associated with the decay of a
$\tilde{N}_2$ produced through a SM-like Higgs decay. However, it might be possible to detect some
lepton-antilepton pairs coming from decays of the form $h_1\rightarrow\tilde{N}_{2}\tilde{N}_{\alpha}$.
In particular we hope that $\mu^{+} \mu^{-}$ pairs coming from such
decays of the lightest $CP$-even Higgs state could be observed at the LHC.

\textbf{Benchmarks~iii, vii, and~viii} can lead to such relatively
energetic muon pairs in the final states of SM-like Higgs decays. Since the Higgs branching
ratios into SM particles are rather suppressed, such decays of the
lightest $CP$-even Higgs state might play an essential role in
Higgs searches.

In addition to the inert Higgs decays, the scenarios considered here imply that
at least two of the inert neutralino states that are predominantly formed from
the fermionic components of the inert Higgs doublet supermultiplets,
as well as one of the inert chargino states, should have masses below
$200$~GeV. Since these states are almost inert Higgsinos they couple rather
strongly to $W$ and $Z$ bosons. Thus at hadron colliders the corresponding inert
neutralino and chargino states could be produced in pairs via off-shell $W$ and $Z$ bosons.
Since they are light their production cross-sections at the LHC would not be negligibly
small. After being produced, inert neutralino and chargino states would sequentially decay into
the LSP and pairs of quarks and leptons resulting in distinct signatures that could be
discovered at the LHC.

\subsection{Neutralino and chargino collider limits}
\label{ref:neuchalim}

The remarkable signatures discussed above raise serious concerns that they could have
already been observed at the Tevatron and/or even earlier at LEP. For example, the light inert
neutralino and chargino states could have been produced at the Tevatron.
The CDF and D0 collaborations have set a stringent lower bound on chargino
masses using supersymmetry searches with a trilepton signal~\cite{Abazov2009,Strologas2010}.
These searches ruled out chargino masses below $164$~GeV. However, this lower
bound on the chargino mass was obtained by assuming that the corresponding chargino
and neutralino states decay predominantly into the LSP and a pair of leptons. In our case,
however, the inert neutralino and chargino states are expected to decay
via virtual $Z$ and $W$ exchange, decaying predominantly into the LSP and
a pair of quarks. As a consequence the lower limit on the mass of charginos that is set
by the Tevatron is not directly applicable to the benchmark scenarios that we consider
here. Instead, in our study, we use the 95\% confidence level lower limit on chargino masses
of about $100$~GeV that was set by LEP~\cite{Kraan2005}.

In principle the LEP experiments also set constraints on the masses and couplings of
neutral particles that interact with the $Z$ boson.
As mentioned above, when the masses of $\tilde{N}_1$ and $\tilde{N}_2$ are below $m_Z/2$
they are almost degenerate and thus the decays of $Z\rightarrow\tilde{N}_{\alpha}\tilde{N}_{\beta}$
all contribute to the invisible width of the $Z$ boson, changing the effective number of
neutrino species $\NeffZ$. The contribution of $\tilde{N}_1$ and $\tilde{N}_2$
to $\NeffZ$ is given by
\be
\Delta \NeffZ= a_{11}+2a_{12}+a_{22},
\label{eq:higgs14}
\ee
where
\be
a\dvp{2}{\alpha\beta} &=& R_{Z\alpha\beta}^2
\left[1-\dfrac{\masss{2}{\tilde{N}_{\alpha}}+\masss{2}{\tilde{N}_{\beta}}}{2 m_Z^2}
-3\dfrac{\mass{\tilde{N}_{\alpha}}\mass{\tilde{N}_{\beta}}}{m_Z^2}
- \dfrac{\left(\masss{2}{\tilde{N}_{\alpha}}-\masss{2}{\tilde{N}_{\beta}}\right)^2}{2 m_Z^4}\right]\nn\\
&&\qquad\qquad\quad\sqrt{\left(1-\dfrac{\masss{2}{\tilde{N}_{\alpha}}+\masss{2}{\tilde{N}_{\beta}}}{m_Z^2}\right)^2
-4\dfrac{\masss{2}{\tilde{N}_{\alpha}} \masss{2}{\tilde{N}_{\beta}}}{m_Z^4}}.
\label{eq:higgs15}
\ee
All three terms in (\ref{eq:higgs14}) contribute to $\Delta\NeffZ$ only if $2|\mass{\tilde{N}_{2}}| < m_Z$.
In the case where only the $Z$ boson decays into $\tilde{N}_1\tilde{N}_1$ are kinematically allowed
$a_{12}$ and $a_{22}$ should be set to zero. If $|\mass{\tilde{N}_{1}}|+|\mass{\tilde{N}_{2}}| < m_Z$
whilst $2|\mass{\tilde{N}_{2}}| > m_Z$ then only $a_{11}$ and $2a_{12}$ contribute.

In order to compare the measured value of $\NeffZexp$ with the effective number of neutrino species
in the $\E$SSM $\NeffZ=3+\Delta \NeffZ$ it is convenient to define the variable
\be
D &=& \dfrac{\NeffZ-\NeffZexp}{\sigma\ud{\rr{exp}}{\NeffZ}}.
\label{eq:higgs16}
\ee
The value of $D$ represents the deviation
between the predicted and measured effective number of neutrinos contributing to the $Z$ boson invisible
width. It is worth pointing out that in the SM, from (\ref{eq:NeffZ}), $D=2$.
In the benchmark scenarios presented in
tables~\ref{tab:tab1}, \ref{tab:tab2}, and~\ref{tab:tab3}
the value of $D$ is always less than 3. The plots in figure~\ref{fig:fig2}
also demonstrate that there is a substantial region
of $\E$SSM parameter space with $\mass{\tilde{N}_{1,2}} < m_Z/2$ and $D < 3$.
This indicates that relatively light inert neutralinos
with masses below $m_Z/2$ are not necessarily ruled out by constraints on the effective number
of neutrinos set by LEP experiments.
Indeed, as argued in section~\ref{ref:novelInert}, the Yukawa couplings $f_{d\alpha\beta}$ and
$f_{u\alpha\beta}$ can be chosen such that $R_{Z\alpha\beta}$ are very small.
The couplings of the lightest and second lightest inert neutralinos to the $Z$ boson
are relatively small anyway because of the inert singlino admixture in these states.
Nevertheless, figure~\ref{fig:fig2} shows that scenarios with light inert neutralinos having
masses below $m_Z/2$ and relatively small couplings to the $Z$ boson can lead to
appropriate dark matter densities consistent with observation.

LEP has set limits on the cross-sections of $e^{+}e^{-}\rightarrow\tilde{N}_2\tilde{N}_1$
and $e^{+}e^{-}\rightarrow \tilde{C}_1^{+}\tilde{C}_1^{-}$,
where predominantly $\tilde{N}_2\rightarrow q\bar{q}\tilde{N}_1$
and $\tilde{C}_1\rightarrow q\bar{q}'\tilde{N}_1$ respectively~\cite{TheOPALcollaboration2004}.
Unfortunately the bounds are not directly applicable for our study because
OPAL limits were set for a relatively heavy $\tilde{N}_2$ or $\tilde{C}_1$ only~---
greater than about 60~GeV. Nevertheless, these bounds demonstrate that
it was difficult to observe light neutralinos with masses less than about 100~GeV
if their production cross-sections
$\sigma(e^{+}e^{-}\rightarrow\tilde{N}_{\alpha}\tilde{N}_{\beta})\lesssim 0.1$--0.3~pb.
Since at LEP energies the cross-sections of colourless particle
production through s-channel $\gamma$/$Z$ exchange are typically a few picobarns, the
lightest and second lightest inert neutralino states in the $\E$SSM could have escaped detection at
LEP if their couplings $R_{Z\alpha\beta}\lesssim 0.1$--0.3.

\subsection{Dark matter direct detection}

Another constraint on the couplings of the lightest inert neutralino comes from experiments
for the direct detection of dark matter.
At the time of the publication of \textbf{paper~II} the most stringent upper limits on the
DMP-nucleon elastic scattering spin-independent cross-section came from the CDMS collaboration~\cite{TheCDMSCollaboration2009}
and from the first analysis of 11.7 days of data from the XENON100 experiment~\cite{Aprile2010}.
In the low DMP mass region relevant for our study, the most stringent of these was the
XENON100 limit. In particular the XENON100 11.7 day analysis produced a limit on the cross-section of
$3.4\times 10^{-44}$~cm$^2$ for a 55~GeV DMP at a confidence level of 90\%. This limit remains fairly constant
for lower DMP masses and does not increase above about $4\times 10^{-44}$~cm$^2$
for even the lowest LSP masses that are consistent with our thermal freeze-out scenario.
Currently the most stringent limits on the spin-independent DMP-nucleon cross-section come from the more recent
analysis of 100.9 days of data from XENON100~\cite{XENON100Collaboration2011}. The best limit
is $7.0\times10^{-44}$~cm$^2$, which is for a 50~GeV DMP, again at a confidence level of 90\%.

Since in the $\E$SSM
the couplings of the lightest inert neutralino to quarks, leptons, and their superpartners are
suppressed, the spin-independent part of the $\tilde{N}_1$-nucleon elastic scattering
cross-section is mediated mainly by t-channel SM-like Higgs boson exchange. Thus, in
the leading approximation the spin-independent part of $\tilde{N}_1$-nucleon cross-section in the
$\E$SSM takes the form~\cite{Ellis2008a,Kalinowski2009}
\be
\sigma_{\rr{SI}} &=& \dfrac{4 m^2_r m_N^2}{\pi v^2 m^4_{h_1}} |X^{h_1}_{11} F^N|^2,
\label{eq:higgs17}
\ee
where $N$ is the nucleon,
\be
m_r &=& \dfrac{\mass{\tilde{N}_1} \mass{N\vphantom{\tilde{N}_1}}}{\mass{\tilde{N}_1}
+\mass{N\vphantom{\tilde{N}_1}}},\quad\mbox{and}\nn\\\nn\\
F^N &=& \sum_{q=u,d,s} f^N_{Tq} + \dfrac{2}{27}\sum_{Q=c,b,t} f^N_{TG},\nn
\ee
with
\be
m_N f^N_{Tq} &=& \langle N | m_{q}\bar{q}q |N \rangle\quad\mbox{and}\nn\\\nn\\
f^N_{TG} &=& 1 - \sum_{q=u,d,s} f^N_{Tq}.\nn
\ee
Here, for simplicity, we assume that the lightest Higgs state has the same couplings
as a SM Higgs boson and ignore all contributions induced by heavy Higgs boson
and squark exchange\footnote{The near degeneracy of the lightest and second to
lightest inert neutralinos could result in the inelastic scattering collisions in which
$\tilde{N}_1$ is upscattered off of a nucleus into $\tilde{N}_2$ and this
could affect the direct
detection of $\tilde{N}_1$ in experiments. However, such processes may take place
only if the LSP-NLSP mass splitting is less than about
100~keV~\cite{Smith2001}. In the $\E$SSM mass splittings of this order
are not expected to be typical. In the benchmark scenarios considered in tables~\ref{tab:tab1},
\ref{tab:tab2}, and~\ref{tab:tab3} the mass splitting is substantially larger
and such inelastic nuclear scattering of $\tilde{N}_1$ does not play a significant role.}.
Due to the hierarchical structure of the active Higgs boson spectrum and the approximate 
$\ZZ{H}$ symmetry this approximation works very well. Using the experimental limits 
set on $\sigma_{\rr{SI}}$ and (\ref{eq:higgs17}) one can obtain upper bounds 
on $X^{h_1}_{11}$~\cite{Cheung2010}.

In tables~\ref{tab:tab1}, \ref{tab:tab2}, and~\ref{tab:tab3} we specify the
interval of variations of $\sigma_{\rr{SI}}$ for each benchmark
scenario. As one can see from (\ref{eq:higgs17}) the value of $\sigma_{\rr{SI}}$ depends
rather strongly on the hadronic matrix elements~--- the coefficients $f^N_{Tq}$ that
are related to the $\pi$-nucleon $\sigma$ term and the spin content of the nucleon.
The hadronic uncertainties in the elastic scattering cross-section of DMPs and
nucleons were considered in ref.~\cite{Ellis2008a}.
In particular, it was pointed out that $f^N_{Ts}$ could vary over a wide range.
In tables~\ref{tab:tab1}, \ref{tab:tab2}, and~\ref{tab:tab3} the lower limit
on $\sigma_{\rr{SI}}$ corresponds to $f^N_{Ts}=0$ while the upper
limit corresponds to $f^N_{Ts}=0.36$ (see ref.~\cite{Kalinowski2009}).
From tables~\ref{tab:tab1}, \ref{tab:tab2}, and~\ref{tab:tab3} and
(\ref{eq:higgs17}) it also becomes clear that $\sigma_{\rr{SI}}$ decreases substantially
when $\mass{h_1}$ grows.

Since in all of the benchmark scenarios presented in tables~\ref{tab:tab1}, \ref{tab:tab2}, and~\ref{tab:tab3}
the lightest inert neutralino is relatively heavy, with $|\mass{\tilde{N}_{1}}|\sim m_Z/2$,
allowing for a small enough dark matter relic density,
the coupling of $\tilde{N}_1$ to the lightest $CP$-even Higgs state is always large,
giving rise to a $\tilde{N}_1$-nucleon spin-independent cross-section that is of the order of,
or larger than, the 90\% confidence level bound of ref.~\cite{Aprile2010}.

The dark matter scenario detailed in the present and previous chapters
is now severely challenged by the most recent XENON100 results~\cite{XENON100Collaboration2011}.
Although these results appear to rule out the $\E$SSM as a model of dark matter,
it should be noted that they do not rule out the model per se. Scenarios similar to those in
tables~\ref{tab:tab1} and~\ref{tab:tab2}, but in which the predicted relic
density is somewhat less than the measured relic density can be consistent with direct
detection experiments, although since such scenarios would not completely explain
the observed dark matter relic density they may be considered less well motivated. In such
scenarios it needs to be assumed that the majority of the observed dark matter is not
composed of $\E$SSM inert neutralino LSPs, but is composed of some extra matter beyond that
of the $\E$SSM.

\section{Summary and Conclusions}
\label{novel:conclusions}

In this work we considered novel decays of the SM-like Higgs boson in the $\E$SSM.
Particular attention was given to the dark matter that the model predicts and
this work also represents a more in-depth study of the inert neutralino and chargino sectors
of the $\E$SSM than the previous study presented in chapter~\ref{chap:ndmin} and \textbf{paper~I}.

To satisfy LEP constraints we restricted our consideration to scenarios with relatively
heavy inert chargino states $m_{\tilde{C}_{1,2}}\gtrsim 100$~GeV. In our
analysis we also required the validity of perturbation theory up to the GUT scale which
sets stringent constraints on the values of the Yukawa couplings at low energies.
Using these restrictions we argued that the lightest and the second lightest inert
neutralinos are always light~--- they typically have masses
below 60--65~GeV. These neutralinos are mixtures of inert Higgsinos and singlinos. In the considered
model the lightest inert neutralino $\tilde{N}_1$ tends to be the LSP and play the
role of dark matter while $\tilde{N}_2$ tends to be the NLSP.
The masses of $\tilde{N}_1$ and $\tilde{N}_2$
decrease with increasing $\tan(\beta) > 1$ and inert chargino masses.

Because the lightest inert neutralino states are predomiantly
inert singlino in nature, their couplings to the gauge bosons, active Higgs bosons, quarks,
and leptons are rather small, resulting in relatively small LSP annihilation
cross-sections and the possibility
of an unacceptably large dark matter density. In the limit where all non-SM states except for the
inert neutralinos and charginos are heavy a reasonable density of dark
matter can be obtained if $|\mass{\tilde{N}_{1,2}}| \sim m_Z/2$,
where the inert LSPs annihilate mainly through an s-channel $Z$ boson.
On resonance an appropriate value of $\Omega_{\chi}h^2$ can be achieved even for a relatively small
coupling of the LSP to the $Z$ boson. In order to achieve plausible scenarios consistent with both
LEP and cosmological observations, requiring $|\mass{\tilde{N}_{1}}| \sim m_Z/2$ if $\tilde{N}_{1}$ contributes to CDM,
$\tan(\beta)$ cannot be too large.

The main message arising from this work is that within the dark matter motivated scenario
although the lightest and the second lightest inert neutralinos can have small couplings to
the $Z$ boson their couplings to the SM-like Higgs state $h_1$ are always large. Indeed, we argued
that in the first approximation the couplings of $\tilde{N}_1$ and $\tilde{N}_2$ to the lightest $CP$-even
Higgs boson are proportional to $|\mass{\tilde{N}_1}|/v$ and $|\mass{\tilde{N}_2}|/v$ respectively.
Since $|\mass{\tilde{N}_{1,2}}|$ must be of order $m_Z/2$ in order for the theory not to predict
too much dark matter,
these couplings are much larger than the corresponding coupling of $b$ quarks to the SM-like Higgs boson.
Thus the SM-like Higgs
boson decays predominantly into the lightest inert neutralino states and has very small branching
ratios (2\%--4\%) for decays into SM particles.

The most recent XENON100 dark matter direct detection limits~\cite{XENON100Collaboration2011}
now place rather stringent constraints on the $\E$SSM inert neutralino dark matter scenario.
As an explanation for all of the observed dark matter relic density the model now looks to
be ruled out. There do, however, exist scenarios in which the $\E$SSM LSP
accounts for only some fraction of the observed dark matter
that are consistent with constraints from colliders and cosmology.

\cleardoublepage

\newpage
\chapter{Dark Matter and Big Bang Nucleosynthesis in the
$\EZ$SSM}
\label{chap:ezssm}

In this chapter, which contains work that was first published in \textbf{paper III},
with the exception of section~\ref{ref:wdm} which contains work that is original to this thesis,
we introduce a new scenario for dark matter in the $\E$SSM
in which the dark matter candidate is just the usual bino. 
At first sight having a bino dark matter candidate seems 
impossible since, as already discussed,
the lightest inert neutralino mass eigenstates,
predominantly the inert singlinos $\tilde{S}_{\alpha}$,
naturally have suppressed masses and it is very difficult to make them even as heavy as 
half the $Z$ mass. 
To overcome this we propose that the inert singlinos are exactly massless and decoupled from the 
bino, which is achieved in practice by setting the Yukawa
couplings $f\dz{(d,u)\alpha\beta}$ to zero.
This is easy to do by introducing 
a discrete symmetry $\ZZ{S}$ under which the inert singlet scalars $S_{\alpha}$
are odd and all other bosonic states are even~--- a scenario we refer to as
the $\EZ$SSM.

In the $\EZ$SSM the inert singlinos $\tilde{S}_{\alpha}$
will be denoted as $\tilde{\sigma}$ in order to emphasise their different (massless and decoupled)
nature.
The stable DMP
is then generally mostly bino and the observed dark matter
relic density can be achieved via a novel scenario in which the
bino inelastically scatters off of
SM matter into heavier inert Higgsinos during the time of thermal freeze-out,
keeping the bino in equilibrium long enough to give the desired relic abundance.
As long as the inert Higgsinos are close in mass to the bino this is always possible to arrange~---
the only constraint being that the inert Higgsinos satisfy the LEP constraint of being heavier than
100~GeV. This in turn implies that the bino must also be heavier than or close to 100~GeV.
These constraints are easy to satisfy and, unlike in the inert neutralino LSP dark matter scenario,
we find that successful relic abundance can be achieved within
a GUT scale constrained version of the model~--- the c$\EZ$SSM~---
assuming a unified soft scalar
mass $m_0$, soft gaugino mass $M_{1/2}$, and soft trilinear mass $A_0$ at
the GUT scale.

It is worth noting that
studies of the c$\E$SSM~\cite{Athron2009a,Athron2009,Athron2011}
have hitherto neglected to study the
full $12 \times 12$ neutralino mass matrix and only considered the $6 \times 6$ mass matrix of the 
USSM~\cite{Kalinowski2009}.
Although the question of dark matter
was addressed in the USSM, the requirement of successful 
relic abundance was not imposed on the c$\E$SSM
in refs.~\cite{Athron2009a,Athron2009,Athron2011}
even though both analyses considered the same $6 \times 6$ neutralino mass matrix.
This is because it was expected that dark matter would arise from the inert
sector of the c$\E$SSM which was not studied.
When cosmological constraints on inert neutralino dark matter are included
in the $\E$SSM certain trilinear Higgs Yukawa couplings relevant to the inert sector
are required to be large as we saw in chapters~\ref{chap:ndmin} and~\ref{chap:nhde}.
In the c$\E$SSM these large couplings strongly affect the RG running
from the GUT scale and we have not been able to show that having inert neutralino LSPs
consistent with CDM constraints can also be consistent with having universal (GUT scale constrained) soft mass
parameters.
Here we shall consider the c$\E$SSM with the full $12 \times 12$ neutralino mass matrix,  
including both the USSM and inert neutralinos, under the
assumption that the fermionic components of the inert SM-singlet supermultiplets,
the two inert singlinos, are forbidden to acquire mass by an extra $\dd{Z}_2$
symmetry of the superpotential. In practice there is then a $10 \times 10$
neutralino mass matrix once the two massless inert singlinos are decoupled.

In summary, the main result of this study is
that bino dark matter, with nearby inert Higgsinos and massless inert singlinos,  
provides a simple and consistent picture of dark matter in the $\E$SSM and is consistent with 
GUT scale unified soft parameters.
We also consider the effect of 
the presence of the two massless inert singlinos in the $\EZ$SSM on the 
effective number of neutrinos contributing to the expansion rate of the universe
prior to BBN, affecting $^4$He production. Current fits to WMAP data~\cite{Komatsu2011}
favour values greater than three, so the presence of additional contributions to the effective number
of neutrinos is another interesting aspect of the $\EZ$SSM.
In practice we find that the additional number of effective neutrino species is 
less than two, due to entropy dilution, depending on the mass of the $Z'$ boson which keeps the
inert singlinos in equilibrium. 

The $\EZ$SSM is introduced and its neutralino sector is explored
in section~\ref{ezssm:model}.
The details of the dark matter calculation are
presented in section~\ref{ezssm:dm}.
$\Neff$ is defined and calculations of its value in the $\EZ$SSM are presented
in section~\ref{ezssm:Neff}. Some benchmark points are presented in
section~\ref{ezssm:benchmarks}. The possibility of inert singlino WDM is
discussed in section~\ref{ref:wdm}
and the conclusions are summarised in
section~\ref{ezssm:conclusions}.

\section{The $\EZ$SSM}
\label{ezssm:model}

In the $\EZ$SSM, as well as being invariant under $\ZZ{M}$ and either $\ZZ{L}$
or $\ZZ{B}$, summarised in tables~\ref{tab:Z2} and~\ref{tab:Z2S},
the superpotential of the $\E$SSM (\ref{eq:w}) is also invariant under an additional exact
$\dd{Z}_2$ symmetry called $\ZZ{S}$. Under this symmetry only
the two inert SM-singlet fields $S_\alpha$ are odd.
The couplings of the forms
$\lambda_{\alpha ij}$ and $\kappa_{\alpha ij}$ are therefore forbidden. This
means that the fermionic superpartners of $S_\alpha$~--- the inert singlinos
$\tilde{\sigma}$~--- are forbidden to have mass
and do not mix with the other neutralinos.
They interact only via their gauge couplings to the $Z'$ boson which
exist since they
are charged under the $U(1)_N$ gauge symmetry.

All of the exact and approximate discrete symmetries relevant to the
$\EZ$SSM superpotential are summarised in table~\ref{tab:Z2S}.

\begin{table}\begin{center}
\begin{tabular}{r|ccccc|}
& $\dd{Z}\ud{M}{2}$ & $\dd{Z}\ud{L}{2}$ & $\dd{Z}\ud{B}{2}$ & $\dd{Z}\ud{H}{2}$ & $\ZZ{S}$ \\\hline
$S\dvp{c}{\alpha}$                                     & $+$ & $+$ & $+$ & $-$ & $-$ \\&&&&&\\
$H\dvp{c}{d\alpha},H\dvp{c}{u\alpha}$                  & $+$ & $+$ & $+$ & $-$ & $+$ \\&&&&&\\
$S\dvp{c}{3},H\dvp{c}{d3},H\dvp{c}{u3}$                & $+$ & $+$ & $+$ & $+$ & $+$ \\&&&&&\\
$\tilde{Q}\dvp{c}{Li},\tilde{d}\ud{c}{Ri},\tilde{u}\ud{c}{Ri}$
                                                       & $-$ & $+$ & $+$ & $-$ & $+$ \\&&&&&\\
$\tilde{L}\dvp{c}{Li},\tilde{e}\ud{c}{Ri},\tilde{N}\ud{c}{i}$
                                                       & $-$ & $-$ & $-$ & $-$ & $+$ \\&&&&&\\
$\bar{D}\dvp{c}{i},D\dvp{c}{i}$                        & $+$ & $+$ & $-$ & $-$ & $+$ \\\hline
\end{tabular}
\caption{The charges of the fields of the $\EZ$SSM superpotential under
various exact and approximate $\dd{Z}\dz{2}$ symmetries that the superpotential may or may
not obey.
$\ZZ{M}$ is already a symmetry due to gauge
invariance.
Either $\ZZ{L}$ or $\ZZ{B}$ is imposed in order to avoid rapid
proton decay. $\ZZ{H}$ is an approximate flavour symmetry.
In the $\EZ$SSM the extra symmetry $\ZZ{S}$ is imposed, forcing the inert singlinos
to be massless and decoupled.
$i\in\{1,2,3\}$ and $\alpha\in\{1,2\}$.\label{tab:Z2S}}
\end{center}\end{table}

One may worry that the effects of the massless inert singlinos would have already been
seen in precision measurements from LEP. The inert singlinos, although not mixing with the
inert Higgsinos as they did in the $\E$SSM inert neutralino dark matter scenario,
still couple to the $Z_1$ mass eigenstate because of the non-zero $Z$-$Z'$ mixing
angle $\alpha_{ZZ'}$ defined in (\ref{eq:alphaZZ}). In the following we neglect the
kinetic term mixing that is expected to be a small effect (see section~\ref{ref:GaugeMixing}).
For a given $m_{Z'} \approx m_{\vphantom{Z'}Z_2}$ in (\ref{eq:ZZ}) $m\ud{2}{ZZ'}$, and hence also
$\alpha_{ZZ'}$, is maximised
in the limit $\tan(\beta) \rightarrow \infty$. For $m_{Z_2} \approx m_{Z'} = 892$~GeV
the maximum value of $m\ud{2}{ZZ'}$ is 3270~GeV$^2$ and the maximum value of
$\alpha_{ZZ'}$ is then $4.15 \times 10^{-3}$. The $Z_1$-$\tilde{\sigma}$-$\tilde{\sigma}$
coupling relative to the $Z$-$\nu$-$\nu$ gauge coupling $R$ is equal to $\alpha_{ZZ'}$.
From (\ref{eq:higgs14}) the change in the effective number of neutrinos contributing to the invisible $Z$ boson width
at LEP due to the presence of massless inert singlinos is then $\Delta\NeffZ = 2R^2 = 2\alpha\ud{2}{ZZ'} = 1.72 \times 10^{-5}$
which is well below the experimental uncertainty $\sigma\ud{\rr{exp}}{\NeffZ} = 8 \times 10^{-3}$.
When the $Z_2$ boson mass is large enough to avoid experimental detection limits the
contributions of massless inert singlinos to the $Z$ boson width and to other
LEP precision measurements are expected to be within the experimental error.

\subsection{The neutralinos of the $\EZ$SSM}
\label{sub:neutralinos}

The chargino sector of the $\EZ$SSM is unchanged from that of the $\E$SSM
without $\ZZ{S}$. The chargino mass matrix is that given in (\ref{eq:cmm}).
The same is true for the active Higgs scalar masses
and mass matrices given in section~\ref{ref:Higgs}. The situation in the neutralino
sector, however, is quite different.

In the present study of the $\EZ$SSM we define the term
`neutralino' not to include the massless inert singlinos
which do not appear in the superpotential and are decoupled.
The neutralino mass matrix $M^N$ in the interaction basis
\be
\tilde{N}_{\rr{int}} &=& \left(\ba{cccccc|cc}
\tilde{B} & \tilde{W}^3 & \tilde{H}^0_{d3} & \tilde{H}^0_{u3} & \tilde{S}^{\vphantom{0}}_3 &
\tilde{B}' & \tilde{H}^0_{d\alpha} & \tilde{H}^0_{u\beta} \ea\right)^T,
\ee
and again neglecting the small bino-bino$'$ mixing,
is then equal to
\be
\left(\ba{cccccc|cc}
M\dvp{\prime}{1} & 0 & -\frac{1}{2}g'v_d & \frac{1}{2}g'v_u & 0 & 0 & 0 & 0 \\
0 & M\dvp{\prime}{2} & \frac{1}{2}gv_d & -\frac{1}{2}gv_u & 0 & 0 & 0 & 0 \\
-\frac{1}{2}g'v_d & \frac{1}{2}gv_d & 0 & -\mu & -\frac{\lambda_{333}v_u}{\sqrt{2}} & Q\ud{N}{d}g_1'v_d &
0 & -\frac{\lambda_{33\beta}s}{\sqrt{2}} \\
\frac{1}{2}g'v_u & -\frac{1}{2}gv_u & -\mu & 0 & -\frac{\lambda_{333}v_d}{\sqrt{2}} & Q\ud{N}{u}g_1'v_u &
-\frac{\lambda_{3\alpha 3}s}{\sqrt{2}} & 0 \\
0 & 0 & -\frac{\lambda_{333}v_u}{\sqrt{2}} & -\frac{\lambda_{333}v_d}{\sqrt{2}} & 0 & Q\ud{N}{S}g_1's &
-\frac{\lambda_{3\alpha 3}v_u}{\sqrt{2}} & -\frac{\lambda_{33\beta}v_d}{\sqrt{2}} \\
0 & 0 & Q\ud{N}{d}g_1'v_d & Q\ud{N}{u}g_1'v_u & Q\ud{N}{S}g_1's & M\ud{\prime}{1} & 0 & 0 \\\hline
0 & 0 & 0 & -\frac{\lambda_{3\alpha 3}s}{\sqrt{2}} & -\frac{\lambda_{3\alpha 3}v_u}{\sqrt{2}} & 0 &
0 & -\frac{\lambda_{3\alpha\beta}s}{\sqrt{2}} \\
0 & 0 & -\frac{\lambda_{33\beta}s}{\sqrt{2}} & 0 & -\frac{\lambda_{33\beta}v_d}{\sqrt{2}} & 0 &
-\frac{\lambda_{3\alpha\beta}s}{\sqrt{2}} & 0
\ea\right),\nn\\\label{eq:nmtx}
\ee
where once again $Q\ud{N}{d,u,S}$ are the $U(1)\dz{N}$ charges of down-type Higgsinos,
up-type Higgsinos, and singlinos respectively, as given in table~\ref{tab:eSSMCharges}.
Typically $g\ud{\prime}{1} \approx g\dvp{\prime}{1}$
all the way down to the low energy scale. If the soft gaugino masses are unified
at the GUT scale then we also have $M\ud{\prime}{1} \approx M\dvp{\prime}{1}
\approx M\dvp{\prime}{2}/2$ (see subsection~\ref{sub:smallGauginoMixing}).

The Yukawa couplings in the off-diagonal blocks, marked out by lines,
are suppressed under the
approximate $\dd{Z}_2^H$. Given the smallness of these couplings, the
inert neutralinos in the bottom-right block are pseudo-Dirac states with
an approximately decoupled mass matrix
\be
-\frac{s}{\sqrt{2}}\left(\ba{cccc}
& & \lambda_{322} & \lambda_{321} \\
& & \lambda_{312} & \lambda_{311} \\
\lambda_{322} & \lambda_{312} & & \\
\lambda_{321} & \lambda_{311} & &
\ea\right)
& \mbox{in the basis} & \left(\ba{cccc}
\tilde{H}^0_{d2} & \tilde{H}^0_{d1} & \tilde{H}^0_{u2} & \tilde{H}^0_{u1}
\ea\right)^{\rr{T}}. \nn
\ee
They are approximately degenerate with the two inert chargino Dirac states.

The top-left block is the USSM neutralino mass matrix (\ref{eq:USSM}) and
contains the states of the MSSM supplemented by the
third generation singlino and the bino$'$.
In the case where $M\dvp{\prime}{1} \approx
M\ud{\prime}{1}$ is small the lightest neutralino mass state will be mostly bino. The
bino$'$ will mix with the third generation singlino giving two mixed states
with masses around $Q\ud{N}{S}g_1's$.
As $M_1 \approx M_1'$ increases, the bino mass will increase relative to both the
third generation Higgsino mass $\mu$ and the inert Higgsino masses given
approximately by the biunitary diagonalisation of
\be
-\f{1}{\sqrt{2}}\lambda_{3\alpha\beta}s.\nn
\ee
At the same time the state mostly
containing the third generation singlino will have a decreasing mass as $M_1'$
increases relative to $Q\ud{N}{S}g_1's$.

\section{Dark Matter in the c$\EZ$SSM}
\label{ezssm:dm}

As discussed previously, in section~\ref{ref:e6g&m},
there is an automatically
conserved $R$-parity under which
the charginos, neutralinos, inert singlinos $\tilde{\sigma}$, and
exotic $\tilde{\bar{D}}$ and $\tilde{D}$ fermions,
along with the squarks and sleptons, are all $R$-parity odd,
\ie~all of the fermions
other than the quarks and leptons are $R$-parity odd.
We shall assume that the
lightest neutralino $\tilde{N}_1$ is the lightest of all of the
$R$-parity odd states
excluding the massless inert singlinos $\tilde{\sigma}$.
However, $\tilde{N}_1$ cannot decay into $\tilde{\sigma}$ via neutralino mixing
since the inert singlinos are decoupled from the neutralino mass matrix.
Furthermore, the potential
decay $\tilde{N}_1 \rightarrow \tilde{\sigma}\sigma$,
allowed by the $\sigma$-$\tilde{\sigma}$-$\tilde{B}'$ supersymmetric $U(1)_N$
gauge coupling,
is forbidden if $\tilde{N}_1$ is lighter than the inert SM-singlet scalars $\sigma$.
In fact, in this case no kinematically viable final states exist that have
the same quantum
numbers as $\tilde{N}_1$. Therefore $\tilde{N}_1$ is absolutely
stable and in the scenario presented is the DMP.
The lightest inert SM-singlet scalar is not stable.
There are no Yukawa couplings
involving $S_\alpha$, but the inert SM-singlet scalars
are able to decay via 
the $\sigma$-$\tilde{\sigma}$-$\tilde{B}'$ supersymmetric $U(1)_N$
gauge coupling.

In the successful dark matter scenario presented in this section $\tilde{N}_1$
is predominantly bino, with at least one of the
two pairs of pseudo-Dirac inert Higgsinos
expected to be close in mass, but somewhat heavier,
in order to achieve the correct relic density.
This is due to a novel scenario in which 
the DMP is approximately the bino and inelastically scatters off of
SM matter into heavier inert Higgsinos during the time of thermal freeze-out,
keeping it in equilibrium long enough to give a successful relic density. 
In this section we discuss in detail how this novel scenario
comes about in this model.

\subsection{The dark matter calculation}

In the considered model the DMP
is not the lightest $R$-parity odd state~--- an inert singlino~--- but
the lightest neutralino $\tilde{N}_1$.
We would like to use (\ref{eq:dotni}) to describe the evolution of $R$-parity
odd states other than the inert singlinos~--- generically $\tilde{\chi}$.
In this case we should also include
in (\ref{eq:dotni}) processes involving $\sigma$ and $\tilde{\sigma}$
particles that change the number of $\tilde{\chi}$ particles by one.
Since such processes necessarily
involve inert SM-singlet scalars $\sigma$,
it is valid to neglect these processes
in the case where these inert SM-singlet bosons have frozen out long before the
freeze-out of dark matter. We will call this \textbf{condition~B}, to go along
with \textbf{condition~A}, defined in subsection~\ref{dm:be}, and it should be satisfied
given our assumption that the inert SM-singlet scalar mass eigenstates are heavier than the DMP,
since they only interact via the heavy $Z'$ boson.
Assuming that both \textbf{conditions~A and~B} are satisfied we can use (\ref{eq:dotn2})
to describe the evolution of the number density $n$ of $R$-parity-odd states
other than inert singlinos $\tilde{\chi}$.
The value of $n$ after the thermal freeze-out of $\tilde{N}_1$ depends on
annihilation cross-sections involving $\tilde{N}_1$ and other $R$-parity odd
states close by in mass and
$n$ will eventually be equal to the number density
of DMPs after other $\tilde{\chi}$ states have decayed to $\tilde{N}_1$.

\subsection{The c$\EZ$SSM dark matter scenario}

In order to carry out the dark matter analysis in the constrained version of
the model we have extended the RG code used for the study in
ref.~\cite{Athron2009}~\cite{cE6SSMCode} to include
the Yukawa parameters and soft masses of the inert sector of the
$\EZ$SSM. The inputs are $\kappa_{3ij}$ and $\lambda_{333}$ at the GUT scale,
$\lambda_{3\alpha\beta}$ at the EWSB scale, $s$, and $\tan(\beta)$,
as well as the known low energy Yukawa couplings and gauge couplings. Given
these inputs and the RGEs the algorithm attempts to find points
with GUT scale unified soft masses $m_0$, $M_{1/2}$, and $A_0$. The low energy
$U(1)_N$ gauge coupling $g_1'$ is set by requiring it to be equal to
the other gauge couplings at the GUT scale, which is calculated.

For consistent points in the $\E$SSM
the lightest non-inert (USSM sector)
supersymmetric particle is typically bino dominated.
For the c$\EZ$SSM we find the same thing. The masses of the inert Higgsino states
depend on $s$ and on the Yukawa couplings $\lambda_{3\alpha\beta}$ and
in the c$\EZ$SSM the lightest neutralino can be either the bino dominated
state or a pseudo-Dirac inert Higgsino dominated state. In the latter case
we find that the pseudo-Dirac inert Higgsino DMPs coannihilate with
full-weak-strength interactions and lead to a too small
dark matter relic density.
In the former case the bino DMP normally annihilates too weakly and yields
a too large dark matter relic density. If, however, there are inert Higgsino
states close by in mass, they contribute significantly to
$\langle\sigma v\rangle$, allowing for the observed amount of dark matter.
This relies on \textbf{condition~A} being satisfied, \ie~the binos being
up-scattered into inert Higgsinos at a large enough rate.

Such points with an appropriate dark matter relic density can be found and three
are presented in section~\ref{ezssm:benchmarks}.
\textbf{Condition~B} is satisfied since
the inert SM-singlet scalars are so much heavier than the DMP and
the $Z_2$ boson mass is so large compared to the regular $Z$ boson mass.
Annihilation and scattering processes involving inert SM-singlets and singlinos
must contain a virtual $Z'$ boson, which is predominantly the $Z_2$ mass eigenstate.

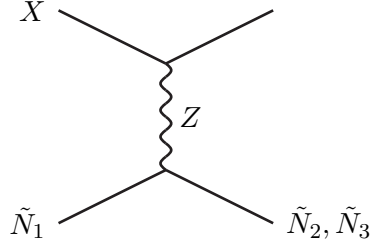
\begin{figure}
\begin{center}
\begin{picture}(120,100)(-60,-60)
\SetWidth{1}
\Line(-40,40)(0,20)
\Line(0,20)(40,40)
\Photon(0,20)(0,-20){2}{4}
\Line(-40,-40)(0,-20)
\Line(0,-20)(40,-40)
\Text(-45,40)[r]{$X$}
\Text(-45,-40)[r]{$\tilde{N}_1$}
\Text(45,-40)[l]{$\tilde{N}_{2},\tilde{N}_{3}$}
\Text(5,0)[l]{$Z$}
\end{picture}
\end{center}
\caption{The form of diagrams for the up-scattering of the bino dominated
DMP $\tilde{N}_1$ off of SM particles $X$ into the pseudo-Dirac inert Higgsino
states $\tilde{N}_2$ and $\tilde{N}_3$.\label{fig:upscattering}}
\end{figure}

To test \textbf{condition~A} let us compare the rate for binos up-scattering into
inert Higgsinos with the inert Higgsino coannihilation rate. We shall label
the mostly bino state $\tilde{N}_1$ and the lightest pseudo-Dirac inert Higgsino
states $\tilde{N}_2$ and $\tilde{N}_3$. The dominant up-scattering diagrams are
of the form shown in figure~\ref{fig:upscattering}.

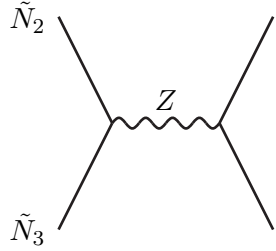
\begin{figure}
\begin{center}
\begin{picture}(120,100)(-60,-50)
\SetWidth{1}
\Line(-40,40)(-20,0)
\Line(-20,0)(-40,-40)
\Photon(-20,0)(20,0){2}{4}
\Line(40,40)(20,0)
\Line(20,0)(40,-40)
\Text(-45,40)[r]{$\tilde{N}_2$}
\Text(-45,-40)[r]{$\tilde{N}_3$}
\Text(0,5)[b]{$Z$}
\end{picture}
\end{center}
\caption{Full-weak-strength coannihilations of the pseudo-Dirac inert Higgsino states
$\tilde{N}_2$ and $\tilde{N}_3$.\label{fig:HiggsinoCo}}
\end{figure}

As in section~\ref{ref:novelInert} we again define $R_{Zij}$ couplings such that the
$Z$-$\tilde{N}_i$-$\tilde{N}_j$ coupling is equal to $R_{Zij}$ times
the $Z$-$\nu$-$\nu$ coupling. In the $\EZ$SSM we can write
\be
R_{Zij} &=& \sum_{D=3,7,9}N_i^DN_j^D - \sum_{U=4,8,10}N_i^UN_j^U,
\ee
where $N_i^a$ is the neutralino mixing matrix element corresponding to
mass eigenstate $i$ and interaction state $a$. $D$ and $U$ index the
down- and up-type Higgsino interaction states respectively. For the
pseudo-Dirac inert Higgsino states we have $m_3 \approx -m_2$ and $R_{Z23} \approx 1$,
allowing for full-weak-strength coannihilations of the form shown in figure~\ref{fig:HiggsinoCo}.

Using the notation from (\ref{eq:dotni}),
the ratio of the rate for the mostly bino state up-scattering into
the lightest mostly inert Higgsino state to the inert Higgsino coannihilation
rate is given approximately by
\be
\Upsilon &=& \frac{\langle\sigma\ud{\prime}{X12}v\dvp{\prime}{1X}\rangle n_1^\eq n_X^\eq}
{\langle\sigma\dvp{\prime}{23}v\dvp{\prime}{23}\rangle n_2^\eq n_3^\eq}.
\ee
To give an idea of the size of this ratio, if the SM particle $X$ is
relativistic and $m_1 \sim m_2 \approx m_3$ then
\be
\Upsilon &\sim& \left(\frac{R_{Z12}}{R_{Z23}}\right)^2 \frac{T^3}{(|m_1|T)^{3/2}\exp(-|m_1|/T)} \nn\\\nn\\
&\approx& R_{Z12}^2x^{3/2}e^{-x},
\ee
where again
\be
x = \f{T}{|m_1|}.\nn
\ee
This ratio is expected to be large because of the overwhelming abundance
of the relativistic SM particle $X$, but it also depends on $R_{Z12}$.
The value of $R_{Z12}$ depends
on the $\ZZ{H}$-breaking couplings that mix the upper-left block of the
neutralino mass matrix in (\ref{eq:nmtx})~--- the USSM states
including the bino~--- with the
inert Higgsino states in the lower-right block. Since this symmetry is
not exact we expect these couplings to be large enough such that we can
still assume $\Upsilon \gg 1$. Explicit examples of this parameter
are included in table~\ref{tab:bm2} in section~\ref{ezssm:benchmarks}.

With \textbf{conditions~A and~B} satisfied we use \texttt{micrOMEGAs}~\cite{Belanger2010}
to calculate
the dark matter relic density for low energy spectra consistent with the
GUT scale constrained scenario.
The observed relic density of dark matter can
arise in this model and examples are shown in table~\ref{tab:bm2} in section~\ref{ezssm:benchmarks}.
The most critical factor is the mass splitting between the bino and the lightest inert
Higgsinos. Too large and there are not enough inert Higgsinos remaining at
the time of the bino's thermal freeze-out to have a significant enough effect.
Too small and $\langle\sigma v\rangle$ is dominated by inert Higgsino
coannihilations, leading to a too small dark matter relic density.

Since in this scenario the DMP is predominantly bino, the spin-independent
DMP-nucleon cross-section $\sigma_{\rr{SI}}$ is not expected to be in the range
that
direct detection experiments are currently sensitive too. The spin-independent
cross-section of a pure bino is suppressed by the squark masses,
but is also sensitive
to the squark mixing angles~\cite{Choi2001}. For each flavour
the cross-section vanishes for zero squark mixing.
Since in practice the DMP will also have non-zero, but small,
active Higgsino components, there are also contributions to $\sigma_{\rr{SI}}$
from t-channel active Higgs scalar exchange via the bino-Higgs-Higgsino
supersymmetric gauge coupling. These contributions, though in fact dominant,
are quite small,
due to the overwhelming bino nature of the DMP. Estimates of
$\sigma_{\rr{SI}}$, using the same proton $f_d$, $f_u$, and $f_s$ parameters
used in the study in ref.~\cite{Gogoladze2011},
are included in table~\ref{tab:bm2} in section~\ref{ezssm:benchmarks}.

\section{The Inert Singlinos and their Contribution to the Effective Number of
Neutrinos prior to BBN}
\label{ezssm:Neff}

In the standard theory of BBN, which happens long after
the thermal freeze-out of dark matter,
the resultant primordial abundances of
the light elements depend on two parameters~--- the effective number of
neutrinos contributing to the expansion rate of the radiation dominated
universe $\Neff$ and the baryon-to-photon ratio $\eta$.

Whilst the
primordial abundance of $^4$He is not the most sensitive measure of $\eta$,
it is much more sensitive to $\Neff$ than the other light element abundances.
This is because prior to nucleosynthesis, when the equilibrium photon temperature
is of order
0.1~MeV, the number of neutrons remaining, virtually all of which are
subsequently incorporated into $^4$He nuclei, is sensitive to the expansion rate
of the universe, which depends on $\Neff$. The greater the expansion rate, the
less time there is for charged current weak interactions to convert neutrons
into protons.

The analysis in ref.~\cite{Izotov2010},
using the more recent neutron lifetime measurement
from ref.~\cite{Serebrov2008},
gives $\Neff = 3.80^{+0.80}_{-0.70}$ at 2-sigma, implying a more-than-2-sigma
tension between the measured $^4$He abundance and the Standard Model prediction
for $\Neff$ of about 3. Although in ref.~\cite{Aver2010} it is suggested
that these errors may be larger,
similar results are also obtained for the effective number of neutrinos
contributing to the expansion rate of the
universe from fits to WMAP data~\cite{Komatsu2011}.

In the $\EZ$SSM the two massless inert singlinos would have decoupled from
equilibrium at an earlier time than the light neutrinos, but nevertheless
would have contributed to $\Neff$. Exactly when the inert singlinos would have
decoupled from equilibrium with the photon depends on
the mass of the $Z_2$ boson which determines the strength of an effective
Fermi-like 4-point interaction vertex that would have been responsible for
keeping the inert singlinos in equilibrium. The various values for $\Neff$
that can be achieved in this model all fit the data better than the SM value.

The implications of extra neutrino-like particles present in the early
universe have long been studied and the methods used
in following analysis rely on relatively simple physics (see for example ref.~\cite{Steigman1979}).

The effective number of degrees of freedom contributing to the expansion rate
of the universe during the run-up to nucleosynthesis is defined by
\be
g_\eff^0 &=& g_\gamma + \nicefrac{7}{8}g_\nu \Neff(\nicefrac{4}{11})^{4/3}\label{eq:411}\nn\\
&=& 2 + \nicefrac{7}{4}\Neff(\nicefrac{4}{11})^{4/3},
\ee
where $g_\eff^0$ is the value of $g\dvp{0}{\eff}$, as defined in (\ref{eq:geff}),
immediately prior to nucleosynthesis.
Here $g_\gamma = 2$ is the number of degrees of freedom of the photon and
$g_\nu = 2$ is the number of degrees of freedom of a light neutrino.
The three SM neutrinos are expected to decouple from equilibrium with
the photon at a temperature above the electron mass whereas nucleosynthesis
does not happen until the temperature is below the electron mass. When the
photon/electron temperature is around the electron mass the electrons
and positrons
effectively disappear from the universe\footnote{A much smaller number of electrons
remains due to the small lepton number asymmetry.}.
Their disappearance heats the photons
to a higher temperature then they would otherwise have had, but the neutrinos,
having already decoupled, would continue to cool at the full rate dictated
by Hubble expansion. Because of the neutrinos' lower temperature
at nucleosynthesis they would then contribute less to $g_\eff^0$ per degree of freedom.
In (\ref{eq:411}) $\Neff$ is defined such that in the SM $\Neff = 3$, for the three
neutrinos decoupling above the electron mass, as we shall see. Extra
particles, such as the $\EZ$SSM inert singlinos, decoupling above the muon
mass would have had even lower temperatures at the time of nucleosynthesis and
would therefore contribute to $g_\eff^0$ even less than light neutrinos
per degree of freedom.

\subsection{The calculation of $\Neff$}

In the c$\EZ$SSM there is a typical scenario in which the massless inert singlinos
$\tilde\sigma$ decouple at a temperature above the colour transition temperature
(when the effective degrees of freedom are quarks and gluons
rather than mesons) and above the strange quark mass,
but below the charm quark mass. This has to do with the strength of the
interactions that keep the inert singlinos in equilibrium which depend heavily
on the mass of the $Z_2$ boson.
If the inert singlinos do decouple in this range then this leads to a definite
prediction for $\Neff$.
We shall explain why the inert singlinos
typically
decouple in this temperature range in the following subsection.
For now we derive the value of $\Neff$ in this scenario as an example.

We shall use the superscript $0$ to denote quantities at some temperature $T^0$
below the electron mass and the superscript $e$ to denote quantities at some
temperature $T^e$ above the electron mass and where all light neutrino
species are still in equilibrium. We shall use the superscript $s$ to denote
quantities at some still higher temperature $T^s$ above the colour transition
and the strange quark mass and where the inert singlinos are still in
equilibrium.

At $T^s$ the effective number of degrees of degrees of freedom contributing to
the expansion rate is
\be
g_\eff^s &=& g_\gamma + g_g + \nicefrac{7}{8}(g_e + g_\mu + g_u + g_d + g_s + 3g_\nu + 2g_{\tilde{\sigma}})\nn\\
&=& 2+16+\nicefrac{7}{8}(4+4+12+12+12+6+4) = 65\nicefrac{1}{4}
\ee
and at $T^e$ it becomes
\be
g_\eff^e &=& 2 + \nicefrac{7}{8}
\left(6+4\left(\frac{T_{\tilde{\sigma}}^e}{T^e}\right)^4\right)
\ee
and at $T^0$ it becomes
\be
g_\eff^0 &=& 2 + \nicefrac{7}{8}
\left(6\left(\frac{T_\nu^0}{T^0}\right)^4+4\left(\frac{T_{\tilde{\sigma}}^0}{T^0}\right)^4\right),
\ee
taking into account that the neutrinos and inert singlinos now have different temperatures.
With no subscript $T$ always refers to the photon temperature, as is the notation
throughout this thesis.

From (\ref{eq:sden}), the entropy within a given volume $V$ due to a
relativistic \{boson, fermion\} with
number of degrees of freedom $g_i$ is given by
\be
S_i &=& \{1,\nicefrac{7}{8}\}g_i\frac{2\pi^2}{45}(T_i)^3V.
\ee
Since we are assuming that the inert singlinos decouple
before the strange quark threshold, in going from $T^s$ to $T^e$
we conserve the entropy in the comoving volume separately for the
inert singlinos and for everything else. Specifically, for the inert singlinos
\be
(T\uvp{s}{\tilde{\sigma}})^3V\uvp{s}{\tilde{\sigma}} &=& (T\ud{e}{\tilde{\sigma}})^3V\uvp{e}{\tilde{\sigma}}
\ee
and for everything else
\be
[g_\gamma + g_g + \nicefrac{7}{8}(g_e + g_\mu + g_u + g_d + g_s + 3g_\nu)](T^s)^3V^s
&=& [g_\gamma + \nicefrac{7}{8}(g_e + 3g_\nu)](T^e)^3V^e \nn\\\nn\\
\Rightarrow\quad 61\nicefrac{3}{4}(T^s)^3V^s &=& 10\nicefrac{3}{4}(T^e)^3V^e.
\label{eq:6134}
\ee
This allows us to write
\be
\frac{(T^s)^3V^s}{(T^e)^3V^e} = \left(\frac{T\ud{e}{{\tilde{\sigma}}}}{T\uvp{e}{\tilde{\sigma}}}\right)^3
&=& \frac{10\nicefrac{3}{4}}{61\nicefrac{3}{4}} = \frac{43}{247}.
\ee
In going from $T^e$ to $T^0$ we conserve the entropy separately for the
neutrinos, for the inert singlinos again, and for everything else, giving
\be
[g_\gamma + \nicefrac{7}{8}g_e](T^e)^3V^e &=& g_\gamma (T^0)^3V^0,\\\nn\\
(T^e)^3V^e &=& (T^0_\nu)^3V^0,\quad\mbox{and}\\\nn\\
(T_{\tilde{\sigma}}^e)^3V^e &=& (T^0_{\tilde{\sigma}})^3V^0.
\ee
This gives us
\be
\left(\frac{T^0_{\nu}}{T^0}\right)^3 &=& \frac{g_\gamma}{g_\gamma + \nicefrac{7}{8}g^e}
= \frac{4}{11}\quad\mbox{and}\\\nn\\
\left(\frac{T^0_{\tilde{\sigma}}}{T^0}\right)^3 &=& \frac{43}{247}\frac{g_\gamma}{g_\gamma + \nicefrac{7}{8}g_e}
= \frac{43}{247}\frac{4}{11}.
\ee
In this case the effective number of neutrinos contributing to the expansion
rate prior to nucleosynthesis (at $T^0$) is then
\be
\Neff &=& 3 + 2\left(\frac{43}{247}\right)^{4/3} \approx 3.194.
\ee

\subsection{The inert singlino decoupling temperature}

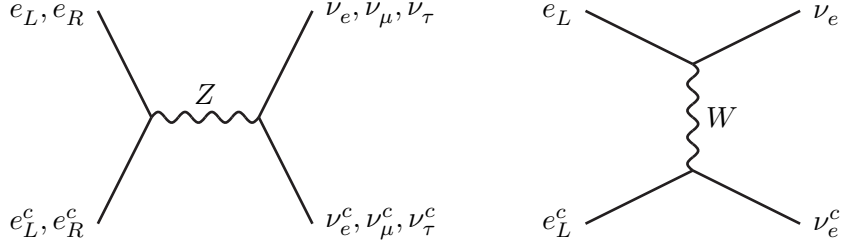
\begin{figure}
\begin{center}
\begin{picture}(150,100)(-75,-50)
\SetWidth{1}
\Line(-40,40)(-20,0)
\Line(-20,0)(-40,-40)
\Photon(-20,0)(20,0){2}{4}
\Line(40,40)(20,0)
\Line(20,0)(40,-40)
\Text(-45,40)[r]{$e\dvp{c}{L},e\dvp{c}{R}$}
\Text(-45,-40)[r]{$e\ud{c}{L},e\ud{c}{R}$}
\Text(0,5)[b]{$Z$}
\Text(45,40)[l]{$\nu\dvp{c}e,\nu\dvp{c}\mu,\nu\dvp{c}\tau$}
\Text(45,-40)[l]{$\nu\ud{c}e,\nu\ud{c}\mu,\nu\ud{c}\tau$}
\end{picture}
\qquad\qquad
\begin{picture}(120,100)(-60,-50)
\SetWidth{1}
\Line(-40,40)(0,20)
\Line(0,20)(40,40)
\Photon(0,20)(0,-20){2}{4}
\Line(-40,-40)(0,-20)
\Line(0,-20)(40,-40)
\Text(-45,40)[r]{$e\dvp{c}L$}
\Text(-45,-40)[r]{$e\ud{c}L$}
\Text(45,40)[l]{$\nu\dvp{c}e$}
\Text(45,-40)[l]{$\nu\ud{c}e$}
\Text(5,0)[l]{$W$}
\end{picture}
\end{center}
\caption{Electroweak interactions responsible for keeping the light neutrinos
in equilibrium in the early universe. For all the light neutrinos there are the
the processes on the left. For the electron neutrinos there is also the
additional process on the right.\label{fig:neudec}}
\end{figure}

The light neutrinos are kept in equilibrium via their electroweak interactions.
The relevant diagrams are shown in figure~\ref{fig:neudec}.

We express the cross-section for processes relevant for keeping
muon and $\tau$ neutrinos in equilibrium as
\be
\langle\sigma_{\nu_\mu,\nu_\tau}v\rangle &=& k_2\frac{T^2}{m_Z^4}\frac{(\nicefrac{5}{3})^2g_1^4}{\sin^4(\vartheta_W)}X^4,
\ee
where $k_2$, like $k_1$ from (\ref{eq:k1}), is a constant defined for convenience and
\be
X^4 &=& \left(\f{1}{2}\left(-\f{1}{2}+\sin^2(\vartheta_W)\right)\right)^2
+ \f{\sin^4(\vartheta_W)}{4} \approx 0.031.
\ee
Note that we are using the GUT normalised $U(1)_Y$ gauge coupling and so
\be
\frac{g_2}{\cos(\vartheta_W)} &=& \sqrt{\frac{5}{3}}\frac{g_1}{\sin(\vartheta_W)}.
\ee
The cross-section for electron neutrinos with their extra diagram is then
\be
\langle\sigma_{\nu_e}v\rangle &=& k_2\frac{T^2}{m_Z^4}\frac{(\nicefrac{5}{3})^2g_1^4}{\sin^4(\vartheta_W)}Y^4,
\ee
where
\be
Y^4 &=& \left(\f{1}{2}\left(\f{1}{2}+\sin^2(\vartheta_W)\right)\right)^2
+ \f{\sin^4(\vartheta_W)}{4} \approx 0.147.
\ee
We express the number densities of all Weyl fermions still in equilibrium with
the photon as
\be
n_{e_L} = n_{e_R} = n_{\mu_L} = n_{\mu_R} = n_{\nu_e} = n_{\nu_\mu} = n_{\nu_\tau} &=& k_3T^3
\ee
and the expansion rate is given by
\be
H &=& k_1\sqrt{g^e_\eff}T^2.
\ee
The neutrino decoupling temperature $T^{\nu}$ can then be approximated by
\be
\langle\sigma_\nu v\rangle n^\nu &=& H \nn\\\nn\\
\Rightarrow\quad (T^{\nu_\mu,\nu_\tau})^3 &=&
K\sqrt{g_e^\eff}m_Z^4\frac{\sin^4(\vartheta_W)}{(\nicefrac{5}{3})^2g_1^4}\frac{1}{X^4}\quad\mbox{and}\\\nn\\
(T^{\nu_e})^3 &=& K\sqrt{g_e^\eff}m_Z^4\frac{\sin^4(\vartheta_W)}{(\nicefrac{5}{3})^2g_1^4}\frac{1}{Y^4},
\ee
with $K = k_1/k_2k_3$. A more detailed calculation finds that in the SM
(with only neutrinos, electrons, and photons contributing to $g^e_\eff$)
$T^{\nu_\mu,\nu_\tau} \approx 3.7$ MeV and
$T^{\nu_e} \approx 2.4$ MeV~--- the muon and $\tau$ neutrinos decoupling earlier.

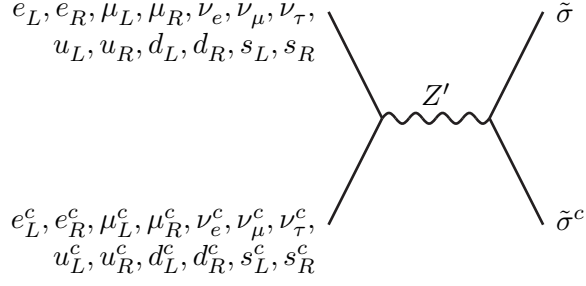
\begin{figure}
\begin{center}
\begin{picture}(120,120)(-60,-70)
\SetWidth{1}
\Line(-40,40)(-20,0)
\Line(-20,0)(-40,-40)
\Photon(-20,0)(20,0){2}{4}
\Line(40,40)(20,0)
\Line(20,0)(40,-40)
\Text(-45,40)[r]{$e\dvp{c}L,e\dvp{c}R,\mu\dvp{c}L,\mu\dvp{c}R,\nu\dvp{c}e,\nu\dvp{c}\mu,\nu\dvp{c}\tau,$}
\Text(-45,27)[r]{$u\dvp{c}L,u\dvp{c}R,d\dvp{c}L,d\dvp{c}R,s\dvp{c}L,s\dvp{c}R$}
\Text(-45,-40)[r]{$e\ud{c}L,e\ud{c}R,\mu\ud{c}L,\mu\ud{c}R,\nu\ud{c}e,\nu\ud{c}\mu,\nu\ud{c}\tau,$}
\Text(-45,-53)[r]{$u\ud{c}L,u\ud{c}R,d\ud{c}L,d\ud{c}R,s\ud{c}L,s\ud{c}R$}
\Text(0,5)[b]{$Z'$}
\Text(45,40)[l]{$\tilde{\sigma}$}
\Text(45,-40)[l]{$\tilde{\sigma}\uvp{c}{L}$}
\end{picture}
\end{center}
\caption{Interaction processes responsible for keeping the inert singlinos
in equilibrium in the early universe.\label{fig:sindec}}
\end{figure}

At temperatures above the strange quark mass the processes relevant for keeping
the inert singlinos in equilibrium are shown in figure~\ref{fig:sindec}.

The part of the Lagrangian containing all of the fermion couplings in figure~\ref{fig:sindec},
illustrating the relevant $U(1)_N$ charges, is
\be
-\left(\begin{array}{cccccc}
L\ud{\dagger}L & e\ud{c\dagger}R & Q\ud{\dagger}L & u\ud{c\dagger}R & d\ud{c\dagger}R & \tilde{\sigma}\uvp{\dagger}{L}
\end{array}\right)i\bar{\sigma}^\mu g\ud{\prime}{1}Z\ud{\prime}{\mu}\frac{1}{\sqrt{40}}\left(\begin{array}{l}
(2)L\dvp{c}L \\ (1)e\ud{c}R \\ (1)Q\dvp{c}L \\ (1)u\ud{c}R \\ (2)d\ud{c}R \\ (5)\tilde{\sigma}
\end{array}\right)
\ee
and the total cross-section taking into account all of these diagrams is then
(neglecting the small $Z$-$Z'$ mixing)
\be
\langle\sigma_{\tilde{\sigma}}v\rangle &=& k_2\frac{T^2}{m_{Z_2}^4}2g_1^4\frac{Z^4}{(40)^2},
\ee
where
\be
Z^4 &=& (5)^2[2(2)^2 + 2(1)^2 + 3(1)^2 + 3(1)^2 + 6(1)^2 + 6(2)^2 + 3(2)^2]\nn\\
&&\quad= 1450,
\ee
leading to an approximate singlino decoupling temperature of
\be
(T^{\tilde{\sigma}})^3 &=& K\sqrt{g^s_\eff}m_{Z_2}^4\frac{1}{g_1^4}\frac{(40)^2}{Z^4}\\\nn\\
\Rightarrow\quad \left(\frac{T^{\tilde{\sigma}}}{T^{\nu_e}}\right)^3
&=& \sqrt{\frac{g^s_\eff}{g^e_\eff}}\left(\frac{m_{Z_2}}{m_Z}\right)^4
\frac{(40)^2(\nicefrac{5}{3})^2}{\sin^4(\vartheta_W)}\frac{Y^4}{Z^4}.
\ee
The only unknown variable here affecting the inert singlino decoupling
temperature is then the $Z_2$ boson mass $m_{Z_2}$. Rearranging we find
\be
m_{Z_2} &\approx& m_Z \left(\frac{T^{\tilde{\sigma}}}{6.60 \mbox{~MeV}}\right)^{3/4}.
\ee

\subsection{$\Neff$ in the $\EZ$SSM}

We now check which values of $m_{Z_2}$ are consistent with our assumption
that the inert singlinos decouple at a temperature between the strange and
charm quark masses. For $T^{\tilde{\sigma}} < m_c$ we find that we require
$m_{Z_2} < 4700$~GeV. For
$m_{Z_2} \sim 1000$~GeV the situation is slightly more complicated. Firstly
the temperature of the QCD phase transition is not accurately known and
secondly the effective number of degrees of freedom is decreased by so much
after the QCD phase transition that even if the inert singlinos were decoupled
beforehand the universe may be expanding slowly enough afterwards that
they could come back into equilibrium. After checking a range of scenarios
we find that for $1300\mbox{~GeV}\lesssim m_{Z_2} < 4700$~GeV our
value of $\Neff = 3.194$ is valid. For $m_{Z_2} \lesssim 950$~GeV the inert
singlinos decouple at a temperature above the muon mass, but below the pion
mass, leading to a larger prediction of $\Neff = 4.373$.
At the time of the publication of \textbf{paper~II} the experimental limit in the $\EZ$SSM was
$m_{Z_2} > 892$~GeV, from ref.~\cite{Accomando2011},
which would allow a $Z_2$ boson
light enough for us to predict a value for $\Neff = 4.373$.
For $Z_2$ masses in between these ranges the value of $\Neff$
depends on the details of the QCD phase transition, but is somewhere between
these predictions. For inert singlinos decoupling above the pion mass,
but after the
QCD phase transition, we have $\Neff = 4.065$. All of these values are within
the 2-sigma measured range $\Neff = 3.80^{+0.80}_{-0.70}$ and closer to the
central value than the SM result $\Neff = 3$.

Since the publication of \textbf{paper~III}
the limit on the $Z_2$ mass in the $\EZ$SSM has increased to around 1350~GeV as discussed in subsection~\ref{sub:ZZ}.
This leads to a concrete prediction of $\Neff = 3.194$ assuming that inert singlinos decouple
at a temperature below the charm quark mass, \ie~$m_{Z_2} < 4700$~GeV.

\section{Benchmark Points}
\label{ezssm:benchmarks}

In the tables~\ref{tab:bm1}, \ref{tab:bm2}, and~\ref{tab:bm3}
we present three benchmark points in the c$\EZ$SSM.
For all three points we fix $\lambda_{322}=0.1$ and $\lambda_{321} =
\lambda_{312} = 0.0001$ at the EWSB scale. For the $\ZZ{H}$-breaking
couplings we also fix $\lambda_{332} = \lambda_{323} = 0.012$ and
$\lambda_{331} = \lambda_{313} = 0.005$ at the EWSB scale. At the GUT scale
we fix $\kappa_{333} = \kappa_{322} = \kappa_{311}$ and $\kappa_{3ij} = 0$ for
$i \ne j$.
The lightest (SM-like) Higgs mass is calculated to second loop order.

\begin{table}
\begin{center}\begin{tabular}{r|ccc|}
\hline
\textbf{Benchmark} & $\bb{1}$ & $\bb{2}$ & $\bb{3}$ \\\hline
$\tan(\beta)$ & 30 & 10 & 3 \\
$s$ [TeV] & 5 & 4.4 & 5.5 \\
$\lambda_{333}$ @ GUT scale & -0.3 & -0.37 & -0.4 \\
$\lambda_{322}$ @ EWSB scale & 0.1 & 0.1 & 0.1 \\
$\lambda_{311}$ @ EWSB scale & 0.0293 & 0.0403 & 0.0399 \\
$\kappa_{3ii}$ @ GUT scale & 0.18 & 0.18 & 0.23 \\\hline
$M_{1/2}$ [GeV] & 590 & 725 & 908 \\
$m_0$ [GeV] & 1533 & 454 & 1037 \\
$A_0$ [GeV] & 1375 & 1002 & 413 \\\hline
\end{tabular}\end{center}
\caption{The input parameters of the three c$\EZ$SSM benchmark points.\label{tab:bm1}}\end{table}

\begin{table}
\begin{center}\begin{tabular}{r|ccc|}
\hline
\textbf{Benchmark} & $\bb{1}$ & $\bb{2}$ & $\bb{3}$ \\\hline
$\mu$ [GeV] & -1086.7 & -1189.5 & -1405.5 \\
$\lambda_{322}s/\sqrt{2}$ [GeV] & 353.55 & 331.13 & 388.91 \\
$\lambda_{311}s/\sqrt{2}$ [GeV] & 103.59 & 125.38 & 155.17 \\\hline
$\tilde{N}_1$ mass [GeV] & 94.07 & 114.49 & 143.50 \\
$\tilde{N}_2$ mass [GeV] & -105.12 & -126.45 & -156.57 \\
$\tilde{N}_3$ mass [GeV] & 105.14 & 126.47 & 156.62 \\
$\tilde{N}_4$ mass [GeV] & 167.05 & 203.19 & 255.47 \\
$\tilde{N}_5$ mass [GeV] & -353.77 & -311.29 & -389.12 \\
$\tilde{N}_6$ mass [GeV] & 353.78 & 311.30 & 389.13 \\
$\tilde{N}_7$ mass [GeV] & -1092.5 & -1194.5 & 1409.6 \\
$\tilde{N}_8$ mass [GeV] & 1093.3 & 1194.8 & -1411.2 \\
$\tilde{N}_9$ mass [GeV] & -1803.2 & -1572.3 & -1964.7 \\
$\tilde{N}_{10}$ mass [GeV] & 1899.7 & 1688.7 & 2109.9 \\\hline
$\tilde{C}_1$ mass [GeV] & 105.04 & 126.41 & 156.52 \\
$\tilde{C}_2$ mass [GeV] & 167.05 & 203.19 & 255.46 \\
$\tilde{C}_3$ mass [GeV] & 353.78 & 311.30 & 389.13 \\
$\tilde{C}_4$ mass [GeV] & -1094.4 & -1196.1 & -1411.3 \\\hline
$m_{Z'}$ [GeV] & 1850.4 & 1628.4 & 2035.4 \\
$\Neff$ & 3.194 & 3.194 & 3.194 \\
$\Omega_{\chi}h^2$ & 0.112 & 0.107 & 0.102 \\
$\Upsilon$ & $1.1 \times 10^8$ & $2.3 \times 10^8$ & $2.3 \times 10^8$ \\
$\sigma_{\rr{SI}}$ [cm$^2$] & $4.9 \times 10^{-48}$ & $2.5\times 10^{-48}$ & $1.2 \times 10^{-48}$ \\\hline
\end{tabular}\end{center}
\caption{The low energy neutralino and chargino masses and associated parameters
of the three benchmark points.
The DMP is the lightest neutralino $\tilde{N}_1$ which is predominantly bino in nature.
There is a nearby
pair of inert neutral Higgsinos $\tilde{N}_2$ and $\tilde{N}_3$ and a chargino $\tilde{C}_1$
into which $\tilde{N}_1$ inelastically scatters
during freeze-out, resulting in a relic density consistent with observation.
The predicted values of $m_{Z_2}$ and $\Neff$ are also shown,
as is the spin-independent $\tilde{N}_1$-nucleon direct detection cross-section $\sigma_{\rr{SI}}$.
\label{tab:bm2}}\end{table}

\begin{table}
\begin{center}\begin{tabular}{r|ccc|}
\hline
\textbf{Benchmark} & $\bb{1}$ & $\bb{2}$ & $\bb{3}$ \\\hline
$h_1$ mass [GeV] & 122.2 & 114.6 & 115.3 \\
$h_2$ mass [GeV] & 1145 & 987.1 & 1522 \\
$h_3$ mass [GeV] & 1890 & 1664 & 2080 \\
$H^\pm$ mass [GeV] & 2106 & 1396 & 1675 \\
$A^0$ mass [GeV] & 2103 & 1393 & 1673 \\\hline
$m_{S_2},m_{S_1}$ [GeV] & 1547 & 518 & 1084 \\
$m_{H_{d2}},m_{H_{d1}}$ [GeV] & 1567 & 611 & 1156 \\
$m_{H_{u2}},m_{H_{u1}}$ [GeV] & 1561 & 599 & 1146 \\\hline
$m_{\tilde{D}_3}$ [GeV] & 1483 & 503 & 1794 \\
$m_{\tilde{D}_2},m_{\tilde{D}_1}$ [GeV] & 1443 & 493 & 1775 \\
$m_{\tilde{\bar{D}}_3}$ [GeV] & 2864 & 2321 & 3065 \\
$m_{\tilde{\bar{D}}_2},m_{\tilde{\bar{D}}_1}$ [GeV] & 2840 & 2318 & 3052 \\\hline
$m_{\tilde{t}_1}$ [GeV] & 1122 & 625.3 & 1110 \\
$m_{\tilde{c}_1}, m_{\tilde{u}_1}$ [GeV] & 1817 & 1774 & 1707 \\
$m_{\tilde{t}_2}$ [GeV] & 1470 & 1069 & 1546 \\
$m_{\tilde{c}_2}, m_{\tilde{u}_2}$ [GeV] & 1838 & 1224 & 1761 \\
$m_{\tilde{b}_1}$ [GeV] & 1434 & 1009 & 1512 \\
$m_{\tilde{s}_1}, m_{\tilde{d}_1}$ [GeV] & 1840 & 1226 & 1763 \\
$m_{\tilde{b}_2}$ [GeV] & 1748 & 1265 & 1818 \\
$m_{\tilde{s}_2}, m_{\tilde{d}_2}$ [GeV] & 1907 & 1278 & 1820 \\
$m_{\tilde{\tau}_1}$ [GeV] & 1500 & 718.8 & 1259 \\
$m_{\tilde{\mu}_1}, m_{\tilde{e}_1}$ [GeV] & 1655 & 731.3 & 1261 \\
$m_{\tilde{\tau}_2}$ [GeV] & 1708 & 949.2 & 1473 \\
$m_{\tilde{\mu}_2}, m_{\tilde{e}_2}$ [GeV] & 1775 & 952.8 & 1474 \\
$m_{\tilde{\nu}_\tau}$ [GeV] & 1705 & 945.6 & 1472 \\
$m_{\tilde{\nu}_\mu}, m_{\tilde{\nu}_e}$ [GeV] & 1774 & 949.5 & 1472 \\\hline
$m_{\tilde{g}}$ [GeV] & 541.3 & 626.9 & 787.7 \\\hline
\end{tabular}\end{center}
\caption{The remaining particle spectrum of the three benchmark points.\label{tab:bm3}}\end{table}

We have chosen three points with quite different values of $\tan(\beta)$~--- 30, 10, and 3.
This illustrates the fact that $\tan(\beta)$ can be quite low in this model since
the SM-like Higgs mass is not constrained to be less than $m_Z|\cos(2\beta)|$
at tree level as it is in the MSSM.

The mass of the bino DMP $\tilde{N}_1$ is not directly constrained to be above above 100~GeV.
However, the lightest pseudo-Dirac inert Higgsino neutralinos $\tilde{N}_2$ and $\tilde{N}_3$ are almost
degenerate with the lightest inert Higgsino chargino $\tilde{C}_1$ and therefore these are constrained to
heavier than 100~GeV in order to be consistent with LEP constraints~\cite{Kraan2005}. Furthermore, the
thermal relic DM
scenario outlined in section~\ref{ezssm:dm} requires $\tilde{N}_2$ and $\tilde{N}_3$ not to be
too much more massive than $\tilde{N}_1$. In practice the $\tilde{N}_1$ is predominantly bino and its
mass cannot be much less
than 100~GeV. In \textbf{benchmark~1}, for example, it is 94~GeV.

Requiring such values for the low energy bino mass $M_1$ and requiring consistent
EWSB in practice means that the SM-singlet VEV $s$ cannot be
too low. This in turn means that the $Z_2$ mass is always
more than about 1.5~TeV, automatically satisfying the most recent experimental lower limit.
In these benchmarks from the constrained scenario the
effective number of neutrinos contributing to the expansion rate of the universe prior to
BBN $\Neff$ therefore takes on the lower value calculated in section~\ref{ezssm:Neff} of around 3.2.
This is more consistent with data than the SM prediction.

In all benchmark points $\tilde{N}_4$ and $\tilde{C}_2$ are predominantly wino. $\tilde{N}_5$,
$\tilde{N}_6$, and $\tilde{C}_3$ are predominantly made up of
the remaining inert Higgsinos states, with masses around $\lambda_{322}s/\sqrt{2}$, whereas
$\tilde{N}_7$, $\tilde{N}_8$ and $\tilde{C}_4$ are predominantly made up of
the active Higgsino states, with masses around $\mu$. $\tilde{N}_9$ and $\tilde{N}_{10}$
are mostly superpositions of the active singlino and bino$'$.

The fact that $\Upsilon \gg 1$ indicates that the inert Higgsino components in the
predominantly bino state $\tilde{N}_1$, though small, are large enough such that
processes involving $\tilde{N}_1$ up-scattering off of a SM particle into $\tilde{N}_2$
happen overwhelmingly more often than neutralino annihilation and coannihilation processes. In this
way the ratios of the number densities of these particles are able to maintain their
equilibrium values.

The spin-independent
DMP-nucleon cross-section $\sigma_{\rr{SI}}$, as estimated using the results
in ref.~\cite{Choi2001}, is quite small for these benchmarks
and is not currently detectable by direct detection experiments.
This is due to the predominantly bino nature of the DMP as well as the large squark masses.

\section{Warm Inert Singlino Dark Matter in the $\E$SSM}
\label{ref:wdm}

A model of dark matter is inconsistent if it predicts a DMP mass that is so light
that the observed structure of the universe would have been erased. Such light
dark matter is known as hot dark matter. Although such hot dark matter is inconsistent with
observation, the dark matter also does not need to be cold, \ie~of a mass such that
it was non-relativistic at freeze-out, to be consistent with current observations~\cite{Viel2005,Boyarsky2009}.
The intermediate scenario is known as warm dark matter.
Limits on WDM from WMAP and Lyman-$\alpha$ forest data require the mass of a warm thermal
relic particle responsible for all of the observed dark matter to be greater than 550~eV~\cite{Viel2005}.
This is of the order of various other keV scale lower bounds on WDM particles~\cite{Boyarsky2009}.

In the $\E$SSM, if the $\ZZ{S}$ symmetry was only approximate then the only stable supersymmetric
particle or particles would be either the lightest of or both of the two light, predominantly inert singlino
states. In this case the inert singlinos could form WDM. It is already known that if all of the observed dark matter
is made up of gravitino WDM decoupling above GeV temperatures with $g\uz{\eff} \sim 100$ then the gravitino
mass would have to be around 100~eV, contradicting the above limit~\cite{Viel2005}. For inert singlinos
decoupling at a temperature $T\uz{s}$ between the strange and charm quark masses the singlinos would have undergone
even less entropy dilution than such gravitinos and their masses would need to be even smaller in order for the
observed dark matter relic density to be predicted.

The number density of a single species of \{boson, fermion\} with temperature
$T\dvp{3}{i}$ is proportional to $s\dvp{3}{i} \propto T\ud{3}{i}$
\be
n\dvp{3}{i} &=& g\dvp{3}{i}\{1,\nf{3}{4}\}\f{\zeta(3)}{\pi\uz{2}}T\ud{3}{i}.
\ee
The number density of all three neutrino species today
\be
3n\ud{0}{\nu} &=& 3\f{3}{2}\f{\zeta(3)}{\pi\uz{2}}\f{4}{11}(T\uvp{0}{\gamma})\uvp{3}{\gamma} = \f{9}{11}n\ud{0}{\gamma}\nn\\
\nn\\
\Rightarrow\quad s\dz{\gamma} + 3s\dz{\nu} &=& \f{20}{11}s\dz{\gamma}.
\ee
Conserving entropy between inert singlino freeze-out and today then gives
\be
n\ud{0}{\tilde{\sigma}}V\uvp{0}{\gamma} &=& n\ud{s}{\tilde{\sigma}}V\uvp{s}{\gamma}
\ee
for inert singlinos and
\be
\Big(n\ud{0}{\gamma}+3n\ud{0}{\nu}\Big)V\uvp{0}{\gamma}
= \f{20}{11}n\ud{0}{\gamma}V\uvp{0}{\gamma} &=& \Big(n\uvp{s}{\gamma} - 2n\ud{s}{\tilde{\sigma}}\Big)V\uvp{s}{\gamma}
\ee
for everything else which means that today
\be
\f{n\ud{0}{\tilde{\sigma}}}{n\ud{0}{\gamma}} &=& \f{20}{11}\f{n\ud{s}{\tilde{\sigma}}}{n\uvp{s}{\gamma} - 2n\ud{s}{\tilde{\sigma}}}
\ee
if the inert singlinos were relativistic at freeze-out.

For the case of two stable inert singlinos with masses $m\dz{1}$ and $m\dz{2}$ the relic density today will be given by
\be
\Omega\dz{\tilde{\sigma}}h\uz{2} &=& \Big(n^0_1m\dvp{0}1 + n^0_2m\dvp{0}2\Big)\f{h\uz{2}}{\rhoCrit}
= \f{n^0_\gamma h^2}{\rhoCrit}\f{n^0_1m\dvp{0}{1}+n^0_2m\dvp{0}{2}}{n^0_\gamma}.
\ee
If $\Omega\dz{\tilde{\sigma}} = \Omega\dz{\rr{DM}}$ and $n^0_1=n^0_2=n^0_{\tilde{\sigma}}$ as derived above
then this can be rearranged to give
\be
m_1+m_2 &=& \Omega\dz{\rr{DM}}h\uz{2}\f{\rhoCrit}{h\uz{2}}\f{1}{n\ud{0}{\gamma}}\f{11}{20}\f{n\uvp{s}{\gamma} - 2n\ud{s}{\tilde{\sigma}}}{n\ud{s}{\tilde{\sigma}}}.
\ee
Using $\rhoCrit = 1.05\times 10\uz{4}h^2$~eVcm$^{-3}$, $n\ud{0}{\gamma} = 410.5$~cm$^{-3}$~\cite{ParticleDataGroupCollaboration2010},
and
\be
\f{n\uvp{s}{\gamma} - 2n\ud{s}{\tilde{\sigma}}}{n\ud{s}{\tilde{\sigma}}}
&=& \f{g\dvp{s}{\gamma} + g\dvp{s}{g} + \nf{3}{4}(g\dvp{s}{e} + g\dvp{s}{\mu} + g\dvp{s}{u} + g\dvp{s}{d} + g\dvp{s}{s} + 3g\dvp{s}{\nu})}
{\vphantom{n\ud{s}{\tilde{\sigma}}}\nf{3}{2}}
= \f{111}{\vphantom{n\ud{s}{\tilde{\sigma}}}3}
\ee
gives
\be
m_1+m_2 &=& 57\mbox{~eV}
\ee
in contradiction with data.

Therefore thermal WDM inert singlinos, like thermal WDM gravitinos, cannot be responsible all of the observed dark matter.
WDM inert singlinos with larger masses could only be responsible for all of the observed dark matter is there
were a significant source of entropy dilution reheating the SM matter, but not reheating the inert singlinos, after the
time of inert singlino freeze-out. Such entropy dilution would lower the number density of inert singlinos today
relative to the known CMB photon number density. Thermal WDM decoupling at $g\uz{\eff} \sim 1000$ could also lead to a successful
WDM scenario, but such a situation is well beyond the framework of the $\E$SSM, requiring, for inert singlino WDM, a much more massive
$U(1)_N$ $Z'$ boson and, more importantly, the existence of many new degrees of freedom, beyond those of the $\E$SSM,
at some high temperature.

However, the $\E$SSM with an approximate $\ZZ{S}$ symmetry does allow for another type of scenario, apart from
having lightest inert neutralinos with masses of order half of the $Z$ boson mass, in which the supersymmetric particles of the $\E$SSM
are responsible for less than the observed dark matter relic density. Such scenarios are consistent with, even if they
do not explain, cosmological observations. If the $\ZZ{S}$ symmetry was only approximate and WDM inert singlinos had
masses significantly less than 57~eV then these inert singlinos would be the only stable supersymmetric particles
and would contribute less than the observed amount of dark matter.

\section{Summary and Conclusions}
\label{ezssm:conclusions}

The difficulty in making the predominantly inert singlinos states predicted by the $\E$SSM
much heavier than 60~GeV makes them
natural dark matter candidates, but has also led to a very tightly constrained scenario
in which inert neutralino LSP dark matter
is now severely challenged by the most recent XENON100 analysis of 100.9
days of data. Furthermore
we have not been able to show that such a scenario could be
consistent with having universal (GUT scale constrained) soft mass parameters.

In this work we discussed a new variant of the $\E$SSM
called the $\EZ$SSM
that involves a novel scenario for dark matter in which 
the DMP is predominantly the bino with a mass 
close to or above 100~GeV which is fully consistent with XENON100 data.
A successful relic density is achieved via
its inelastic up-scattering into nearby heavier inert Higgsinos during the time of thermal freeze-out.
The model
also predicts two massless inert singlinos which contribute to the
effective number of neutrino species at the time of BBN, depending
on the mass of the $Z_2$ boson which keeps them in equilibrium. For 
$m_{Z_2} > 1300$~GeV we find $N_\eff \approx 3.2$.

We presented a few benchmark points
in the c$\EZ$SSM to illustrate this new scenario.
The benchmark points show that it is easy to find consistent points that satisfy the correct
relic abundance as well as all other phenomenological constraints. The points also show that 
the typical $Z_2$ mass is expected to be around 2~TeV, with the gluino having a mass
around 500--800~GeV and squarks and sleptons typically having masses around 1--2~TeV.
The DMP-nucleon spin-independent direct detection cross-sections
are well below current sensitivities.

Although very light inert singlinos in the $\E$SSM provide a candidate for WDM,
in order to account for all of the observed dark matter thermal WDM inert singlinos
would need to be too light~--- lighter than would be consistent with other cosmological observations.
Inert singlino WDM contributing less than the observed dark matter relic density would, however, provide
another scenario in which the $\E$SSM predicts less than the observed amount of dark matter and
is consistent with all observations.

\cleardoublepage

\newpage
\chapter{Summary and Conclusions}
\label{chap:conclusions}

In chapter~\ref{chap:ndmin} the first study of the inert neutralino sector of the
$\E$SSM is presented. It was found that in the $\E$SSM the dark matter naturally arises
from this approximately decoupled sector. The inert neutralino dark matter scenario was studied both
analytically and numerically. It was found that in order for the inert neutralino LSP
not to be too light and singlino dominated, leading to too large a dark matter relic
density, certain trilinear Higgs Yukawa couplings relevant to the inert sector
should be large and the ratio of Higgs VEVs $\tan(\beta)$ should be relatively close to unity.
If the LSP mass is allowed to increase to around half of the $Z$ boson mass
then the LSP also contains larger inert Higgsino components and can annihilate
more efficiently in the early universe, leading to a reduced dark matter relic density.
Imposing that the LSP has a mass greater that half of the $Z$ boson mass, to avoid potential
conflict with LEP data, and accounts for all of the observed dark matter implies that
$\tan(\beta)$ should be less than about 2, depending on the sizes of various Yukawa couplings
that one is willing to allow. The inert neutralino dark matter scenario relies
mostly on parameters that only affect the inert sector physics. As a result the parameter
space of the MSSM-like sector of the $\E$SSM is less constrained compared that of the MSSM since
in the $\E$SSM these parameters are not constrained from dark matter considerations.
The exception is $\tan(\beta)$ which strongly affects the LSP mass, with the
LSP mass being approximately proportional to $\sin(2\beta)$.

In chapter~\ref{chap:nhde} a more in-depth study of the inert neutralino and chargino sectors of the $\E$SSM,
with a particular focus on physics
relating to the Higgs boson, is presented.
The condition that Yukawa couplings remain perturbative up to the GUT scale
is imposed and the LSP and NLSP masses cannot then be made greater than about 60~GeV.
Scenarios where the LSP and NLSP masses are around half of the $Z$ boson mass are found
that produce less than or equal to the observed amount of dark matter.
It is found that inert neutralino masses below half of the $Z$ boson mass can be
consistent with LEP data provided that $\tan(\beta)$ is not too large.
In plausible scenarios consistent with observations from both cosmology and LEP
it is found that the couplings of the lightest
inert neutralinos to the SM-like Higgs boson are always rather large.
This means that the SM-like Higgs boson has a large branching ratio into invisible
final states and this has major implications for
Higgs boson collider phenomenology. The branching ratio into SM particles is reduced
to around 2--4\%.
It also leads to large spin-independent LSP-nucleon cross-sections
and because of this scenarios in which $\E$SSM inert neutralino LSPs account for all of the observed dark matter are
now severely challenged by recent dark matter direct detection experiment analyses.

In chapter~\ref{chap:ezssm} a new variant of the $\E$SSM called the $\EZ$SSM is presented in which the dark matter scenario
is very different to the inert neutralino CDM scenario and in which the presence of
supersymmetric massless states in the early universe modifies the expansion rate of the universe prior to BBN.
In the dark matter scenario the DMP is the bino and a successful relic density is achieved via its inelastic
up-scattering into nearby heavier inert Higgsinos during the time of thermal freeze-out. The nearby pair of inert Higgsino neutralinos
form a pseudo-Dirac pair with masses approximately equal the corresponding inert charged Higgsino Dirac mass and cannot have masses
below about 100~GeV. In the $\EZ$SSM the two inert singlino states are exactly massless and contribute to
the effective number of neutrino species at the time of BBN,
depending on the mass of the $Z_2$ boson which keeps them in equilibrium. For
$m_{Z_2} > 1300$~GeV we find $\Neff \approx 3.2$. The dark matter scenario is consistent
with having universal (GUT scale constrained) soft mass parameters and the DMP-nucleon spin-independent direct
detection cross-sections are well below current sensitivities.

In the $\E$SSM light inert singlinos contribute too much CDM
if they are non-relativistic at freeze-out~--- more than the observed dark matter relic density.
However, in section~\ref{ref:wdm} we showed if the inert singlinos
have masses less than around 50~eV then they will contribute WDM less than
the observed dark matter relic density.

In the future it would be interesting to study more theoretical aspects of
the $\E$SSM and $\EZ$SSM such as how much fine-tuning these models involve
and what the effects are of potential non-renormalisable terms in the
superpotential. At the same time, now that the LHC is taking data it is
important to study the collider phenomenological predictions of these models.
Gluino cascade decays in which the gluino sequentially decays
into the DMP, giving off pairs of fermions at each stage, is the subject of a
further paper currently in preparation~\cite{gluinoDecays}.
In this paper we try to identify how the $\E$SSM could be
distinguished from the MSSM at the LHC.

\cleardoublepage

\appendix 
\newpage
\chapter{Weyl, Majorana, and Dirac Spinors in $3+1$ Dimensions}
\label{ap:spinors}

We represent the Dirac gamma matrices in the Weyl basis
\be
\gamma^\mu &=& \left(\ba{cc}0 & \sigma^\mu\\\bar{\sigma}^\mu & 0\ea\right),
\ee
where $\sigma^0 = \bar{\sigma}^0 = 1$ and $\bar{\sigma}^i = -\sigma^i$ and write a
general Dirac spinor
\be
\Psi &=& \left(\ba{c}\psi_L\\\psi_R\ea\right),
\ee
with $\psi_L$ a LH Weyl spinor and $\psi_R$ a RH Weyl spinor.
We write a general infinitesimal Lorentz transformation on a Dirac spinor $\Psi$ as
\be
\Lambda_{1/2}(\uu{\rr{d}\vartheta},\uu{\rr{d}\beta})\left(\ba{c}\psi_L\\\psi_R\ea\right)
&=& \left(\ba{c}
\Big(1 + \nf{1}{2}i\uu{\rr{d}\vartheta}.\uu{\sigma} + \nf{1}{2}\uu{\rr{d}\beta}.\uu{\sigma}\Big)\psi_L\\
\Big(1 + \nf{1}{2}i\uu{\rr{d}\vartheta}.\uu{\sigma} - \nf{1}{2}\uu{\rr{d}\beta}.\uu{\sigma}\Big)\psi_R\ea\right).
\ee
Using the mathematical identity
\be
\sigma^2\uu{\sigma}^* = -\uu{\sigma}\sigma^2\label{eq:identity}
\ee
we can see that $\sigma^2\psi^*_L$ transforms as RH spinor and $\sigma^2\psi^*_R$
transforms as a LH spinor
\be
\Lambda_{1/2}(\uu{\rr{d}\vartheta},\uu{\rr{d}\beta})\sigma^2\psi^*_L
&=& \sigma^2\Big(\Big(1 + \nf{1}{2}i\uu{\rr{d}\vartheta}.\uu{\sigma} + \nf{1}{2}\uu{\rr{d}\beta}.\uu{\sigma}\Big)\psi_L\Big)^*\nn\\
&=& \Big(1 + \nf{1}{2}i\uu{\rr{d}\vartheta}.\uu{\sigma} - \nf{1}{2}\uu{\rr{d}\beta}.\uu{\sigma}\Big)\sigma^2\psi^*_L\quad\mbox{and}\nn\\\nn\\
\Lambda_{1/2}(\uu{\rr{d}\vartheta},\uu{\rr{d}\beta})\sigma^2\psi^*_R
&=& \sigma^2\Big(\Big(1 + \nf{1}{2}i\uu{\rr{d}\vartheta}.\uu{\sigma} - \nf{1}{2}\uu{\rr{d}\beta}.\uu{\sigma}\Big)\psi_R\Big)^*\nn\\
&=& \Big(1 + \nf{1}{2}i\uu{\rr{d}\vartheta}.\uu{\sigma} + \nf{1}{2}\uu{\rr{d}\beta}.\uu{\sigma}\Big)\sigma^2\psi^*_R.
\ee
We therefore define the charge conjugation operation acting on a Dirac spinor $\Psi$ to be
\be
\Psi^c &=& \left(\ba{c}\omega\dvp{*}R\sigma^2\psi^*_R\\\omega\dvp{*}L\sigma^2\psi^*_L\ea\right).
\ee
Since $-\sigma^2\sigma^{2*} = 1$, applying the charge conjugation operation twice yields
$\Psi^{cc}=\Psi$ as long as we define $\omega\dvp{*}R\omega^*_L = \omega\dvp{*}L\omega^*_R = -1$.
We define $\omega_R = -\omega_L = -\omega$ implying that $|\omega|^2=1$.
We define the $CP$ conjugation operation acting on a LH Weyl spinor so that
the RH spinor
\be
\psi^c_L = \omega\sigma^2\psi^*_L,
\ee
and on a RH Weyl spinor so that the LH spinor
\be
\psi^c_R = -\omega\sigma^2\psi^*_R.
\ee

The gauge and Lorentz invariant part of the Lagrangian
for a massive Dirac spinor with mass $m$
\be
\cc{L}\dvp{\dagger}{D} &=& \Psi\ud{\dagger}{D}\gamma^0\Big(i\gamma^\mu\cc{D}_\mu-m\Big)\Psi\dvp{\dagger}{D}\\
&=& \psi^\dagger_Li\bar{\sigma}^\mu\cc{D}_\mu\psi\dvp{\dagger}{L}
+ \psi^\dagger_Ri{\sigma}^\mu\cc{D}_\mu\psi\dvp{\dagger}{R}
- m\psi\ud{\dagger}{R}\psi\dvp{\dagger}{L} - m\psi\ud{\dagger}{L}\psi\dvp{\dagger}{R}.
\ee
This may be rewritten in terms of the two LH Weyl spinors $\psi\dvp{c}L$ and $\psi\ud{c}{R}$ as
\be
\cc{L}\dvp{\dagger}{D} &=& \psi^\dagger_Li\bar{\sigma}^\mu\cc{D}_\mu\psi\dvp{\dagger}{L}
+ \psi^{c\dagger}_Ri\bar{\sigma}^\mu\cc{D}_\mu\psi\ud{\vphantom{\dagger}c}{R}
- \Big(m\psi\ud{cc\dagger}{R}\psi\dvp{\dagger}{L} +\mbox{c.c.}\Big)\label{eq:sdvsdfvbd}
\ee
up to a total derivative, revealing that a Dirac spinor with mass $m$ is formed from two Weyl spinors
of the same handedness with a mass matrix
\be
\left(\ba{cc}0 & m\\m & 0\ea\right).\label{eq:mmtrsdc}
\ee

The covariant derivative acting on $\psi\ud{c}{R}$ in (\ref{eq:sdvsdfvbd}) is the complex conjugate of
covariant derivative acting on $\psi\dvp{c}{R}$ so that if $\psi\dvp{c}{R}$ is in some
representation $r$ of some gauge group such that
\be
\cc{D}\dvp{a}\mu\psi\dvp{c}{R} &=& \Big(\partial\dvp{a}\mu - igA\ud{a}\mu T\ud{a}{r}\Big)\psi\dvp{c}{R}
\ee
then
\be
\cc{D}\dvp{a}\mu\psi\ud{c}{R} &=& \Big(\partial\dvp{a}\mu + igA\ud{a}\mu T\ud{a*}{r}\Big)\psi\ud{c}{R}\nn\\\nn\\
&=& \Big(\partial\dvp{a}\mu - igA\ud{a}\mu T\ud{a}{\bar{r}}\Big)\psi\ud{c}{R}
\ee
and $\psi\ud{c}{R}$ is in the conjugate representation $\bar{r}$.

The Lagrangian for a single LH Weyl spinor $\psi$ with mass $m$
\be
\cc{L}_{M} &=& \psi^\dagger i\bar{\sigma}^\mu\cc{D}_\mu\psi - \f{m}{2}\Big(\psi^{c\dagger}\psi + \mbox{c.c.}\Big)
\ee
may be written in terms of a Majorana spinor
\be
\Psi_M &=& \left(\ba{c}\psi\\\omega\sigma^2\psi\ea\right)
\ee
as
\be
\cc{L}\dvp{\dagger}{M} &=& \nf{1}{2}\Psi\ud{\dagger}{M}\gamma^0\Big(i\gamma^\mu\cc{D}_\mu-m\Big)\Psi\dvp{\dagger}{M}.
\ee
The Majorana spinor is nothing but a Dirac spinor that is self-charge-conjugate.

The Weyl or Majorana mass matrix for a Dirac particle (\ref{eq:mmtrsdc}) is diagonalised to
\be
\left(\ba{cc}-m & 0\\0 & m\ea\right)
\ee
in the Weyl or Majorana mass eigenstate basis. Conversely, if two mass eigenstate Majorana spinors
have equal and opposite masses and all other quantum numbers equal then together they form a Dirac spinor.
If the masses of two such Majorana spinors are opposite, but not quite equal then together they are
said to form a pseudo-Dirac state.

\cleardoublepage

\newpage
\chapter{The Pseudoreality of the Spinor Representation of $SU(2)$}
\label{ap:pseudoreality}

Let the field $\varphi$ form the representation $(2,r)$ under the gauge group $SU(2)\otimes G$
so that an infinitesimal gauge transformation acting on $\varphi$ can be written
\be
\Xi(\rr{d}\alpha,\rr{d}\beta)\varphi &=& \Big(1 - i\rr{d}\alpha^a\tau^a - i\rr{d}\beta^bT^b_r\Big)\varphi,
\ee
where $\tau^a = \sigma^a/2$ are the generators of 2 and $T^a$ are the generators of $r$.
We also define a field $\bar{\varphi}$ that is in the representation $(\overline{2},\bar{r})$ so that
\be
\Xi(\rr{d}\alpha,\rr{d}\beta)\bar{\varphi} &=& \Big(1 + i\rr{d}\alpha^a\tau^{a*} + i\rr{d}\beta^bT^{b*}_r\Big)\bar{\varphi}.
\ee

The field $\bar{\varphi}$ may be redefined as the equivalent field $2\omega\tau^2\bar{\varphi}$.
Using (\ref{eq:identity}) again, this time in the form $\tau^2\tau^{a*} = -\tau^a\tau^2$, we see that this
field transforms with
\be
\Xi(\rr{d}\alpha,\rr{d}\beta)2\omega\tau^2\bar{\varphi}
&=& 2\omega\tau^2\Big(1 + i\rr{d}\alpha^a\tau^{a*} + i\rr{d}\beta^bT^{b*}_r\Big)\bar{\varphi}\nn\\
&=& \Big(1 - i\rr{d}\alpha^a\tau^{a} + i\rr{d}\beta^bT^{b*}_r\Big)2\omega\tau^2\bar{\varphi},
\ee
meaning that that it is in the representation $(2,\bar{r})$. We have somehow
managed to redefine the field so that it transforms in the 2 rather than $\overline{2}$ representation
of $SU(2)$. Therefore, even though the spinor representation is not real, the 2 and $\overline{2}$
representations are somehow equivalent. This representation is sometimes called pseudoreal.

Antidoublet representations of $SU(2)$ can always be redefined to be doublet representations.
If two fields are in the doublet representation of $SU(2)$, a gauge invariant bilinear may
be formed by transforming one of the two fields such that it is in the antidoublet representation.
We thus define the gauge invariant product of two doublet representations of
$SU(2)$
\be
\left(\ba{c}\uparrow\dz{1} \\ \downarrow\dz{1}\ea\right).
\left(\ba{c}\uparrow\dz{2} \\ \downarrow\dz{2}\ea\right)
&=&
\left(\ba{cc}\uparrow\dz{1} & \downarrow\dz{1}\ea\right)i\sigma^2
\left(\ba{c}\uparrow\dz{2} \\ \downarrow\dz{2}\ea\right)\nn\\\nn\\
&=& \downarrow\dz{1}\uparrow\dz{2} - \uparrow\dz{1}\downarrow\dz{2}.\label{eq:pseudoreality}
\ee

\cleardoublepage

\cleardoublepage
\addcontentsline{toc}{chapter}{\numberline{}Bibliography}
\bibliographystyle{JHEP-2}
\bibliography{library}

\end{document}